\documentclass{SciPost}

\hypersetup{
    colorlinks,
    linkcolor={red!50!black},
    citecolor={blue!50!black},
    urlcolor={blue!80!black}
}

\usepackage[bitstream-charter]{mathdesign}
\DeclareSymbolFont{usualmathcal}{OMS}{cmsy}{m}{n}
\DeclareSymbolFontAlphabet{\mathcal}{usualmathcal}

\fancypagestyle{SPstyle}{
\fancyhf{}
\lhead{\colorbox{scipostblue}{\bf \color{white} ~SciPost Physics }}
\rhead{{\bf \color{scipostdeepblue} ~Submission }}

\fancyfoot[C]{\textbf{\thepage}}
}

\usepackage{graphicx}
\usepackage{subcaption}

\def\ee{\end{equation}}
\def\eea{\end{eqnarray}}
\def\be{\begin{equation}}
\def\bea{\begin{eqnarray}}
\begin{document}

\pagestyle{SPstyle}

\begin{center}{\Large \textbf{\color{scipostdeepblue}{
Massive scalar clouds and black hole spacetimes in Gauss-Bonnet gravity\\
}}}\end{center}

\begin{center}\textbf{
Iris van Gemeren\textsuperscript{1$\star$},
Tanja Hinderer\textsuperscript{2$\dagger$} and
Stefan Vandoren\textsuperscript{3$\ddagger$}
}\end{center}

\begin{center}
{\bf 1} Institute for Theoretical Physics,
Utrecht University, Princetonplein 5, 3584 CC Utrecht, The Netherlands
\\[\baselineskip]
$\star$ \href{mailto:email1}{\small i.r.vangemeren@uu.nl}\,,\quad
$\dagger$ \href{mailto:email2}{\small t.p.hinderer@uu.nl}\,,\quad
$\ddagger$ \href{mailto:email2}{\small s.j.g.vandoren@uu.nl}
\end{center}

\section*{\color{scipostdeepblue}{Abstract}}
\boldmath\textbf{%
We study static black holes in scalar-Gauss-Bonnet (sGB) gravity with a massive scalar field as an example of higher curvature gravity.  
The scalar mass introduces an additional scale and leads to a strong suppression of the scalar field beyond its Compton wavelength.
We numerically compute sGB black hole spacetimes and scalar configurations and also compare with perturbative results for small couplings, where we focus on a dilatonic coupling function. 
We analyze the constraints on the parameters from requiring the curvature singularity to be located inside the black hole horizon $r_h$ and the relation to the regularity condition for the scalar field. For scalar field masses $m r_h \gtrsim 10^{-1}$, this leads to a new and currently most stringent bound on sGB coupling constant $\alpha$ of $\alpha/r_h^2\sim10^{-1}$ in the context of stellar mass black holes. Lastly, we look at several properties of the black hole configurations relevant for further work on observational consequences, including the scalar monopole charge, Arnowitt–Deser–Misner mass,  curvature invariants and the frequencies of the innermost stable circular orbit and light ring.    
}

\vspace{\baselineskip}

\noindent\textcolor{white!90!black}{%
\fbox{\parbox{0.975\linewidth}{%
\textcolor{white!40!black}{\begin{tabular}{lr}%
  \begin{minipage}{0.6\textwidth}%
    {\small Copyright attribution to authors. \newline
    This work is a submission to SciPost Physics. \newline
    License information to appear upon publication. \newline
    Publication information to appear upon publication.}
  \end{minipage} & \begin{minipage}{0.4\textwidth}
    {\small Received Date \newline Accepted Date \newline Published Date}%
  \end{minipage}
\end{tabular}}
}}
}


\makeatletter
\def\l@subsubsection#1#2{}
\makeatother
\vspace{10pt}
\noindent\rule{\textwidth}{1pt}
\tableofcontents
\noindent\rule{\textwidth}{1pt}
\vspace{10pt}


\section{Introduction}
 General Relativity (GR) as the theory of gravity has passed all empirical tests to date~\cite{Will:2014kxa,Turyshev:2008dr,Berti:2015itd}. Yet modern theoretical developments suggest that modifications of Einstein's gravity are required at some level. This has motivated a significant research effort in high-energy physics to develop a theory of quantum gravity. However, modifications to GR may already arise at intermediate, lower energy scales than the full quantum-gravity regimes. Such modifications have been constrained by high-precision tests of gravity in tabletop experiments~\cite{Adelberger:2003zx}, the solar system~\cite{Will:2014kxa}, and binary pulsars~\cite{Wex:2014nva}. However, the genuinely nonlinear regimes of gravity remain largely unexplored and have only recently started to become accessible to measurements, for instance, with gravitational waves~\cite{LIGOScientific:2016lio,LIGOScientific:2019fpa,LIGOScientific:2020tif}. This opens new opportunities to test modified theories where corrections to GR only become relevant in high-curvature regimes. One such family of theories is scalar-Gauss-Bonnet (sGB) gravity, where the gravitational action of GR is augmented by adding a quadratic-in-curvature contribution involving the topological Gauss-Bonnet invariant dynamically coupled with a scalar field. Because of the topological nature of the higher curvature term, the theory is ghost-free and the equations of motion are still second order in the fields~\cite{Nojiri:2018ouv} and thus a dynamical system whose mathematical well-posedness was proved in~\cite{Kovacs:2020pns,Kovacs:2020ywu,R:2022tqa}. The sGB form of the gravitational action also has motivations from the low energy limit of quantum gravity paradigms~\cite{Zwiebach:1985uq,Gross:1986mw,Boulware:1985wk}. \\ 
In this paper, we focus on consequences of sGB gravity for static spherical symmetric black holes when including a nonvanishing scalar field mass. 
Black holes are clean testbeds for precision tests of higher-curvature gravity as they are devoid of any matter and solely involve curved spacetime. In GR, black holes are conjectured to have 'no-hair': their exterior spacetime can be entirely described by only three parameters: their mass, spin, and electromagnetic charge~\cite{Israel:1967wq,Israel:1967za, Carter:1971zc,Wald:1971iw,Bekenstein:1971hc}. This also implies that black holes cannot be dressed with any nontrivial scalar, vector, or spinor fields~\cite{Chase, Teitelboim:1972ps, Bekenstein:1972ny, Bekenstein:1971hc,Hartle:1971qq}, even when considering more complex potentials for the fields~\cite{Bekenstein:1995un}. The no-hair property of black holes also extends to several classes of modified gravity theories
 such as Brans-Dicke theories~\cite{Hawking:1972qk} and more generalized scalar-tensor theories~\cite{Sotiriou:2011dz}. Yet for many other classes of theories, including sGB gravity, the no-hair properties no longer hold. Instead, depending on the parameters, the scalar field can develop a nontrivial profile around black holes that extends through the horizon~\cite{Kanti:1995vq, Pani:2009wy, Sotiriou:2013qea,Benkel:2016rlz, Antoniou:2017hxj, Antoniou_2018,Papageorgiou:2022umj,Prabhu:2018aun,Saravani:2019xwx, R:2022tqa} or spontaneous (de-)scalarization can arise~\cite{Silva:2017uqg, Dima:2020yac,Herdeiro:2020wei,Berti:2020kgk,Collodel:2019kkx, Doneva:2020nbb}, see the review articles~\cite{Doneva:2022ewd,Herdeiro:2015waa} for a detailed discussion. 
 
 The scalarization of black holes in sGB strongly depends on properties of the coupling function $f(\varphi)$ between the scalar field and the quadratic curvature terms. When $f(\varphi)$ has a non-vanishing first derivative for all values of $\varphi$, which is often referred to as type I and includes dilatonic couplings $f(\varphi)\sim e^{\gamma \varphi}$, with $\gamma$ a numerical coefficient~\cite{Kanti:1995vq,Pani:2009wy,Ripley:2019irj} and linear functions $f(\varphi)\sim \varphi$  leading to shift-symmetric sGB theories~\cite{Sotiriou:2013qea}, only scalarized black hole solutions exist. 
 Studies showed explicitly that black holes evade the no-hair theorem~\cite{Kanti:1995vq, Antoniou_2018, Papageorgiou:2022umj} and obtained static~\cite{Julie:2019sab, Sotiriou:2014pfa, Sullivan:2019vyi}, slowly rotating~\cite{Pani:2009wy, Ayzenberg:2014aka,Maselli:2015tta} and rapidly rotating~\cite{Kleihaus:2011tg,Kleihaus:2014lba,Kleihaus:2015aje,Kleihaus:2016dui} black hole solutions. They found that requiring regularity of the scalar field at the horizon leads to an analytical bound in the parameter space beyond which no physical solutions exist~\cite{Kanti:1995vq}. Additionally, the resulting sGB black hole solutions generally have a curvature singularity at a finite radius~\cite{Sotiriou:2014pfa, Julie:2022huo}. For a fixed sGB coupling and smaller black hole masses, the singularity moves farther away from the origin and closer to the horizon. Requiring the absence of naked singularities thus leads to a minimum mass for the domain of existence of black holes. 
For type II coupling functions whose derivative vanishes for some values of $\varphi$, such as quadratic $f(\varphi)\sim \varphi^2$~\cite{Berti:2020kgk,Collodel:2019kkx,Silva:2017uqg} and Gaussian $f(\varphi)\sim e^{\gamma \varphi^2}$~\cite{Doneva:2017bvd, Cunha:2019dwb} couplings, the quadratic scalar field term acts as an effective scalar mass. As the effective mass term can be negative, the black hole solution can become unstable and the presence of scalar condensates becomes favored and results in scalarized black holes.\\
While black holes in sGB theories with a massless scalar field have been extensively studied as discussed above, the effects of including a scalar field mass remain less explored. Including a mass term in the action is natural from a theoretical perspective and represents the lowest order self-interaction. Accounting for a mass of the scalar field is further motivated by the only scalar field measured to date, the Higgs boson, and common in scalar models for other sectors of particle physics such as the proposed QCD axion and ultralight dark matter candidates~\cite{Peccei:1977hh, Hui:2016ltb, Ferreira:2020fam, Arvanitaki:2010sy, Kodama:2011zc}. A mass term leads to an exponential suppression of effects of the scalar field at scales larger than its Compton wavelength instead of having an infinite extent as in the massless case. \\
The phenomenology of massive scalar fields around compact objects has been considered in several contexts, including studies of charged black holes~\cite{Boyadjiev:2002en,Horne:1992bi}, black hole superradiance~\cite{Barsanti:2022vvl, Brito:2015oca, Herdeiro:2014goa}, neutron stars in scalar tensor gravity~\cite{Staykov:2018hhc,Ramazanoglu:2016kul}, and type II sGB black holes~\cite{Macedo:2019sem}. 
For black holes in type I sGB with a massive scalar field, previous work has numerically calculated black hole solutions~\cite{Doneva:2019vuh}, included a scalar potential and cosmological constant~\cite{Bakopoulos_2020}, and studied the dynamics of a massive scalar field with self- interaction in the decoupling limit, i.e. on a fixed Schwarzschild spacetime, via a numerical relativity code~\cite{Zhang:2022kbf}. Observational consequences of a massive scalar field in the context of compact objects have also been considered. While the exponential suppression of the field at large distances reduces the size of several of the observational signatures compared to the massless case it may also lead to novel features due to the additional scale involved, as found for gravitational waves from superradiant ultralight boson clouds~\cite{Brito:2017zvb}. Several previous studies further showed that gravitational waves are promising probes for detecting or setting stringent constraints on theories involving massive scalar fields based on effects of scalar dipolar radiation losses in compact-object binary systems. For example,~\cite{Alsing:2011er} considered binary neutron stars in scalar-tensor gravity,~\cite{Maselli:2020zgv} analyzed extreme mass ratio inspirals,~\cite{Chen:2024ery} analyzed probing massive fields in the context of multiband detection, and~\cite{Yamada:2019zrb} placed the first empirical gravitational-wave constraints on massive sGB.\\

In this paper, we go beyond previous work on static black holes in massive sGB~\cite{Doneva:2019vuh,Bakopoulos_2020,Zhang:2022kbf} by (i) combining perturbative and numerical analyses to gain deeper insights into the behavior of the spacetime and scalar field and (ii) performing a systematic study of the solutions and resulting observables over a wide parameter space. This differs from the scope of the work in~\cite{Doneva:2019vuh}, which developed details of the theoretical framework and performed systematic numerical studies of solutions focused on extracting the horizon radius and consequences for thermodynamics.  
Specifically, in this paper, we numerically compute black hole solutions and, for the first time, also calculate perturbative solutions for small sGB couplings to trace behaviors of the metric functions and scalar field configurations. Together, these two methods enable us to study features of curvature invariants of the spacetime and its energetics such as the gravitational mass and scalar-induced energy density of the configurations from different perspectives. We also analyze the parameter dependencies of the bounds on maximum scalar field at the horizon based on requiring the absence of naked singularities, as obtained from numerical solutions, and regularity of the scalar field at the horizon, as obtained from an analytical bound. This lead to a theory bound on the coupling constant of the gravitational theory. In addition, we calculate the parameter dependencies of observables such as the shifts in the ISCO and light ring away from the GR values. We discuss the relevance of our results as a first step towards making connections with measurements such as the black hole shadows, tidal effects close to the black holes, and as a baseline for computing gravitational wave imprints beyond the leading-order dipole radiation losses. The latter would contribute to the recent ongoing efforts of constructing the gravitational waveforms for black hole binary systems in sGB gravity~\cite{Shiralilou:2020gah, Shiralilou:2021mfl, vanGemeren:2023rhh}. Our findings also identify interesting mass ranges for the sGB scalar condensate within the broader context of proposed scalar fields in the universe, and highlight interesting qualitative characteristics and parameter ranges for further studies.\\
In this paper we use Greek indices to denote tensor components in standard Einstein notation. However we use Latin superscripts to assign orders in the small coupling expansion. 
\section{Black holes in scalar-Gauss-Bonnet Gravity}

\subsection{Action}\label{sec:action}
\noindent We consider the following action for sGB gravity\footnote{For the numerical prefactor of the kinetic and potential \eqref{eq:masspotential} scalar field terms, we follow the standard convention also considered for massless sGB, see e.g.~\cite{Julie:2019sab, Shiralilou:2020gah}. However there is a discrepancy in how these factors are defined in the literature on the massive scalar field extension, specifically between~\cite{Bakopoulos_2020} and~\cite{Doneva:2019vuh}. We follow here the convention of~\cite{Doneva:2019vuh}, which means that our results of the field equations, metric and scalar field solutions will differ in numerical factors from~\cite{Bakopoulos_2020}. }
\begin{equation}
\label{eq:actionsGB}
\begin{aligned}
    S_{sGB} =& \frac{c^4}{16 \pi G}\int_M d^4 x \sqrt{-g}[R-2g^{\mu\nu}\partial_{\mu}\varphi \partial_{\nu}\varphi - V(\varphi)+ \alpha f(\varphi)\mathcal{R}_{\rm GB}^2]\ .
\end{aligned}
\end{equation}
Here $R$ denotes the Ricci scalar on manifold $M$ with metric $g_{\mu\nu}$. The scalar field $\varphi$ has potential $V(\varphi)$ and is non-minimally coupled to the Gauss-Bonnet invariant
\begin{equation}
    \mathcal{R}_{G B}^2=R^2-4 R^{\mu \nu} R_{\mu \nu}+R^{\mu \nu \rho \sigma} R_{\mu \nu \rho \sigma}\ .
\end{equation}
via a dimensionless coupling function $f(\varphi)$ and a coupling constant $\alpha$ with dimension length squared. In this work we focus on the simplest potential for a massive scalar field
\begin{equation}\label{eq:masspotential}
    V(\varphi) = 2m^2 \varphi^2\ ,
\end{equation}
where 
\be
\label{eq:scalarmassparam}
m=\frac{m_{\varphi} c}{\hbar}\ ,
\ee
denotes the scalar field mass parameter having the dimension of inverse length with $m_{\varphi}$ the scalar field mass in kilograms. 
While much of our analysis is general for any coupling function $f(\varphi)$, our case studies of static black hole solutions specialize to type I coupling functions of the form $f(\varphi) = \beta e^{\gamma \varphi}$. This choice is inspired by the low-energy effective action of certain string theories, with the choice of $\beta$ and $\gamma$ corresponding to different string models~\cite{Gross1987,Zwiebach1985, Moura:2006pz}. We will focus here on $f(\varphi) = \frac{1}{4} e^{2 \varphi}$ corresponding to Einstein-dilaton-Gauss-Bonnet (EdGB) gravity~\cite{ Kanti:1995vq,Pani:2009wy,Julie:2019sab}. Other choices for $\gamma$ will lead to qualitatively the same behavior for black hole spacetimes~\cite{Doneva:2019vuh}.

The massless scalar field theory with dilatonic coupling has most constraints on the coupling constant $\alpha$ compared to the other types of sGB theories, altough often obtained in the weak field limit where the exponential coupling function is approximated by the leading order linear term in the scalar field. The constraints are coming from different types of observations ranging from solar system test to binary pulsar observations, for an overview see \cite{Saffer:2021gak, Bertotti:2003rm, Yagi:2012gp, Yagi:2015oca, Pani:2011xm, East:2022rqi, Yordanov:2024lfk, Xu:2021kfh}. The strongest current observational constraints on the coupling constant come from a Bayesian analysis of the data from the first three observing runs from the LIGO-Virgo-KAGRA (LVK) detector network to $\sqrt{\alpha}\lesssim 0.8-1.33$km~\cite{Perkins:2021mhb, Lyu:2022gdr, Wang:2021jfc}. For massive scalar field sGB, a first observational constraint based on data from the first two observing runs of LVK obtained $\sqrt{\alpha}\lesssim 2.47$km~\cite{Yamada:2019zrb}. A weaker bound in the massive case is consistent with expectations, as the mass causes a suppression of the scalar field effect on large scales. 

\subsection{Relevant length scales}
\begin{figure}[!h]
\centering
\begin{subfigure}{0.4\textwidth}
   \includegraphics[width=1\linewidth]{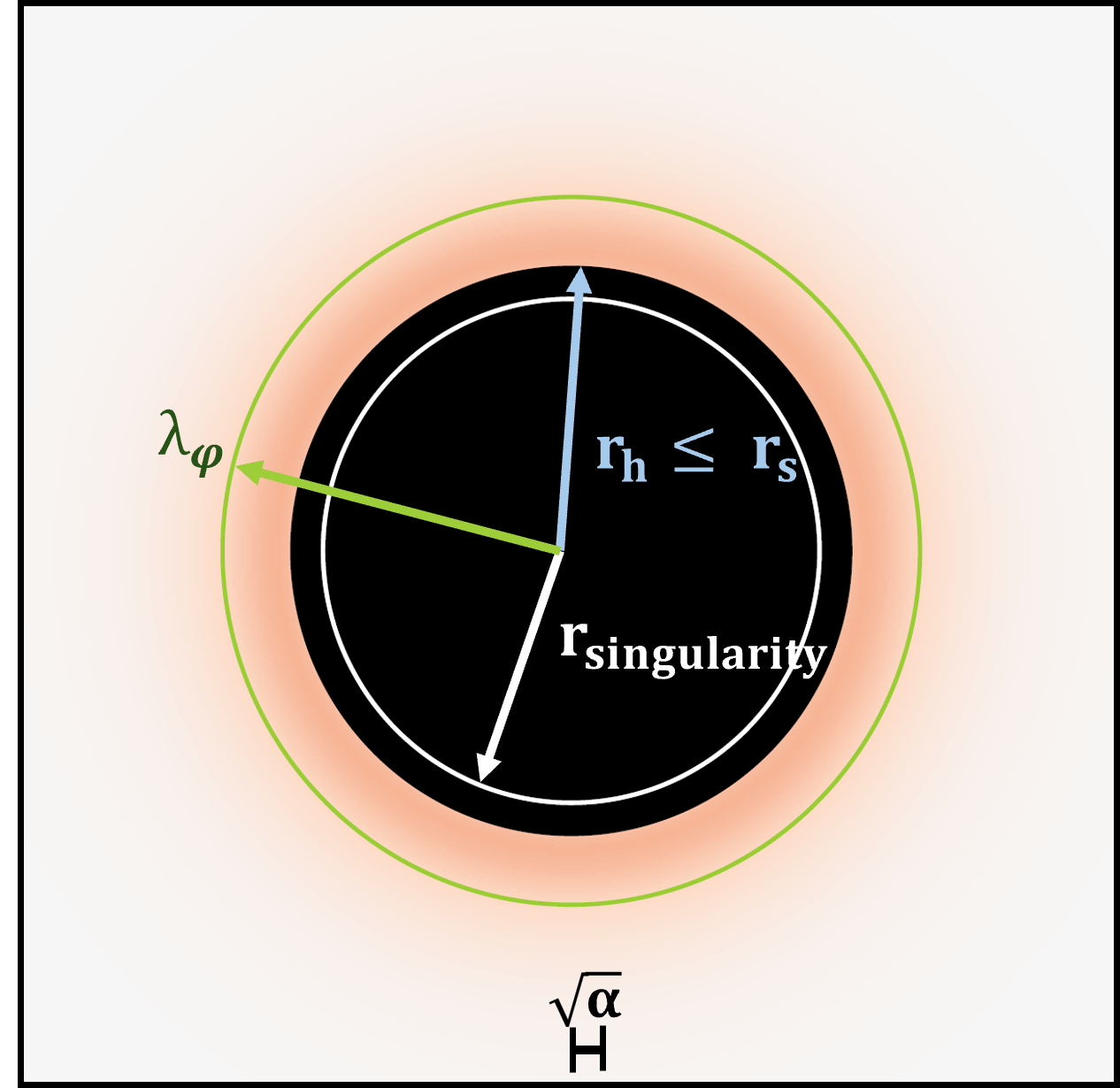}
       \centering
\end{subfigure}
\caption[]{Sketch of the black hole horizon (black region) and the scalar condensate around the black hole (red) with the relevant length scales and their hierarchy in an example of a small scalar field mass.}
\label{fig:lengthscales}
\end{figure}

Before discussing the technical details of computing static black hole solutions in massive sGB, we give an overview of the key length scales and their hierarchy, which has important consequences for qualitative features of the solutions and for defining perturbative approximations. Figure~\ref{fig:lengthscales} illustrates these scales for an example of a black hole and scalar condensate.  
We consider a static, spherically symmetric black hole of horizon radius $r_h$ which is of the order (but slightly smaller~\cite{Doneva:2019vuh}) of the Schwarzschild radius 
\begin{equation}
\label{eq:rsdef}
r_h \sim r_S = \frac{2 G M}{c^2}\ ,
\end{equation}
with $M$ the mass of the black hole. The black hole is surrounded by a massive scalar field cloud that extends inside the horizon. The characteristic size of the cloud is related to the mass of the scalar field. The cloud is exponentially suppressed for distances beyond the Compton wavelength $\lambda_{\varphi}$ which is inversely proportional to the scalar field mass $m$
\begin{equation}\label{eq:comptonw}
    \lambda_{\varphi} \sim 1/m\ . 
\end{equation}
Hence in the small-mass limit the scalar field cloud stretches out further to infinity, approaching the massless sGB solution. By contrast, for larger masses, the scalar field becomes more confined to the vicinity of the horizon, and for $m\to \infty$ the scalar field decouples and the solution approaches the Schwarzschild black hole. In Fig.~\ref{fig:lengthscales} we show the Compton wavelength length scale for small masses. Here, small masses refers to the Compton wavelength being larger than the black hole horizon.

The last length scale is set by the coupling constant $\sqrt{\alpha}$ which determines the strength of the higher curvature contributions. When we apply perturbation theory in section~\ref{sec:pertsol}, we assume the dimensionless version of the coupling to be small
\begin{equation}\label{eq:alphahat}
    \hat{\alpha} \equiv \frac{\alpha}{r_h^2}\ .
\end{equation}
Assuming the current observational bound is saturated $\sqrt{\alpha}= 2.47$ km and considering black holes in the mass range $5 M_{\odot} \lesssim M \lesssim 10^{10} M_{\odot}$, the dimensionless coupling lies in the range $10^{-11}\lesssim \hat{\alpha}\lesssim 0.2$, validating the assumption of working in the small coupling regime. 
The perturbation theory we set up is exact in $m$, i.e. we do not assume any restriction on the scalar field mass. Expanding both in the small mass and coupling limit resulted in non-regular solutions for the scalar field at the black hole horizon. On the other hand, when discussing the numerical solution to the field equations, no restrictions on the length scales related to both the mass and coupling are assumed. However it turns out that requiring the scalar field to be regular at the horizon does give a restriction on the value of the coupling and scalar field mass depending on the black hole mass and amount of scalar field at the horizon. This restrictions ensures that the curvature singularity at $r\neq0$\footnote{When working in Schwarzschild coordinates.} lies within the horizon and hence prevents a naked singularity, see also Fig.\ref{fig:lengthscales}.

\subsection{Field equations} \label{sec:fieldequations}
\noindent Varying the action~\eqref{eq:actionsGB} with respect to the metric $g_{\mu\nu}$ results in the following field equations
\begin{equation}\label{eq:FE1}
\begin{aligned}
    G_{\mu\nu} &= T_{\mu\nu}\ ,
    \end{aligned}
\end{equation}
with $G_{\mu\nu}$ the Einstein tensor and $T_{\mu\nu}$ the 'effective' energy momentum tensor which includes contributions from the scalar field and the higher curvature terms, 
\begin{equation}\label{eq:EMtensor}
\begin{aligned}
    T_{\mu\nu}=&2\partial_{\mu}\varphi\partial_{\nu}\varphi - g_{\mu\nu}\partial_{\rho}\varphi\partial^{\rho}\varphi - g_{\mu\nu}m^2\varphi^2 - 4\alpha {}^{*}R^*_{\alpha \mu \nu \beta}\nabla^{\alpha}\nabla^{\beta}f(\varphi)\ .
\end{aligned}
\end{equation}
Here ${}^{*}R^*_{\alpha \mu \nu \beta}$ is the double dual Riemann tensor defined as ${}^{*}R^*_{\alpha \mu \nu \beta}=\frac{1}{4} \epsilon_{\alpha \mu }^{\quad\gamma\sigma} R_{\gamma\sigma \rho \varepsilon} \epsilon_{\quad \nu\beta}^{\rho\varepsilon}$ with $\epsilon_{\alpha\mu\gamma\sigma}$ the anti-symmetric Levi-Civita tensor. 
The scalar field equation is given by
\begin{equation}\label{eq:FE2}
    \Box \varphi = m^2\varphi - \frac{1}{4} \alpha f'(\varphi)\mathcal{R}_{\rm GB}^2\ ,
\end{equation}
with $\square \equiv g^{\alpha \beta} \nabla_\alpha \nabla_\beta$ the d'Alembertian operator.
One can check that the field equations~\eqref{eq:FE1},~\eqref{eq:FE2} are invariant under the rescaling of the coordinates with a generic factor $c$ which leaves the fields invariant together with redefining $m\xrightarrow{}m/c$, $\alpha \xrightarrow{} c^2 \alpha$.

\subsection{Metric and asymptotic behavior}\label{sec:explFEsbound}
In this work we focus on static, spherically symmetric black hole solutions for which the general metric is given by
\begin{equation}\label{eq:SSSmetric}
ds^2 = -e^{A(r)}dt^2 + e^{B(r)} dr^2 + r^2( d\theta^2 + \sin^2{\theta}d\phi^2) .
\end{equation}
We assume the same symmetries for the scalar field, hence $\varphi = \varphi(r)$. Substituting this and the general metric~\eqref{eq:SSSmetric} in the field equations~\eqref{eq:FE1} and~\eqref{eq:FE2}, we obtain the components of~\eqref{eq:FE1} given explicitly in~\eqref{eq:GmunuTmunu} in Appendix~\ref{AppA}. The scalar field equation is given explicitly by~\eqref{eq:scalareomexpl}.
To obtain the desired black hole and condensate solutions to the equations of motion for the metric functions $A(r)$, $B(r)$ and the scalar field $\varphi(r)$ requires imposing the correct boundary conditions at the black hole horizon and at spatial infinity. The black hole horizon is defined in Schwarzschild coordinates by a vanishing time component of the metric and a diverging radial component. Furthermore we require the scalar field to remain regular at the horizon. Hence we have the following conditions approaching the black hole horizon $r_h$
\begin{equation}\label{eq:horcond}
\begin{aligned}
&A(r)\to-\infty\ ,\\
&B(r)\to\infty\ ,\\
&\varphi'(r), \varphi''(r)\mathrm{\,finite}\ .
\end{aligned}
\end{equation}
Furthermore at infinite radial distance we require the solution to be asymptotically flat and approach Minkowski spacetime. Therefore, at spatial infinity, the scalar field sourcing the metric equations should vanish as well and we have
\begin{equation}\label{eq:asympcond}
\begin{aligned}
A(r)&\to0\ ,\\
B(r)&\to0\ ,\\
\varphi(r)&\to0\ .
\end{aligned}
\end{equation}
To capture the nontrivial fall-off behavior of the scalar field near infinity, we substitute the asymptotic metric functions~\eqref{eq:asympcond} in the scalar field equation~\eqref{eq:scalareomexpl} and obtain
\begin{equation}
    2 r^2 \varphi''(r) + 4 r \varphi'(r) - 2 m^2 r^2 \varphi(r)=0\ .
\end{equation}
Solving this differential equation for $\varphi(r)$ yields the asymptotic solution 
\begin{equation}\label{eq:asymplimsf}
\varphi(r) \to  c_1\frac{e^{-mr}}{r} + c_2 \frac{e^{mr}}{2 m r}\ ,
\end{equation}
 with $c_1$, $c_2$ two integration constants. For an asymptotically flat solution we require $c_2=0$ while the remaining coefficient $c_1$ is determined by matching to the near-horizon solutions and depends on the coupling as we show in Sec.~\ref{sec:MADMQ}. 
The expression~\eqref{eq:asymplimsf} with $c_2=0$ quantifies the qualitative behavior alluded to earlier: the scalar field mass causes the field configuration to be constrained to the vicinity of the black hole and exponentially suppressed beyond the scale of the Compton wavelength~\eqref{eq:comptonw}. In the limit $m\to 0$, the exponential in~\eqref{eq:asymplimsf} becomes unity and the falloff of the field is much slower $\sim 1/r$, consistent with calculations in the massless case~\cite{Julie:2019sab,Sotiriou:2014pfa}. 

\section{Perturbative black hole solutions for small coupling}\label{sec:pertsol}
Before we compute the exact metric and scalar field solutions by solving the field equations~\eqref{eq:FE1},~\eqref{eq:FE2} numerically, we analyze the solution in the small coupling expansion to gain further insights into the behavior of the solution. We expand in the dimensionless coupling constant $\hat\alpha$ defined in~\eqref{eq:alphahat}. It is convenient to define a dimensionless radial coordinate 
\be
\label{eq:udef}
u=\frac{r_h}{r}\ ,
\ee
so the horizon always lies at $u=1$ and spatial infinity at $u=0$. Furthermore we introduce the dimensionless mass 
\be
\label{eq:mhatdef}
\hat{m} = r_h m\ .
\ee
We expand the metric components and the scalar field for small coupling $\hat \alpha \ll 1$. At this stage it is more convenient to reparameterize the metric functions 
\begin{equation}\label{eq:rescaleAB}
\begin{aligned}
e^{A(u)}&\to\bar{A}(u)\ ,\\
e^{B(u)}&\to\frac{1}{\bar{B}(u)}\ ,
\end{aligned}
\end{equation}
 as it makes the expansion more straightforward. Then the small-coupling expansion is given by the ansatz
\begin{eqnarray}\label{eq:expABphi}
&&\bar{A} = \sum_{i=0}^\infty\bar{A}^{i}\,\hat{\alpha}^i\ , \nonumber \\
&& \varphi = \sum_{i=0}^\infty\varphi^{i}\,\hat{\alpha}^i\ ,
\end{eqnarray}
where we omit here and in the following the explicit expansion of $\bar{B}$ as it is similar to~\eqref{eq:expABphi}. We substitute this ansatz into~\eqref{eq:GmunuTmunu},~\eqref{eq:scalareomexpl} and solve order by order in $\hat{\alpha}$. At $\mathcal{O}(\hat{\alpha}^0)$ we need to obtain the Schwarzschild solution as the limit of $\alpha\xrightarrow{}0$ should recover GR. Therefore we can already impose
\begin{equation}\label{eq:A0B0phi0}
\begin{aligned}
\bar{A}^{0}&=\bar{B}^{0}=1-u\ ,\\
\varphi^{0}&=0\ .\\
\end{aligned}
\end{equation}
To recover the Schwarzschild solution at zeroth order in the coupling, in the context of the perturbative solution $r_h$ in~\eqref{eq:udef},~\eqref{eq:mhatdef} and~\eqref{eq:alphahat} is equal to $r_S$~\eqref{eq:rsdef}. However we defined the variable $u$, mass and coupling parameters in terms of the general horizon radius so they can be used in the context of the exact solution in Sec.~\ref{sec:fullsol} as well.

\subsection{Equations of motion at linear order in the coupling}\label{sec:linsol}
Before analyzing in detail the expansion of the field equations, we can already gain insights into the scalings of different contributions with the coupling by considering the field equations~\eqref{eq:FE1} with the expansion~\eqref{eq:expABphi}. At linear order in the coupling, there is a correction to the scalar field as the source term in~\eqref{eq:FE2} is linear in the coupling. Next, 
analyzing the source of the metric equations of motion~\eqref{eq:EMtensor} we find that the energy momentum tensor consists of terms quadratic in the scalar field and a contribution linear in the coupling times $\nabla^{\alpha}\nabla^{\beta}f(\varphi) = \nabla^{\alpha}(f'(\varphi)\partial^{\beta}\varphi)$ which is at least linear in the scalar field.  
As the scalar field to lowest order is linear in the coupling, $T_{\mu\nu}$ is quadratic and higher order in $\hat \alpha$. Consequently, the corrections to the field equations for the metric potentials~\eqref{eq:FE1} will only appear $O(\hat\alpha^2)$.
At linear order in $\hat\alpha$, the metric remains the Schwarzschild metric and we need to compute the solution to the linearized scalar field equation in a Schwarzschild background. This is summarized in the second row of Table~\ref{tab:dependencies}.
In particular, to solve for the linear solutions in $\hat \alpha$, we substitute the small coupling expansion for the scalar field and metric components~\eqref{eq:expABphi} in~\eqref{eq:FE2} and use~\eqref{eq:A0B0phi0} for the zeroth order coefficients and $\bar{A}^{1}=\bar{B}^{1}=0$ as discussed above. 
This leads to the linearized scalar field equation in the radial coordinate $u$ defined in~\eqref{eq:udef}
\begin{equation}\label{eq:eomphilin}
(u-1)\varphi^{1}{}^{\prime\prime}(u) + \varphi^{1}{}^{\prime}(u) + \frac{\hat{m}^2}{u^4}\varphi^{1}(u)=3u^2 f'(\varphi^{0})\ .
\end{equation}

\subsubsection{Near-horizon and asymptotic behavior of the linearized field}\label{sec:linfieldNHasmp}
To capture the solution of \eqref{eq:eomphilin} near the horizon, we expand around the horizon radius 
\be
\label{eq:epsdef}
\epsilon = u-1\ . 
\ee
 This leads to a double expansion of the fields in $\hat\alpha$ and $\epsilon$, where each coefficient in the $\hat \alpha$ expansion in~\eqref{eq:expABphi} is further expanded in a Taylor series in $\epsilon$. For the $O(\hat\alpha)$ coefficient we have
\be
\label{eq:Taylorhor}
\varphi^{1} = \varphi_h^{1} + \epsilon \varphi_h^{1\prime} + \mathcal{O}(\epsilon^2)\ .
\ee
For the $O(\hat\alpha)$ terms, solving the differential equation~\eqref{eq:eomphilin} order by order in $\epsilon$ and using~\eqref{eq:A0B0phi0} determines $\varphi_h^{1\prime}$ in terms of $\varphi_h^{1}$ via
\begin{equation}\label{eq:phi1prime}
    \varphi_h^{1\prime} = 3 f'(0) - \hat{m}^2\varphi_h^{1}\ . 
\end{equation}
The coefficient $\varphi_h^{1}$ corresponds to the amount of scalar field at the horizon at linear order in the coupling and $f'(0)$ is a constant.
One can reason that the solution to~\eqref{eq:eomphilin} has to be a monotonically increasing solution (see Appendix \ref{AppB} for the detailed arguments) and therefore the first derivative at the horizon has to be positive \cite{Horne:1992bi}. This leads to the following constraint on the amount of (linearized) scalar hair at the horizon and the mass of the scalar field
\begin{equation}\label{eq:constraint1}
\varphi_h^{1} < \frac{3 f'(0)}{\hat{m}^2}\ .
\end{equation}
In the linearized case, we thus find a constraint on the amount of scalar field at the horizon. In massless sGB, similar arguments result in an expression for the scalar field derivative at the horizon (in this case for the full theory)~\cite{Kanti:1995vq} given by \begin{equation}\label{eq: boundphihmassless}
\varphi'_h= \frac{r_h}{4 \alpha f'(\varphi_h)}\left(-1 \pm \sqrt{1-\frac{24 \alpha^2 f'(\varphi_h)^2}{r_h^4}}\right)\ .
\end{equation} 
Requiring the square root to be positive yields the constraint  \begin{equation}\label{eq:mlconstralpha}
 f'(\varphi_h)^2<\frac{r_h^4}{24 \alpha^2}\ . 
\end{equation} 
For a fixed coupling function and constant, this bound~\eqref{eq:mlconstralpha} determines the maximum amount of allowed scalar hair at the horizon depending on the size of the black hole. Conversely, given a certain amount of scalar field at the horizon, the constraint~\eqref{eq:mlconstralpha} sets a lower bound on the black hole mass that can sustain this hair.  
Using the definition of $\hat{m}$ from~\eqref{eq:mhatdef} in~\eqref{eq:constraint1} shows that the maximum amount of scalar hair at the horizon depends both on the scalar field mass and the black hole mass. In Sec.~\ref{sec:propertiessol} below we study the effect of the scalar mass on these quantities with full black hole solutions and establish a more meaningful comparison to the massless results~\eqref{eq:mlconstralpha}.\\ 

As at linear order in the coupling the background is still Schwarzschild spacetime, the asymptotic limit of the scalar field at this order follows~\eqref{eq:asymplimsf} to first order in the asymptotic expansion in $u$. To write it in the notation introduced in this section
\begin{equation}\label{eq:asymplimsflin}
\varphi^1(u) =  \varphi^{1\prime}_{\infty}e^{-\hat{m}/u} u + \bar{\varphi}^{1\prime}_{\infty} \frac{e^{\hat{m}/u}}{2 \hat{m} } u + \mathcal{O}(u^2)\ ,
\end{equation}
where we absorbed the factor $r_h$ in the first term in the constant $ \varphi_{\infty}^{1\prime}$.

\subsubsection{Numerical solution for the linearized field}\label{sec:numsollin}
The solution to~\eqref{eq:eomphilin} has to be calculated numerically. It is computed by defining an initial value problem at an infinitesimal distance from the black hole horizon $u=1-10^{-5}$, with~\eqref{eq:Taylorhor} as initial condition, and integrating to spatial infinity $u=0$.
We keep the description and discussion of the numerical methods needed on top of a numerical integrator general. For the numerical integration we specify to an 8th order explicit Runge Kutta scheme with a machine and working precision of 30 digits to acquire the needed numerical precision. For more details we refer to the last section of Appendix \ref{AppC}. \\

In~\eqref{eq:Taylorhor} $\varphi_h^{1}$ is the constant that needs to be determined by matching to the asymptotic limit~\eqref{eq:asymplimsflin}. A difficulty is to ensure that the asymptotic solution~\eqref{eq:asymplimsflin} obeys the desired fall-off conditions at infinity, with $\bar{\varphi}^{1\prime}_{\infty}=0$ to eliminate the growing mode. If this condition is not exactly fulfilled, the growing mode always takes over at some large distance from the horizon. Furthermore, any small numerical error in the initial condition that results in an inexact match to $\varphi^{1\prime}_{\infty}$ finite and $\bar{\varphi}^{1\prime}_{\infty}$ zero in~\eqref{eq:asymplimsflin} immediately leads to a diverging solution. Therefore, finding the exact exponentially decaying solution numerically is a challenge. However, solutions close to the desired solution can be computed using the bisection method described in~\cite{Horne:1992bi} and in Appendix~\ref{AppC}. This method is based on identifying the domain of existence of the exponentially decaying solution in the range of input guesses $\varphi_h^{1}$ for which, when integrating the solution outwards, the behavior at infinity switches from positively to negatively diverging for too large or too small guesses respectively. Decreasing this range for $\varphi_h^{1}$ through several iterations leads to a narrow range of guesses that approach the 'right' value for $\varphi_h^{1}$ such that the solution only decays. The more cycles in this bisection method, the more accurate the guess for $\varphi_h^{1}$ and the farther the diverging behavior is pushed out to larger distances. This is shown in Fig.~\ref{fig:estimmethod} below.
\begin{figure}[h]
\centering
\begin{subfigure}{0.55\textwidth}
   \includegraphics[width=1\linewidth]{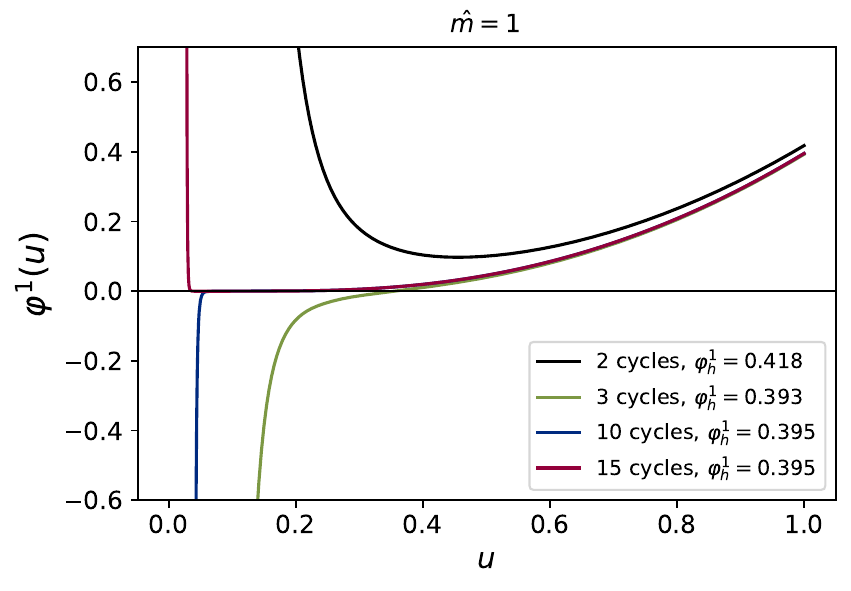}
       \centering 
\end{subfigure}
\caption[]{\emph{Solution for the linearized scalar field with $\hat{m}=1$ for different numbers of cycles of the bisection method.} The legend also shows the value of $\varphi_h^{1}$ corresponding to each curve. The integration starts at the horizon $u=1-10^{-5}$ and proceeds outwards to infinity $u=0$. }
\label{fig:estimmethod}
\end{figure}
For solving~\eqref{eq:eomphilin} we apply this bisection method for 15 cycles, where the difference in $\varphi_h^{1}$ from the value of the previous cycle is $\sim 10^{-14}$. We compare these small-coupling results for $\varphi_h^{1}$ to the values obtained in the full solution in Sec.~\ref{sec:fullsol}.

\subsection{Higher order corrections in $\hat \alpha$}\label{sec:higherordercorr}

\begin{table*}[]
\centering
\begin{tabular}{|l|l|l|l|}
\hline
Order & $ tt$ field equation & $rr$ field equation &    scalar field equation \\ \hline
$\hat{\alpha}^0$ &
  $\bar{B}^{0}$, $\varphi^{0}$ &
  $\bar{A}^{0}$, $\bar{B}^{0}$, $\varphi^{0}$ &
  $\bar{A}^{0}$, $\bar{B}^{0}$, $\varphi^{0}$ \\ \hline
$\hat{\alpha}^1$        & -                              & -                                                                      & $\varphi^{1}$                  \\ \hline
$\hat{\alpha}^2$        & $\bar{B}^{2}$, $\varphi^{1}$     & $\bar{A}^{2}$, $\bar{B}^{2}$, $\varphi^{1}$& $\varphi^{2}$                  \\ \hline
$\hat{\alpha}^3$ &
  $\bar{B}^{3}$, $\varphi^{1}$, $\varphi^{2}$ &
  $\bar{A}^{3}$, $\bar{B}^{3}$, $\varphi^{1}$, $\varphi^{2}$ &
  $\bar{B}^{2}$, $\bar{A}^{2}$, $\varphi^{1}$, $\varphi^{2}$, $\varphi^{3}$ \\ \hline
\end{tabular}
\centering
\caption{\emph{Dependencies of the equations of motion~\eqref{eq:GmunuTmunu} and~\eqref{eq:scalareomexpl} in the small-coupling limit on the expansion coefficients at each order $n$ in $\hat{\alpha}$.} At orders $n>0$, the dependencies listed in the table are those obtained after substituting the lower order solutions.}
\label{tab:dependencies}
\end{table*}
As discussed in Sec.~\ref{sec:linsol}, the corrections to the Schwarzschild metric first appear at order $\hat{\alpha}^2$. At each $n$-th order in the perturbative expansions in $\hat \alpha$ with $n\geq 2$, the field equations~\eqref{eq:GmunuTmunu} together with the background and linearized solutions discussed above 
depend on the metric coefficients at orders $\leq n$ as well as the scalar field corrections up to one lower order $\leq n-1$. The scalar field equation~\eqref{eq:FE2} becomes dependent on the metric corrections only at $O(\hat{\alpha}^3)$. Therefore we focus on obtaining the perturbative solution to that order so as to capture all the different dependencies of the solutions and compare with the full solution in the next section. We summarize these dependencies of the field equations at the different orders in Table~\ref{tab:dependencies}. 
The approach to assemble all the inputs to compute solutions is similar to the linearized case: after obtaining the system of equations order by order in $\hat \alpha$ from the small-coupling expansion of the field equations, the next step is to analyze their asymptotic and near-horizon limits. 

\subsubsection{Near horizon and asymptotic limit of the higher order correction solutions}\label{sec:pertboundaries}

For the near horizon limit, we expand all functions $\epsilon$ defined in~\eqref{eq:epsdef} as in the linearized case~\ref{sec:linsol}. Specifically, we make the ansatz 
\begin{equation}\label{eq:nearhoranz}
\begin{aligned}
&\bar{A}^{i}  =\bar{A}_h^{i} + \epsilon \bar{A}_h^{i\prime}{}+ \epsilon^2 \bar{A}_h^{i\prime\prime}+ \mathcal{O}(\epsilon^3) \ ,\\
&\varphi^{i}= \varphi_h^{i} + \epsilon \varphi_h^{i\prime} + \epsilon^2 \varphi_h^{i\prime\prime}  + \mathcal{O}(\epsilon^3)\ ,
\end{aligned}
\end{equation}
and similarly for $\bar{B}^{i}$, where we focus on $i=2,3$ for the quadratic and cubic orders in the coupling respectively. We substitute this ansatz into the $tt$ and $rr$ components of the field equations~\eqref{eq:FE1} and the scalar equation of motion~\eqref{eq:FE2}, expand for $\epsilon\ll 1$ and solve order by order. \\
To capture the asymptotic behavior at spatial infinity, we first note that as discussed above, the corrections to the scalar field equation of motion from the metric enter only at $O(\hat \alpha^3)$. Thus, at $O(\hat \alpha^2)$, the asymptotic behavior of $\varphi^2$ is still given by~\eqref{eq:asymplimsf}. By contrast, the metric field equations~\eqref{eq:FE1} at $O(\hat\alpha^2)$ and higher depend on the scalar field one order lower in $\hat{\alpha}$ (see Table~\ref{tab:dependencies}). Thus, near spatial infinity they involve contributions from a quadratic combination of the scalar field asymptotics~\eqref{eq:asymplimsf} with $c_2\to 0$. In turn, this implies that at $O(\hat\alpha^3)$, the asymptotic scalar field involves a cubic combination of~\eqref{eq:asymplimsf}. Based on these considerations, we include the expected number of factors of the exponential from~\eqref{eq:asymplimsf} in our ansatz for the expansion of the functions near spatial infinity, specifically
\begin{equation}\label{eq:asympanz}
\begin{aligned}
&\bar{A}^{i} = e^{-2 \hat{m}/u }\left(\bar{A}_{\infty}^{i} + u \bar{A}_{\infty}^{i\prime}  + u^2 \bar{A}_{\infty}^{i\prime\prime} + u^3 \bar{A}_{\infty}^{i\prime\prime\prime}+ \mathcal{O}(u^4)\right) \ ,\\
&\varphi^{3}=e^{-3 \hat{m}/u }\left(\varphi_{\infty}^{3} + u \varphi_{\infty}^{3\prime} + u^2 \varphi_{\infty}^{3\prime\prime}+ u^3 \varphi_{\infty}^{3\prime\prime\prime}+ \mathcal{O}(u^4)\right)\ ,
\end{aligned}
\end{equation}
and similarly for $\bar{B}^{i}$ again focussing on $i=2,3$. With the dependencies on the exponentials captured in the ansatz, one can factor them out in the field equations to the lowest orders in $u$ (here up to $u^3$). Factoring out the exponentials is important to be able to proceed, as otherwise the field equations do not have a series expansion around $u=0$ since $e^{1/u}$ remains large in this limit. In~\eqref{eq:asympanz} we only kept terms up to $O(1/u^3)$, which we found to give sufficient accuracy for our purposes. However, the method can be extended to include higher orders by altering the ansatz in such a way that the dependencies on exponentials can be factored out in the equations of motion. \\
We substitute the ansatz~\eqref{eq:asympanz} into the $tt$, $rr$ components of the field equations and the scalar one at each $O(\hat \alpha^i)$ and solve order by order in $u$ for the coefficients. We find that, as expected based on the scaling considerations discussed above, these coefficients depend on the scalar field integration constants up to one order lower in $\hat \alpha$. 

\subsubsection{Numerical solutions with higher order corrections}

With the asymptotics near the horizon and spatial infinity in hand, we turn to solving the field equations over the entire spatial domain order by order in $\hat \alpha$. We first note the simplifying fact that at quadratic order in $\hat\alpha$, the $tt$ component of the field equations at $O(\hat \alpha^2)$ depends only on the $\bar{B}$ correction and the scalar field at $O(\hat{\alpha})$ (see Table \ref{tab:dependencies}). We can therefore first solve the $tt$ component of the field equations at $O(\hat \alpha^2)$ for $\bar{B}$ by substituting the numerical solution of the linearized scalar field as described in Sec.~\ref{sec:linsol} and solving the equation numerically by starting the integration from an infinitesimal distance outside of the horizon $u=1-10^{-5}$ and integrating towards $u=0$ using the same specifications for the numerical integrator as mentioned in Sec.~\ref{sec:numsollin}. As discussed in Sec.~\ref{sec:linsol} and Appendix~\ref{AppC}, the divergent behavior of the linearized solution, which enters into all subsequent calculations at higher orders in $\hat \alpha$, can numerically only be suppressed out to a small but finite $u$. This implies that the higher order solutions can only be computed up to a slightly larger value of $u$, as the onset of the divergence must be pushed outside the domain of integration. For a given accuracy of the linearized solution, this leads to a deterioration in accuracy at each higher order in $\hat \alpha$. \\
For the initial conditions of the integration we use~\eqref{eq:nearhoranz} to linear order in $\epsilon$. For $\bar{B}^2$ this is given by
\begin{equation}\label{eq:horlimB2}
   \bar{B}^{2} \sim \bar{A}_h^{2} + \epsilon \left((\hat{m}\varphi_h^{1})^2 + \bar{A}_h^{2}\right)\ .
\end{equation}
The coefficient $\bar{A}_h^{2}$ needs to be determined by matching to the asymptotic limit~\eqref{eq:asympanz}. The asymptotic solution of $\bar{B}^{2}$ in~\eqref{eq:asympanz} is given by
\begin{equation}\label{eq:asymplimB2}
   \bar{B}^{2} \sim e^{\frac{-2 \hat{m}}{u}}\left(\hat{m} \varphi^{1\prime}_{\infty} u + \frac{1}{2} u^2 \left(2 (\varphi^{1\prime}_{\infty}){}^2 - \hat{m}(\varphi^{1\prime}_{\infty}){}^2\right)\right)\ ,
\end{equation}
with $\varphi^{1\prime}_{\infty}$ the integration constant of the asymptotic limit of the linearized scalar field~\eqref{eq:asymplimsflin} and is thus completely determined by the scalar field solution at linear order in the coupling. 
We compute the numerical solution having this desired asymptotic behavior by using a shooting method. This is based on obtaining the solution for $\bar{B}^2$ for different guesses of $\bar{A}_h^{2}$ and evaluating these solutions at infinity until these values agree with the values at infinity of~\eqref{eq:asymplimB2}. 
In Appendix \ref{AppC} we describe details of the implementation of the shooting method in this context by giving the explicit example for computing $\bar{B}^{2}$.

Having solved the $tt$ component of the field equations, we use the resulting numerical solution for $\bar{B}^{2}$ together with $\varphi^1$ in the $rr$ field equation and solve for $\bar{A}^{2}$ using the shooting method described above and in Appendix \ref{AppC}. This completes the computation of the metric functions at $O(\hat \alpha^2)$.  The solution for the scalar field expansion coefficient $\varphi^{2}$ at that order can be determined separately, as its equation of motion does not involve any metric corrections (see Table~\ref{tab:dependencies}). Therefore we can use the same bisection method as for the linearized scalar field. Finally, the metric and scalar field  corrections at $O(\hat \alpha^3)$ can be determined via the same procedure and methods as described for the second order corrections. 

\section{Full numerical black hole solutions}\label{sec:fullsol}

To check to what extent the perturbative solution captures the behavior of the black hole spacetime correctly and compute results including non-perturbative effects, we solve the field equations~\eqref{eq:GmunuTmunu},~\eqref{eq:scalareomexpl} without approximations using numerical methods. We follow the methodology of~\cite{Doneva:2019vuh} for a specific choice of coupling function, however, our analysis in Sec.~\ref{sec:propertiessol} has a different focus and therefore complements the results in \cite{Doneva:2019vuh}.\\
 For solving the full field equations, it is more convenient to work with a different setup from that used for the small-coupling approximations described above. In particular, we work with the parameterization of the metric potentials in terms of $A$ and $B$ instead or $\bar{A}$ and $\bar{B}$ and rewrite~\eqref{eq:GmunuTmunu},~\eqref{eq:scalareomexpl} as follows~\cite{Sotiriou:2014pfa, Bakopoulos_2020}. We use the $rr$-component to eliminate the $B(r)$ and $B'(r)$ contributions to the field equations and cast the $rr$-component~\eqref{eq:GmunuTmunu} as a quadratic equation in $e^{B(r)}$
\begin{equation}
\label{eq:Bquadratic}
e^{2 B(r)} \rho(r) + e^{B(r)}\beta(r) + \gamma(r)=0\ ,
\end{equation}
where 
\begin{equation}
\begin{aligned}
    \rho(r) &= 4\left(1-(mr\varphi(r))^2\right)\ ,\\
    \beta(r) &= -4\left(1+r A'(r) + 2\alpha A'(r)f'(\varphi)\varphi'(r) -r^2\varphi'(r)^2\right)\ ,\\
    \gamma(r) &= 24 \alpha A'(r) f'(\varphi)\varphi'(r)\ .  
\end{aligned}
\end{equation}
The solution to the quadratic equation~\eqref{eq:Bquadratic} is given by
\begin{equation}\label{eq:expBexpl}
e^{B(r)} = \frac{-\beta(r) + \sqrt{\beta(r)^2 - 4 \rho(r)\gamma(r)}}{2\rho(r)}\ .
\end{equation}
Here, we chose the solution with the positive sign as it gives the desired asymptotic limit\footnote{Substituting the asymptotic behavior for $A(r)$ and $\varphi(r)$ assuming both fall off to $0$ as $\sim1/r$ and $\sim e^{-m r}/r$ respectively , which we discuss in Sec.~\ref{sec:asymplim}, leads to $\beta(r) \xrightarrow{} -4$. Then the positive sign solution gives $e^{B(r)}\xrightarrow{}1$ which is the desired asymptotically flat result.} defined by~\eqref{eq:asympcond}.  
Furthermore, the expression for $B'(r)$ is given by the derivative of~\eqref{eq:expBexpl}. The remaining field equations can then be rewritten as two second order differential equations for $A(r)$ and $\varphi(r)$  given explicitly by
\begin{equation}\label{eq:ddAddphi}
\begin{aligned}
A''(r) =& f(r, \varphi(r),\varphi'(r), A'(r))\ ,\\
\varphi''(r) =& h(r, \varphi(r),\varphi'(r), A'(r))\ .
\end{aligned}
\end{equation}
Here $f$ and $h$ are functions of the corresponding variables in their arguments, which are given in by~\eqref{eq:f},~\eqref{eq:h}. We note that in obtaining~\eqref{eq:ddAddphi} we focused on rewriting the $\theta\theta$ and scalar field equations~\eqref{eq:GmunuTmunu},~\eqref{eq:scalareomexpl}, however the final solutions of the metric function $A(r)$ and $\varphi$ are independent of this choice. In practice, finding the black hole solution requires solving~\eqref{eq:ddAddphi} for $A(r)$ and $\varphi(r)$ as a boundary value problem corresponding to~\eqref{eq:horcond} and~\eqref{eq:asympcond}. 

\subsection{Near-horizon and asymptotic behavior of the exact solutions}
\label{subsec:asympt}
 As our goal to obtain the spherically symmetric black hole solution has been reduced to solving the boundary value problem corresponding to~\eqref{eq:ddAddphi}, we study in this section the behavior of the metric functions and scalar field approaching these boundaries in more detail following~\cite{Sotiriou:2014pfa, Bakopoulos_2020}.  
\subsubsection{Asymptotic limit}\label{sec:asymplim}
From our estimate in section~\ref{sec:explFEsbound}, by substituting in this limit the Minkowski metric in the field equations, we found that the scalar field falls of as $\sim e^{- m r}/r$ to first order in $1/r$. We also limit the expansion of the asymptotic limit for the full solution to first order in the $1/r$. This is motivated by the perturbative results of Sec.~\ref{sec:pertboundaries}, which showed that higher order corrections in $1/r$ occur together with higher order powers of the exponent, hence these corrections are strongly suppressed. For the order $1/r$ correction to the metric functions, we can make the following argument. As the scalar field falls of exponentially, at spatial infinity the scalar field has decreased to zero. In the case of zero scalar field, the higher curvature corrections to the field equations vanish as well, see~\eqref{eq:GmunuTmunu}. This can also be reasoned from the action~\eqref{eq:actionsGB}, where for a vanishing scalar field, the prefactor of the GB invariant is constant and because the term is a topological invariant it becomes a boundary term and its contribution to the dynamics vanishes. The asymptotic behavior at order $1/r$ of the metric function $e^{A(r)}$ and $e^{B(r)}$ therefore correspond to the Schwarzschild metric. Again we know from the perturbative case that in this regime, higher orders in $1/r$ are strongly suppressed. The asymptotic behavior of the functions in~\eqref{eq:ddAddphi} is then given by
\begin{equation}\label{eq:asymplim}
\begin{aligned}
e^{A(r)}&\to \frac{A^{\prime}_{\infty}}{r}+\mathcal{O}(1/r^2)\ , \\
\varphi(r)&\to \frac{\varphi_{\infty}^{\prime} e^{-m r}}{r}+\mathcal{O}(1/r^2)\ .
\end{aligned}
\end{equation}
The integration constants $A^{\prime}_{\infty}$ and $\varphi_{\infty}^{\prime}$ are proportional to the system's ADM mass and scalar monopole charge respectively and are fixed by matching the solution to the near horizon limit detailed below.

\subsubsection{Near horizon limit}
The behavior of the metric functions and the scalar field at the horizon is given by~\eqref{eq:horcond}. The divergence in the function $A(r)$ implies $A'(r)\xrightarrow{}\infty$. Thus, $1/A'(r)\to 0$ and we expand the field equations~\eqref{eq:expBexpl} in $1/A'(r)$, which leads to
\begin{equation}\label{eq:expBexp}
\begin{aligned}
e^{B(r)}=&\frac{2 \alpha f'(\varphi) \varphi^{\prime}(r)+r}{\left(1-r^2 m^2 \varphi(r)^2\right)} A^{\prime}+\left[2 \alpha f'(\varphi) \varphi^{\prime}(r)\left(2-3 m^2 r^2\varphi(r)^2 +r^2\varphi'(r)^2\right)\right.\\
&\left.+r\left(r^2 \varphi'^2-1\right)\right]/\left.[\left(r^2m^2\varphi(r)^2-1\right)\left(2  \alpha f'(\varphi) \varphi^{\prime}(r)+r\right)\right]+\mathcal{O}\left(\frac{1}{A^{\prime}}\right)\ .
\end{aligned}
\end{equation}

Substituting the expanded expression~\eqref{eq:expBexp} in~\eqref{eq:ddAddphi} and expanding the equations in the same limit gives
\begin{subequations}\label{eq:ddAddphiexp}
\begin{eqnarray}
A^{\prime \prime}(r) & =&\frac{a}{b} A(r)^2+\mathcal{O}\left(A^{\prime}\right)\ , \\
\varphi^{\prime \prime}(r) & =&\frac{c}{b}\left(2 \alpha f'(\varphi) \varphi^{\prime}(r)+r\right) A^{\prime}(r)+\mathcal{O}(1)\ ,\label{eq:ddphihorizon}
\end{eqnarray}
\end{subequations}
where $a$, $b$ and $c$ are given by \eqref{eq:abc}. For $\varphi''(r)$ to remain finite as $A'(r)\xrightarrow{} \infty$, we require the coefficient of $A^\prime$ in~\eqref{eq:ddphihorizon} to vanish at a rate equal or faster than $A'$ diverges. 
However, comparing~\eqref{eq:ddAddphiexp} with~\eqref{eq:expBexp}, we see that letting $(2 \alpha f'(\varphi) \varphi^{\prime}(r)+r)$ vanish would also make the divergent term $\sim A^\prime$ of $e^{B(r)}$ vanish, which is inconsistent with the horizon condition~\eqref{eq:horcond}. 
Therefore, to impose regularity of the scalar field near the horizon requires $c\xrightarrow{}0$ and $b\neq 0$.

At the black hole horizon we can rewrite $c=0$ using the explicit expression~\eqref{eq:abc} as a condition on $\varphi'(r_h)=\varphi'_h$ given by
\begin{equation}\label{eq:phihprimem}
\varphi_h^{\prime}=-\frac{A \pm\left(1-m^2 r_h^2 \varphi_h{}^{2}\right) \sqrt{C}}{B}\ ,
\end{equation}
with $A$, $B$ and $C$ given by~\eqref{eq:AB} and~\eqref{eq:fullconstr}.
Only the minus solution converges to~\eqref{eq: boundphihmassless} in the small mass limit and to~\eqref{eq:phi1prime} in the small coupling limit. We also note that the square root in~\eqref{eq:phihprimem} adds an additional requirement as it should be positive definite, imposing an inequality which gives a further restriction on the parameters $\varphi_h, \, r_h$.

Next, considering the near-horizon expansion of the field equations and substituting the minus solution of~\eqref{eq:phihprimem} in~\eqref{eq:ddAddphiexp} yields
\begin{equation}
\begin{aligned}
& A^{\prime \prime}=-(A^\prime)^2+\mathcal{O}\left(A^{\prime}\right)\ , \label{eq:Aprimeprimehor}\\
& \varphi^{\prime \prime}=\mathcal{O}(1)\ .
\end{aligned}
\end{equation}
Integrating~\eqref{eq:Aprimeprimehor} yields a logarithmic function and fixing the integration constant such that the solution diverges to minus infinity at $r_h$ leads to the derivative $
A^{\prime}(r)\sim \frac{1}{r-r_h}$. Combining this with~\eqref{eq:expBexp} we obtain the near-horizon behavior of the metric components and scalar field
\begin{equation}\label{eq:horlim}
\begin{aligned}
e^{A(r)}&=A^{\prime}_h\left(r-r_h\right)+\mathcal{O}(r-r_h)\ , \\
\varphi(r)&=\varphi_h+\varphi_h^{\prime}\left(r-r_h\right)+\mathcal{O}(r-r_h)\ ,
\end{aligned}
\end{equation}
where $\varphi_h'$ is given by~\eqref{eq:phihprimem}. Then $A_h^{\prime}$, $\varphi_h$ are the only free integration constants which get fixed by matching with the asymptotic solution.

\subsection{Numerical computation of the full solution}\label{sec:numfullsol}
We use an initial value formulation to solve the second order differential equations~\eqref{eq:ddAddphi} for $A(r)$ and $\varphi(r)$ simultaneously again using the same specifications for the numerical integrator as mentioned in Sec.~\ref{sec:numsollin}. Note that $\alpha$, $m$ and $r_h$ are all input parameters in this initial value problem. The solution for $B(r)$ can be recovered by substituting these solutions in~\eqref{eq:expBexpl}. We start the integration at an infinitesimal distance ($r/r_h=1+10^{-3}$) outside the event horizon using the near-horizon solutions~\eqref{eq:horlim} and~\eqref{eq:phihprimem} as initial conditions. The amount of scalar field at the horizon $\varphi_h$ and the coefficient $A_h^{\prime}$ are determined by matching to the right asymptotic behavior. The unstable nature of the scalar field solution poses a challenge for solving~\eqref{eq:ddAddphi} simultaneously with the right asymptotic behavior. It turns out that the scalar field solution and approximation for $\varphi_h$ are not sensitive to the estimation for $A_h^{\prime}$. One can therefore obtain an educated guess for $\varphi_h$ independent of $A_h^{\prime}$ and use this guess to solve the system simultaneously. The scalar field solution up to some finite value of $r$ then already behaves as the exponentially decaying solution and a numerical root finding routine is then able to extract the initial conditions corresponding to the right asymptotic behaviors. 

More explicitly, we implement these considerations as follows. After defining the system of differential equations~\eqref{eq:ddAddphi} as functions of the initial values $\varphi_h$, $A_h^{\prime}$, we use the bisection method described in Sec.~\ref{sec:linsol} and Appendix~\ref{AppC} to obtain an educated guess for $\varphi_h$, setting $A_h^{\prime}$ temporarily to $1$. Looking at the scalar field solution with these initial conditions, we define the maximum $r$ for which the solution is still exponentially decaying as $r_{\infty}$, where for $r>r_\infty$ the exponentially growing mode takes over. We set up a shooting method routine similar to the methodology described in Sec.~\ref{sec:higherordercorr} and Appendix \ref{AppC} to find the initial conditions that match the solution to the asymptotic behavior~\eqref{eq:asymplim} at $r_{\infty}$. We justify matching the solutions to the asymptotic limit for some finite $r_{\infty}\neq \infty$ by similar arguments as for the higher order perturbative solutions. In brief, $r_\infty$ is the maximum distance where the scalar field has essentially fallen off to $0$. 
For a vanishing scalar field the metric is the Schwarzschild solution as described in Sec.~\ref{sec:asymplim}, hence we can already require the metric function $A(r)$ and scalar field to follow~\eqref{eq:asymplim} at $r_{\infty}$. Additionally as the constants $A^{\prime}_{\infty}$, $\varphi^{\prime}_{\infty}$ are unknown, we define our shooting method in terms of the ratios $e^{A}/(e^{A})'$ and $\varphi/\varphi'(r)$ as functions of the initial conditions to match 
\be
\frac{e^{A(r)}}{{e^{A(r)}}^{'}}\to-r\ , \qquad  \frac{\varphi(r)}{\varphi'(r)}\to -\frac{r}{(1+mr)}\ ,
\ee
 and determine $A^{\prime}_{\infty}$, $\varphi^{\prime}_{\infty}$ afterwards. We achieve this by defining a function of the difference between the metric solution with the initial conditions found as described above and the asymptotic limit in~\eqref{eq:asymplim}, and similarly for the scalar field solution, as a function of $A^{\prime}_{\infty}$ and $\varphi^{\prime}_{\infty}$ respectively. \\
To match the coefficients, we integrate over the absolute difference between the solution and the asymptotic limit and determine the constants $A^{\prime}_{\infty}$ and $\varphi^{\prime}_{\infty}$ that minimize the integral over a small region in $r$. For $A^{\prime}_{\infty}$ the small region was determined around $r_{\infty}$ and for $\varphi^{\prime}_{\infty}$ the region is based on integer multiples of the Compton wavelength. For each choice of parameters, we require that the minimized integral is $\lesssim 10^{-9}$ as criterion for a good match, where $A^{\prime}_{\infty}$ is approximately constant and thus less sensitive to the choice of integral range than $\varphi^{\prime}_{\infty}$, which requires matching two functions that are rapidly decaying.\\
Additionally, we are interested in the solution for the spacetime inside the horizon to see if the scalar field extends inside the horizon and to analyze the singular behavior of the spacetime inside the black hole. 
We therefore use an extension of the metric~\eqref{eq:SSSmetric} as done in~\cite{Sotiriou:2014pfa} by defining a coordinate patch inside the horizon described by similar metric potentials as in~\eqref{eq:SSSmetric} but the opposite signs. With this convention, we capture the sign flip that occurs for the time and radial components of the metric in Schwarzschild coordinates inside the horizon, for which the time coordinate becomes spacelike and vice versa. This switch is then incorporated in the additional minus sign and therefore the solution to the metric corrections itself can retain the same sign in- and outside the horizon. With this setup, we calculate numerical solutions to~\eqref{eq:ddAddphi} by integrating from a small distance inside the event horizon to $r=0$. An important assumption in this process needed to set the initial value of the scalar field is that the limit of the scalar field approaching the horizon from both sides exists and can be glued together smoothly. This implies that the same initial conditions and coefficients $\varphi_h$, $A_h^{\prime}$ apply as for the outside solution. However, the metric functions are discontinuous in this setup, for instance, the solution for $A(r)$ diverges to minus infinity on both sides of the horizon.\\
A caveat is that the solution inside the horizon in Schwarzschild coordinates is not very meaningful, for example, there is no intuitive interpretation of the coordinates. However we can nevertheless use this solution to show that the scalar field extends to the inside of the black hole and to analyze the behavior of curvature invariants inside the horizon. We compute and discuss these curvature scalars in Sec.~\ref{sec:propertiessol}. As these quantities contain coordinate independent information, the conclusions of our analysis are valid more generally beyond the particular choice of interior coordinates. \\
In this way, we construct the full numerical solution for $A(r)$ and $\varphi(r)$ in and outside the horizon. We compare this to the solution for a massless scalar field and the perturbative solution. 

\section{Comparison of the solutions}
For the numerical results discussed in this section, we specialize to the coupling function $f(\varphi)= e^{2\varphi}/4$. As mentioned in Sec.~\ref{sec:action} we focus on $\gamma=2$ as parameter in the exponent in the coupling function corresponding to EdGB. We fix this choice to keep the amount of free parameters tractable. Additionally, the effect of varying $\gamma$ on the properties of the BH solution 
was studied in \cite{Doneva:2019vuh} and shown to result in qualitatively similar behavior.
In general the effect of larger values of $\gamma$ correspond to an enlarged effect from the scalar field, leading for example to stronger constraints on the coupling~\cite{Yordanov:2024lfk}. \\
In this section we first show the result of the perturbative and exact numerical solution for the metric function $A(r)$ and the scalar field $\varphi(r)$ for different values of the coupling and scalar field mass. We discuss the dependency of these two parameters on the solutions and compare the massless and massive solutions to the perturbative case. Secondly, we show an overview of different sGB black hole solutions analysing the difference in horizon radius as compared to the Schwarzschild radius as function of the black hole mass for different choices of the coupling and scalar field mass. \\

\subsection{Comparison between massless, massive, and perturbative solutions}\label{sec:showsol}
 Figure~\ref{fig:Asol} shows different results for the metric function $A(r)$ defined in~\eqref{eq:SSSmetric} and Fig.~\ref{fig:phisol} displays the corresponding scalar field. The pink curves correspond to the full solution with vanishing scalar field mass, while black curves are the results for a mass of $\hat{m}=0.1$. The left panels are for a larger value of the coupling than the right ones. For the perturbative and Schwarzschild solutions we only show the curves outside the black hole horizon. 

Before discussing the results, we note an important point regarding comparisons between the perturbative and exact solutions. The perturbative solutions are computed in terms of $u=r_S/r$ and similarly for the Schwarzschild solution. These need to be rescaled to compare with the full solution shown here in terms of $r_h/r$. We choose to compare black holes with the same ADM mass\footnote{For sGB and Schwarzschild black holes with the same ADM mass, the global mass generally differs due to the contributions from the scalar field in sGB~\cite{Julie:2019sab}}, which implies for the asymptotic limit of the full sGB solutions~\eqref{eq:asymplim} that $A^{(1)}_{\infty}/r = r_S/r$. 
Next, we rescale the radial coordinate of the full solution such that $r_h=1$. The ratio between the Schwarzschild and sGB horizons can be obtained via $r_S/r_h = A^{(1)}_{\infty}/1$ and is used to rescale the perturbative and Schwarzschild solution in the figures below. 

\subsubsection{Massless case: code check and singularity}

First, we focus on the massless case as it has been more comprehensively studied in previous literature. For an independent check of our results, we compare the pink curves in Fig.~\ref{fig:Asol} with a corresponding result in Fig.~1 of \cite{Sotiriou:2014pfa} and verify a similar qualitative behavior, up to small differences arising from different choices of coupling functions and -constant. 
Next, we analyze the features of the metric potential in Fig.~\ref{fig:Asol} and corresponding scalar profile in Fig.~\ref{fig:phisol}. At large distances, they show the expected asymptotic behavior $A(r)\to 0$ and an exponential decay for the scalar field. Near the horizon (black vertical line), the scalar field remains finite while $A(r)$ diverges to minus infinity when approaching from the outside. For the coupling $\hat \alpha \sim 0.2$ shown in the left panel of Fig.~\ref{fig:Asol}
the divergence for $r<r_h$ occurs very close to the horizon $r/r_h\sim0.99$ (pink dashed line). This is due to the presence of a finite radius singularity, which is a well known phenomenon for massless sGB black holes~\cite{Sotiriou:2014pfa,Sullivan:2019vyi,Julie:2022huo}. We will make a concrete identification between this divergence in $A(r)$ and a genuine curvature singularity in Sec.~\ref{sec:curvscal}. We see from the right panel of Fig.~\ref{fig:Asol} that for a smaller coupling $\hat \alpha$, the singularity moves further to the interior, as expected based on recovering the GR limit for zero coupling. 
 This implies that the maximum value of $\hat \alpha$ for which a black hole exists is determined by the singularity coinciding with the horizon; higher values of $\hat \alpha$ will lead to a naked singularity.

 \subsubsection{Effect of the scalar mass}
Qualitatively, the features of the solutions for finite scalar field mass are similar to the massless case. For the metric functions outside the horizon, the mass has a very small effect, as seen in Fig.~\ref{fig:Asol}, while for the scalar field in Fig.~\ref{fig:phisol} the differences are more noticeable. The singularity for the massive case occurs at $r/r_h=0.97$ for a coupling of $\hat{\alpha}=0.2$. A larger mass of the scalar field thus shifts the singularity further inwards, as also expected from the infinite mass limit, where the scalar condensate disappears and the black hole reduces to Schwarzschild with a singularity at $r/r_h=0$.
This implies that the maximum value of the coupling for which black hole solutions exist increases for larger scalar field masses, consistent with the results of~\cite{Doneva:2019vuh}. \\
 From a computational perspective we directly identify the maximum $\hat \alpha$ for black hole solutions based on the fact that for any value exceeding it, the near-horizon initial conditions and the asymptotically flat limit can no longer be connected by a smooth numerical solution. 

\subsubsection{Performance of the perturbative small-coupling solutions and comparison to\\ Schwarzschild}\label{sec:showsolpert}
\begin{figure}[!h]
\centering
\begin{subfigure}{0.49\linewidth}
   \includegraphics[width=\linewidth]{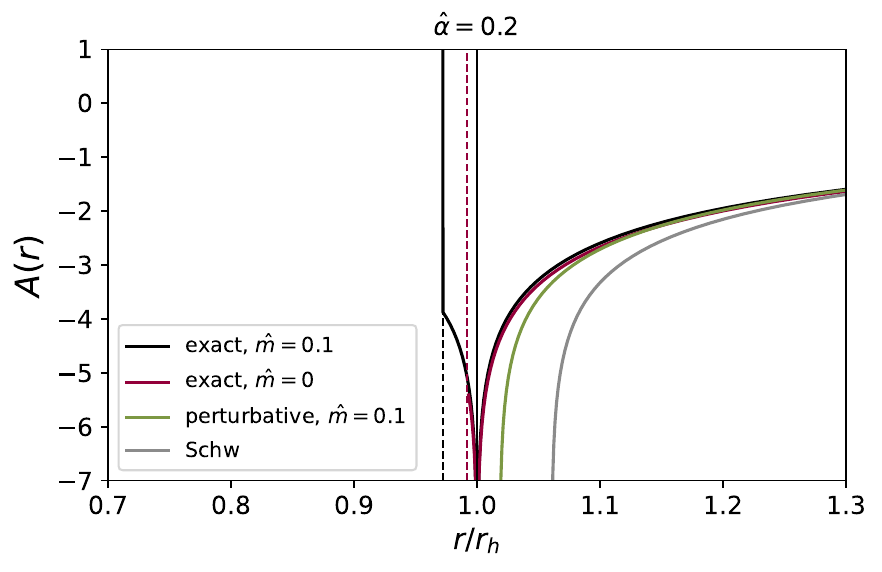}
       \centering
\end{subfigure}
\begin{subfigure}{0.49\linewidth}
    \includegraphics[width=\linewidth]{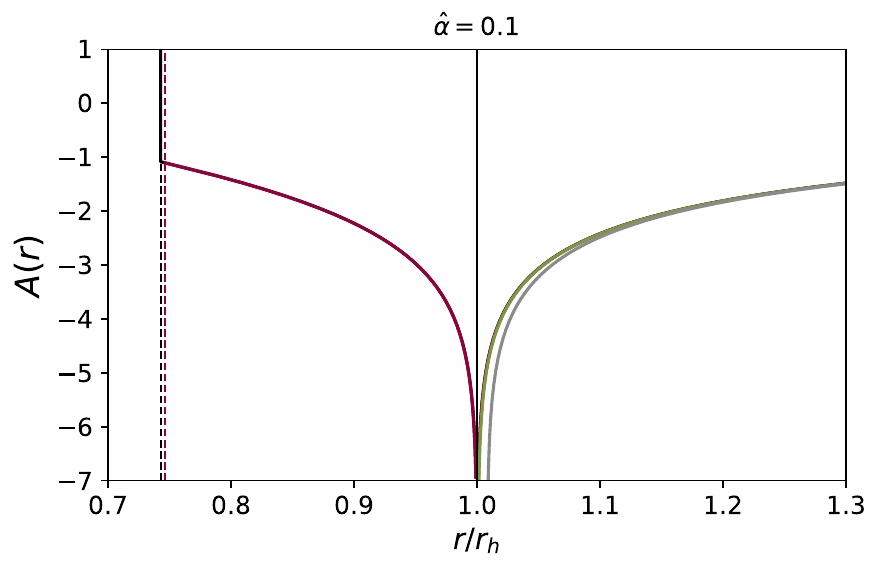}
        \centering
 \end{subfigure}
\caption[]{\emph{Behavior of the metric function $A(r)$ characterizing the time-time component of the metric} for couplings of $\hat{\alpha}=0.2$ (left panel) and $\hat{\alpha}=0.1$ (right panel). Black curves show the full solution for a scalar field mass $\hat{m}=0.1$, pink curves the massless case, green curves represent the perturbative solution including corrections to $O(\hat{\alpha}^3)$ and grey curves show the Schwarzschild solution for comparison. For the latter two only the solutions outside the horizon are shown. The black vertical line denotes the horizon radius and the vertical dashed curves the singularities. }
\label{fig:Asol}
\end{figure}

\begin{figure}[h!]
\centering
\begin{subfigure}{0.49\textwidth}
   \includegraphics[width=\linewidth]{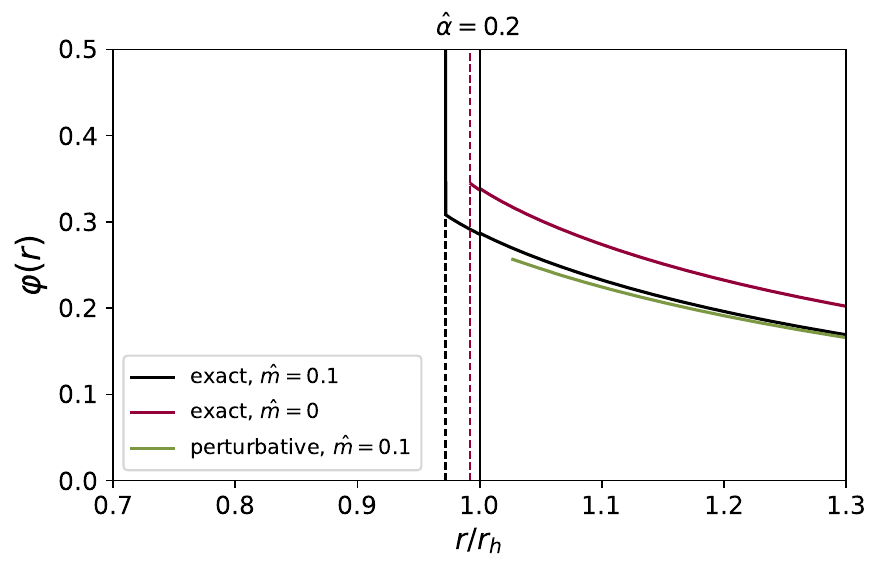}
       \centering
   \label{fig:boundaryfreq1} 
\end{subfigure}
\begin{subfigure}{0.49\textwidth}
    \includegraphics[width=\linewidth]{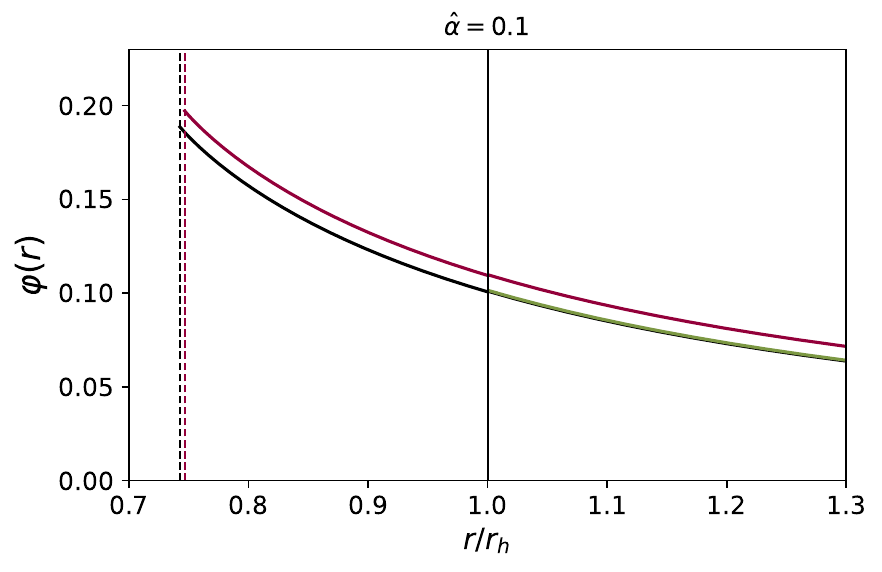}
    \label{fig:boundaryfreq2}
        \centering
 \end{subfigure}
\caption[]{\emph{Scalar field profile} for couplings of $\hat{\alpha}=0.2$ (left panel) and $\hat{\alpha}=0.1$ (right panel). Black curves show the full solution for a scalar field mass $\hat{m}=0.1$, pink curves are for the massless case, and green curves represent the perturbative solution including corrections to $O(\hat{\alpha}^3)$ only showing the solution outside the horizon. The black vertical line denotes the horizon radius and the vertical dashed curves the singularities.}
\label{fig:phisol}
\end{figure}

Another interesting feature illustrated in Figs.~\ref{fig:Asol} and~\ref{fig:phisol} is the quality of the perturbative solutions to $O(\hat{\alpha}^3)$ corresponding to the green curves. We see that near the horizon for the larger value of the coupling (left panel) the perturbative solution differs appreciably from the full solution. This is most noticeable when comparing the locations of the horizon, which for the perturbative and Schwarzschild solution lie at larger radial coordinate than the full solution, as indicated by the divergence of $A$ to $-\infty$. While the near-horizon behavior of the perturbative solution is based on expanding around the Schwarzschild horizon~\eqref{eq:nearhoranz}, the actual black hole horizon in this case is determined by the root of $\bar{B}$. After appropriately rescaling coordinates as described in the beginning of the section, this leads to the horizon locations indicated in the plots. As expected, for larger couplings the differences between the perturbative and exact solutions become larger, which is especially noticeable near the horizon. As mentioned, for larger couplings the singularity lies close to the horizon, and it is reasonable to expect non-perturbative effects to be important in its vicinity. In the large $r/r_h$ limit, the perturbative and numerical solutions coincide as the curvature effects become less and less significant. We also see that for the smaller coupling shown in the right panel, the perturbative solution agrees much better with the full solutions near the horizon, as it is also farther from the singularity and the horizon moves closer to $r_h$. 
In Appendix~\ref{Apppert} we give some additional analysis on the perturbative solution comparing also the solutions up to different orders in the coupling. Together with this analysis we conclude that the perturbative solution deviates less from the exact solution in the large scalar mass regime. As expected as for large scalar masses the scalar field decreases and decouples in the limit of the mass to infinity. 
Furthermore, we find no particular behavioural change comparing the solution up to quadratic and cubic order, for which the metric corrections to scalar field come in, see Table~\ref{tab:dependencies}. Lastly we find the difference of the perturbative solution in the near horizon region to be largest. However even with the finite radius singularity lying close to the horizon for larger values of the coupling, when restricting to the regimes away from the immediate vicinity of the divergence, we find no sign of qualitatively new non-perturbative behaviour that would not be approximately captured by adding higher small coupling corrections to the solution.

\subsection{Comparing horizon radii and mass for different BH solutions}
\begin{figure}[h!]
\centering
\begin{subfigure}{0.6\textwidth}
   \includegraphics[width=\linewidth]{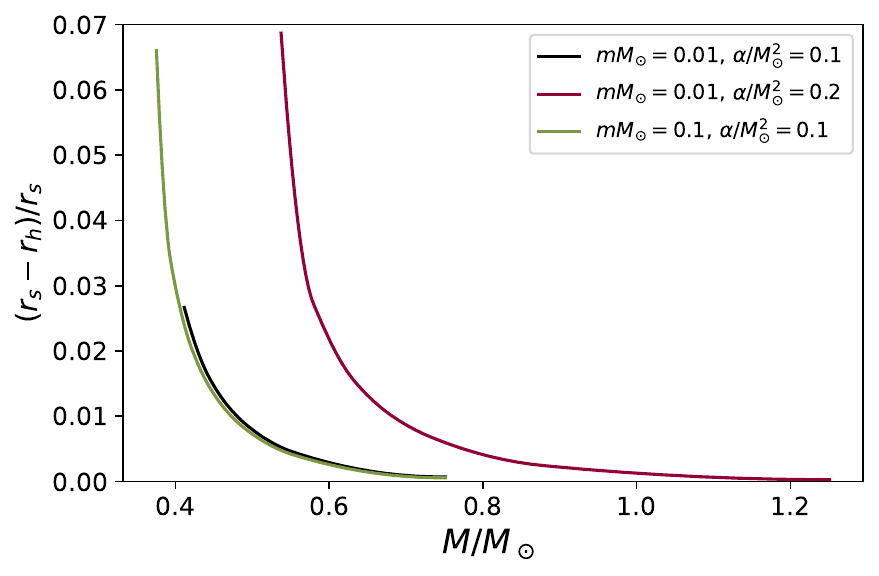}
       \centering
\end{subfigure}
\caption[]{\emph{Fractional difference of the horizon radius with the Schwarzschild radius} as function of the black hole mass for different values of the scalar field mass and coupling constant. Note the scalar field mass and coupling constant are rescaled by $M_{\odot}$ instead of the horizon radius as in \eqref{eq:mhatdef} and \eqref{eq:alphahat}. The curves end when the difference with the Schwarzschild radius becomes smaller than the initial offset to the horizon radius when computing the numerical solution, see Sec.~\ref{sec:numfullsol}.  }
\label{fig:rhM}
\end{figure}

To give an overview of different sGB black hole solutions compared to Schwarzschild black holes, we find the solution to \eqref{eq:ddAddphi} for the metric functions and scalar field for different black hole masses. As mentioned in Sec.~\ref{sec:asymplim} and Sec.~\ref{sec:numfullsol}, the ADM mass of the black hole, corresponding to half the $1/r$ coefficient of the metric function in the asymptotic limit, is extracted after computing the metric solution.\\ 
We rescaled the coordinates and correspondingly the scalar field mass and coupling constant, see Sec.~\ref{sec:fieldequations}, with $M_{\odot}$ instead of the horizon radius, working in natural units. Consequently, $r_h$ becomes, together with the scalar field mass and coupling, an input parameter of the initial value problem instead of being equal $1$ in rescaled coordinates $r/r_h$. For two choices of the rescaled scalar field mass and coupling constant we then obtained the metric and scalar field solutions as described in Sec.~\ref{sec:numfullsol} for a range of black hole radii and determined the corresponding black hole mass. We show the results in Fig.~\ref{fig:rhM}. Depending on the choice of the parameters we find a minimum horizon radius for which no numerical solution can be found connecting the near horizon and asymptotic limit for the fields. In Sec.~\ref{sec:phihanalysis} we find this corresponds to the smallest radius for which the singularity is still censored by the horizon. Additionally, the curves stop for a finite value of the black hole mass, this is a consequence of limited numerical accuracy. As mentioned in Sec.~\ref{sec:numfullsol}, to solve \eqref{eq:ddAddphi} requires the integration to start at a small offset to the horizon radius ($r/r_h = 1+10^{-3}$). When the difference between the horizon radii shrinks and becomes smaller than this offset, this difference can no longer be accurately obtained. Hence, we stopped the computation at the corresponding black hole mass. Qualitatively in the large black hole mass limit the horizon radius approaches the Schwarzschild radius. \\
A comparable diagram was shown in Fig. 5 of \cite{Doneva:2019vuh}, for a broader range of scalar field masses and also varying the coefficient in the coupling function. Note the coordinates and parameters in this work are rescaled by $\sqrt{\alpha}$. Fig.~\ref{fig:rhM} complements this diagram by showing the effect of the mass and coupling on the black hole horizon separately. \\

Comparing the black and green curve in Fig.~\ref{fig:rhM} one can see the effect of increasing the scalar field mass for a constant coupling function. The minimum black hole mass slightly decreases for a larger scalar field mass. This corresponds to our findings of the solutions inside the horizon in Fig.~\ref{fig:Asol},~\ref{fig:phisol}. For the massive case, the radius for which the functions diverge are shifted inwards compared to the massless solution. In Sec.~\ref{sec:curvscal} we'll show this point of divergence corresponds to a curvature singularity. Hence, for larger scalar field masses the singularity moves inwards and the black hole is allowed to be smaller for the singularity to stay inside the horizon, corresponding to smaller black hole masses. Additionally, we find that for a larger scalar field mass the difference with the Schwarzschild radius increases for black hole masses close to the minimum allowed mass. Thus as the black hole is allowed to be smaller because of the inwards shifting singularity, for larger scalar field masses the offset to $r_S$ increases accordingly. \\
We can obtain similar conclusions comparing the black and pink curves varying the coupling. As found in Fig.~\ref{fig:Asol} the singularity moves closer to the horizon by increasing the coupling constant. The minimum black hole mass therefore increases for the pink curve corresponding to a larger coupling. The increase is much more apparent for the compared coupling constants then the decrease for the two compared scalar field masses. We do find an opposite relation for the offset of the horizon radius compared to $r_S$ near the minimum black hole mass. Even though the black holes have to be larger to contain the singularity inside the horizon for a larger coupling value, the offset with $r_S$ slightly increases for a larger value of the coupling. This increase is much smaller than for the two compared scalar field masses.\\

\section{Properties of the solutions }\label{sec:propertiessol}
Having constructed the full and perturbative numerical solutions for a static black hole in massive sGB, we analyse the properties of these solutions. We start by studying the spacetime curvature in and outside the horizon and recover how properties such as the amount of scalar field on the horizon or the scalar monopole charge depend on the parameters of the theory.
The analysis in this section complements the discussion of \cite{Doneva:2019vuh} which focused on the horizon radius, amount of scalar field at the horizon, black hole surface, entropy, and temperature as function of the black hole mass for different coupling functions and scalar field masses. Note that the rescalings in~\cite{Doneva:2019vuh} to obtain dimensionless variables are different from those used in this paper, in particular, we rescale based on the horizon radius, while~\cite{Doneva:2019vuh} rescaled by the coupling constant. In all further analysis we specify to a dilatonic coupling function $f(\varphi)= e^{2\varphi}/4$.

\subsection{Characterizing the curvature and field density}
Before we analyze more specifically how certain properties of the black hole solutions depend on the parameters of the theory, we first consider the curvature scalars and energy density around the black hole to gain more intuition for the solutions.

\subsubsection{Curvature invariants and singularity}\label{sec:curvscal}
To characterize the curvature we analyze the curvature invariants. Here we focus on the Kretschmann scalar 
\be
\mathcal{K} = R_{\mu\nu\rho\sigma}R^{\mu\nu\rho\sigma}\ ,
\ee
and its cousin; the fully contracted Weyl tensor squared 
\be
\mathcal{C} = C_{\mu\nu\rho\sigma}C^{\mu\nu\rho\sigma}\ .
\ee
In vacuum in GR these two invariants coincide. We calculate them using the full numerical solution for a coupling of $\hat{\alpha}=0.2$ and for masses of $\hat{m}=0.1$ and $\hat{m}=1$. 
\begin{figure}[h!]
\centering
\begin{subfigure}{0.49\textwidth}
   \includegraphics[width=\linewidth]{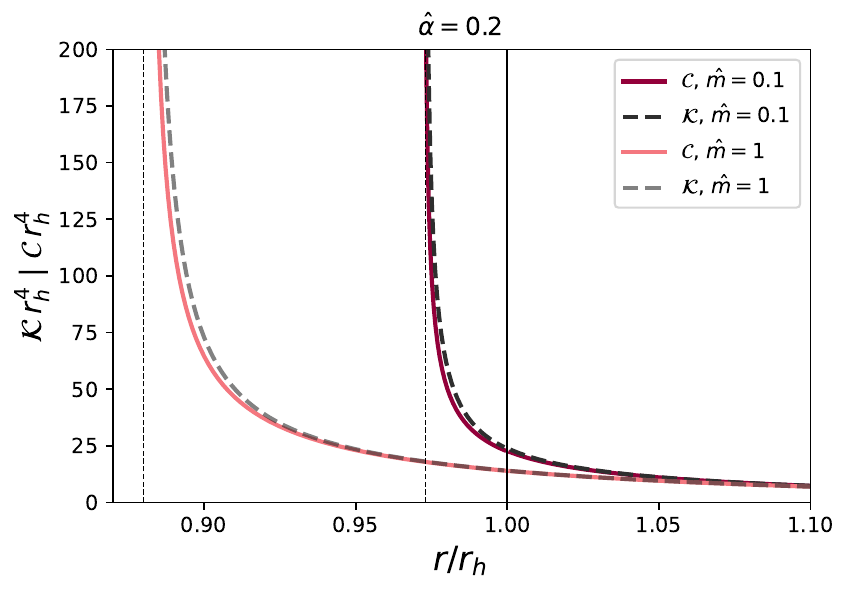}
       \centering
\end{subfigure}
\begin{subfigure}{0.49\textwidth}
    \includegraphics[width=\linewidth]{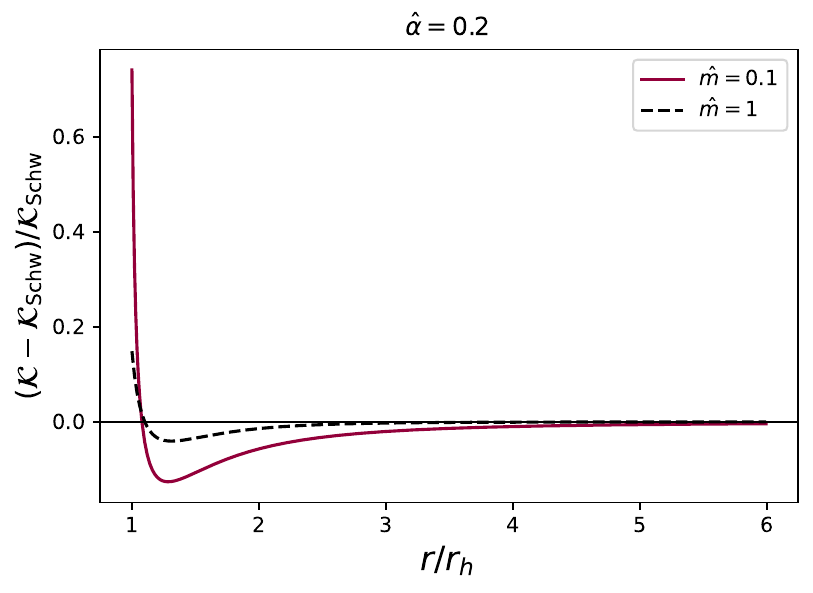}
        \centering
 \end{subfigure}
\caption[]{left panel: \emph{Curvature invariants} of the Kretschmann scalar $\mathcal{K}$ (dashed lines) and the contracted Weyl tensor $\mathcal{C}$ (pink and purple curves) in- and outside the event horizon (black vertical line) for two values of the scalar field mass (lighter colors for larger mass). The vertical dashed black lines denote the location of the finite radius singularity. right panel: The \emph{percent difference} of $\mathcal{K}$ in massive sGB and Schwarzschild for two values of the scalar field mass only for the spacetime outside the horizon. } 
       \label{fig:curv}
\end{figure}
The results are illustrated in Fig.~\ref{fig:curv}, where the right panel shows the percent difference of the Kretschmann scalar for a massive sGB 
and Schwarzschild black hole. Here we make the same choice as for Fig.~\ref{fig:Asol}, comparing to a Schwarzschild black hole with the same ADM mass. We see that the curvature invariants blow up for $r/r_h\sim 0.88$ and $r/r_h\sim 0.97$ for $\hat{m}=1$ and $\hat{m}=0.1$ respectively. For $\hat{m}=0.1$ this corresponds to the same location as the divergences in $A$ and $\varphi$ seen in the left panels in Fig.~\ref{fig:Asol} and~\ref{fig:phisol}, which corroborates the identifications between these divergences and genuine singularities already mentioned in Sec.~\ref{sec:fullsol}. We also note from comparing the solid curves corresponding to ${\cal C}$ and the dashed lines illustrating the results for ${\cal K}$ in Fig.~\ref{fig:curv} that while for most regions outside the horizon the two kinds of curvature invariants coincide, they differ slightly in its immediate vicinity and the interior. \\
Looking at the right panel of Fig.~\ref{fig:curv} we see that close to the horizon up to $r/r_h \sim 1.1$, the curvature in sGB gravity is larger than for the Schwarzschild black hole. Interestingly, however, in the region $1.1 \lesssim r/r_h \lesssim 5$ the curvature in sGB is weaker than Schwarzschild, with the fractional difference attaining its largest negative value around $r/r_h\sim 1.3$. 
In the large-$r$ limit the curvature invariants coincide, as expected. With increasing scalar field mass, the curvature decreases. Hence, the massless limit leads to the strongest curvature and thus largest deviation from Schwarzschild. 
The distinguishability of the curvature up to $r/r_h\sim 5$ could have interesting consequences, for instance, for tidal effects.

\subsubsection{Energy density}
\begin{figure}
        \centering
        \begin{subfigure}[b]{0.49\textwidth}
            \centering
            \includegraphics[width=1\textwidth]{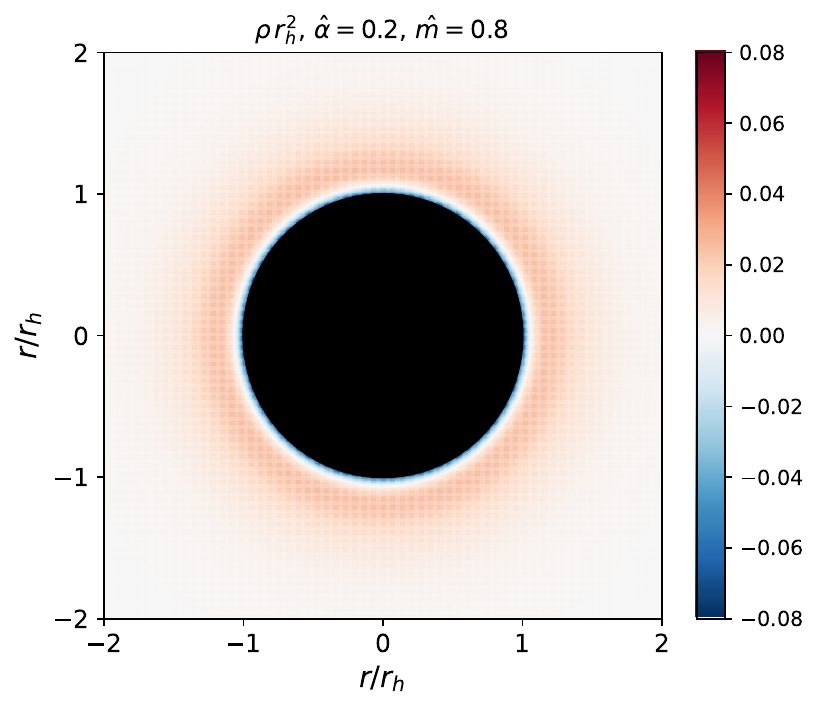}
        \end{subfigure}
        \begin{subfigure}[b]{0.49\textwidth}   
            \centering 
            \includegraphics[width=\textwidth]{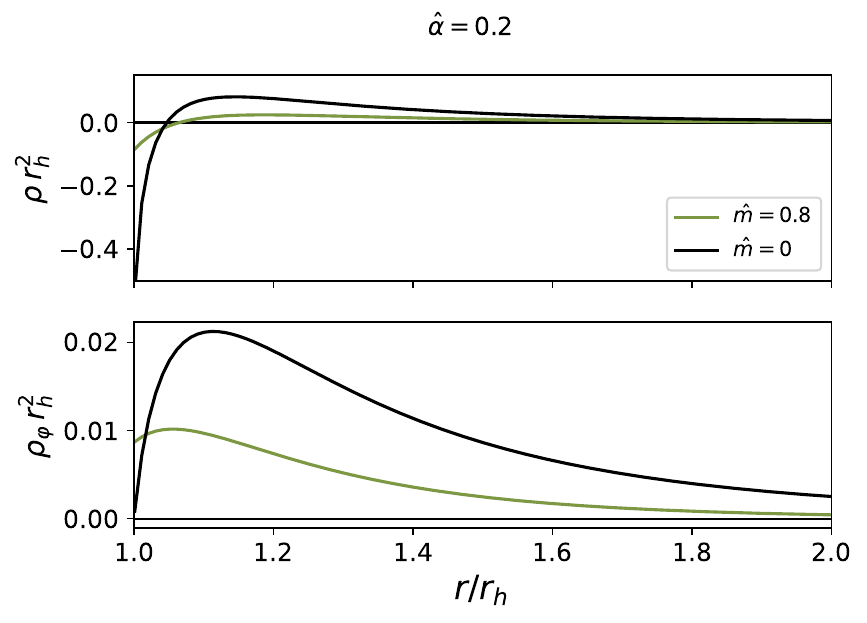}
        \end{subfigure}
        \caption{left panel: \emph{2D figure of the energy density }around the black hole shown as the black disk for a system with $\hat{\alpha}=0.2$, $\hat{m}=0.8$. right panels: \emph{Energy density (top) and scalar part of the energy density (bottom)} as function of $r$ for a massless and $\hat{m}=0.8$ scalar field.}
        \label{fig:energydensity}
\end{figure}

The energy density of the spacetime is given by $T_t^t=-\rho$ in~\eqref{eq:GmunuTmunu}. Additionally we define $\rho_{\varphi}$ as the pure scalar contributions of $T_t^t$ which can be obtained by setting $\alpha\to 0$ in~\eqref{eq:GmunuTmunu}. The results for the energy densities for a case with the maximum coupling for a massless sGB black hole are illustrated in Fig.~\ref{fig:energydensity}. The left panel shows the full energy density including the higher curvature contributions for a scalar mass $\hat m=0.8$, while the right panels show the corresponding radial profiles for that case (green curves) and the massless one (black curves). 

The upper panel of Fig.~\ref{fig:energydensity} shows that $\rho$ is concentrated close to the horizon and becomes more dilute further away from the black hole. Around $r/r_h\sim 2$ the energy density has already fallen off to essentially zero. From the right panels of Fig.~\ref{fig:energydensity}, we see that for the full energy density (top), the same behavior occurs in the massless case, also around the same values. However, the pure scalar field contribution to the energy density (bottom) has very different features, namely for the massless field configuration, the falloff to zero is much slower, as expected based on the asymptotic behavior of the field~\eqref{eq:asymplim} indicating the scalar field is supressed for distances larger than the Compton wavelength. Specifically, the percent difference in $\rho_{\varphi}$ between one and two times the Compton wavelength  ($\lambda_{\varphi}=1.25$) for the massive case is $99\%$. For comparison, in the massless case, the falloff of the density between the same radial distances is only $94\%$. Another interesting feature is that while the scalar field contribution is always positive, the full energy density is not. The reason is that the higher curvature contributions can have different signs, which leads to a negative total energy density close to the black hole horizon. The fact that the energy density can become negative is one of the reasons black holes both in massless and massive sGB can evade the no hair theorem \cite{Bekenstein:1995un, Kanti:1995vq}. 

\subsection{Scalar hair, regularity constraint and bound on the coupling}\label{sec:phihanalysis}
As explained in Sec.~\ref{subsec:asympt}, requiring the scalar field solution to be regular at the horizon leads to a constraint for the derivative of the scalar field at the horizon, c.f.~\eqref{eq:phi1prime},~\eqref{eq: boundphihmassless} and~\eqref{eq:phihprimem} for the linear-in-coupling, massless and massive full theory respectively.  For the linearized case we showed in Sec.~\ref{sec:linfieldNHasmp} and Appendix~\ref{AppB} that the solution for the scalar field should be a monotonically decreasing (or increasing in terms of $u$) function and therefore that the scalar field derivative at the horizon should be negative. For the scalar field background solution for general coupling we expect to find ground state type behaviour as well corresponding to a monotonically decreasing solution, as sudden bumps or wells are expected for exited states. This assumption is also build upon the fact that the full exact solution should reduce to the linearized solution in the small coupling limit. Also numerically, the solutions to \eqref{eq:ddAddphi} found to connect the near horizon and asymptotic limits of the scalar field for generic coupling, see Fig.~\ref{fig:phisol}, show this type of function too. Hence we state the scalar field at the horizon for generic coupling ought negative to be able to find a solution to \eqref{eq:ddAddphi} connecting the near horizon and asymptotic limits. 

For the linearized case requiring $\varphi_h'<0$ is accomplished via~\eqref{eq:constraint1} and in the massless full theory case this is done by imposing the square root to be real via~\eqref{eq:mlconstralpha}. However, in the massive case requiring the square root to be real by imposing $C>0$ does not ensure $\varphi_h'<0$~\eqref{eq:phihprimem}. Therefore in this case both $C>0$ and $\varphi_h'<0$ need to be imposed to ensure an asymptotically flat solution. All of these inequalities depend on the parameters $\varphi_h$, $\hat{\alpha}$ and $\hat{m}$. The dependence on $r_h$ is encapsulated in the dimensionless parameters $\hat{\alpha}$, $\hat{m}$. In this section, we study these inequalities imposed near the horizon to determine how $\varphi_h$ depends on the theory parameters.

\begin{figure}[h]
\centering
\begin{subfigure}{0.49\textwidth}
   \includegraphics[width=1\linewidth]{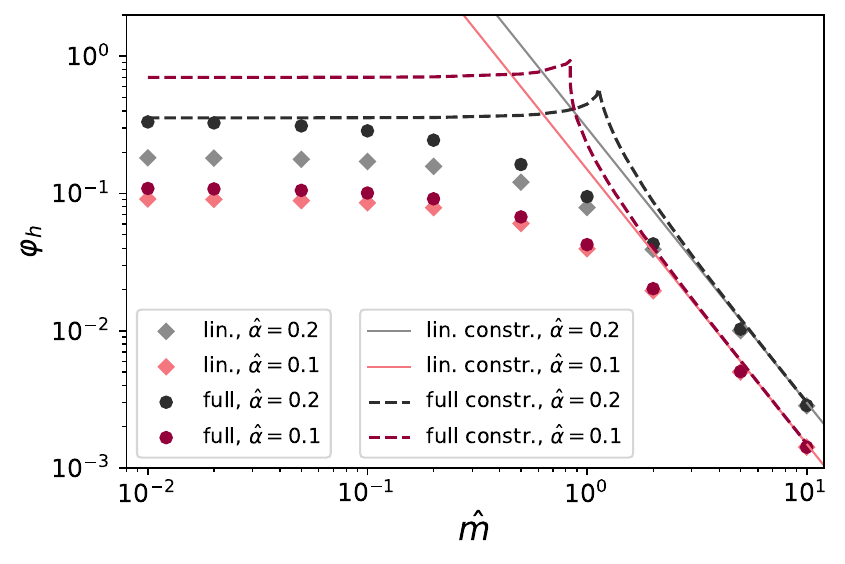}
       \centering 
\end{subfigure}
\begin{subfigure}{0.5\textwidth}
   \includegraphics[width=1\linewidth]{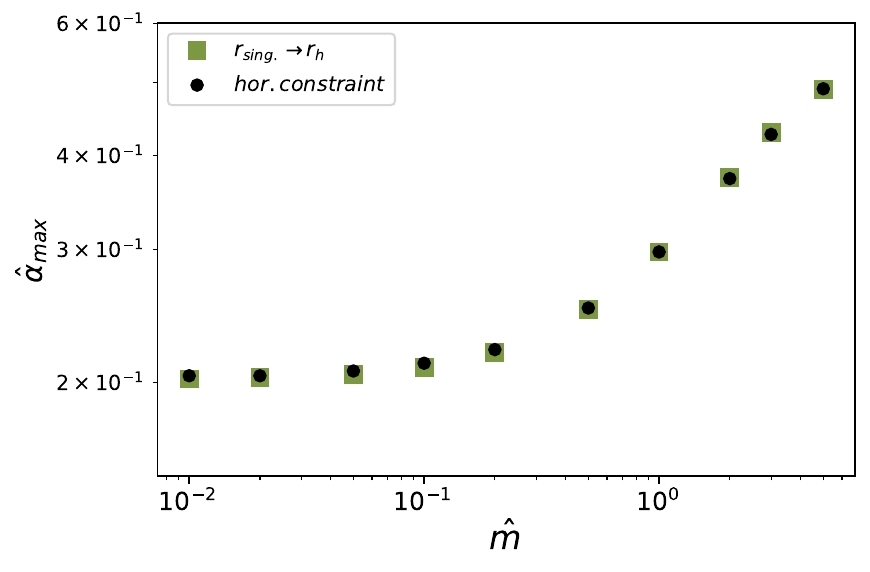}
       \centering 
\end{subfigure}
\caption[]{left panel: The amount of \emph{scalar field at the horizon $\varphi_h$} as function of the dimensionless scalar field mass $\hat{m}$. The panel shows the values for the linearized (diamonds) and full solution (dots) of the scalar field equation for two different values of the coupling (grey and pink shades). Additionally, the linearized~\eqref{eq:constraint1} (solid lines) and full theory constraints~\eqref{eq:phihprimem} (dashed lines) are shown for two different values of the coupling.
right panel: \emph{maximum allowed value for the coupling constant} as a function of the scalar field mass from the requirement of preventing a naked singularity (squares) and the near horizon constraint~\eqref{eq:phihprimem} (dots).} 
\label{fig:phim}
\end{figure}

\noindent The left panel of Fig.~\ref{fig:phim} compares the following near horizon results as function of the scalar field mass
\begin{itemize}
    \item \emph{solid lines}: linearized theory near horizon constraint \eqref{eq:constraint1},
    \item \emph{dashed curves}: full theory near horizon constraint \eqref{eq:phihprimem},
    \item \emph{diamonds}: value $\varphi_h$ from the linearized numerical solution,
    \item \emph{dots}: value $\varphi_h$ from the full theory numerical solution,
\end{itemize}
all for two values of the coupling constant in gray and pink shades for respectively $\hat{\alpha}=0.2$, $\hat{\alpha}=0.1$.\\
We see that $\varphi_h$ and the difference of this parameter between the linearized and full theory is largest for a larger coupling, as expected. For scalar field masses larger than $\hat{m}>1$, where in dimensionfull parameters the Compton wavelength lies inside the black hole horizon, the scalar hair is severely suppressed (note the logarithmic scale of the plot). In the large $\hat m$ limit, the linearized and full theory result coincide as for $\hat{m}\rightarrow\infty$ the scalar field should decouple and black holes should have no hair.\\
In the same panel we also show the linearized inequality~\eqref{eq:constraint1} as the pink and gray lines. We obtain the full theory constraint by selecting the largest value for $\varphi_h$ allowed for which~\eqref{eq:phihprimem} is real and negative for each choice of $\hat{\alpha}$ and $\hat{m}$.\\
For increasing $\hat m$, we find that beyond a certain coupling-dependent threshold that coincides with the cusp in the dashed curves in Fig.~\ref{fig:phim}, two branches of values for $\varphi_h$ arise for which $\varphi_h'<0$. For one branch the values of $\varphi_h$ becomes larger for larger mass while for the other branch they become smaller, which we identify as the desired physical solution. We therefore selected the largest possible $\varphi_h$ in the physical branch.\\
From Fig.~\ref{fig:phim} we see that the linearized and full constraints coincide in the large mass limit as required. The values for $\varphi_h$ obtained from the numerical solutions are always below the curves from the near-horizon constraints. In the small-mass limit, the matching to the asymptotic falloff fixes $\varphi_h$ to smaller values than allowed by the near-horizon constraints. In the zero-mass limit and largest possible coupling in the massless theory is $\hat{\alpha}\sim 0.2$, we find indeed that when approaching the massless limit $\varphi_h$ approaches the largest allowed value by the near-horizon constraint. Similarly, in the large-mass limit, the numerical solution for $\varphi_h$ approaches the maximum allowed value by the corresponding near-horizon constraint.   \\

\noindent The literature on the massless theory suggests that the near-horizon constraint~\eqref{eq: boundphihmassless} prevents the finite surface singularity from extending outside the black hole horizon. We analyze the link between the singularity and the near-horizon constraint in the massive theory in the right panel of Fig.~\ref{fig:phim}. \\
The green squares correspond to the maximum value for the coupling constant for which the singularity lies on the horizon radius. We obtained these values by computing the solution for the scalar field and metric functions and increasing the coupling constant up to when the singularity came to lie on the horizon. This was repeated for each value of the scalar field mass as shown in Fig.~\ref{fig:phim}.\\
The black dots correspond to the maximum values of the coupling for which the near horizon constraint~\eqref{eq:phihprimem} is still real and negative; $\varphi_h'<0$. These values for the coupling are found by substituting the values for $\varphi_h$, found when computing the solutions to obtain the green square values, in~\eqref{eq:phihprimem} and solving the inequality $\varphi_h'<0$ for $\hat{\alpha}$ for each value of the scalar field mass. For the more detailed discussion on this procedure and a short analysis on the maximum allowed $\varphi_h$ we refer to Appendix~\ref{Appphimax}.\\
From this right panel of Fig.~\ref{fig:phim} we find that for the studied mass range the near horizon constraint coupling values agree with the singularity constraint maximum coupling. Hereby we thus conclude that indeed the near horizon constraint ensuring regularity of the scalar field at the horizon prevents the finite radius singularity from lying outside the BH horizon. Additionally, these numerically obtained maximum values give a theoretical constraint on the coupling values for the considered scalar mass range. 
Note that the constraint from Fig.~\ref{fig:phim} is on the dimensionless coupling constant \eqref{eq:alphahat}, hence rescaled by the horizon radius. One can invert the argument by rescaling with $M_{\odot}$ and freeing the $r_h$ parameter. When fixing the coupling constant this leads to a theoretical constraint on the minimum black hole mass instead, which we showed in Fig.~\ref{fig:rhM}. 

\subsubsection{Implications in relation to the coupling and scalar field mass}\label{sec:impl}
The results from the right panel on Fig.~\ref{fig:phim} show the theoretical bound on the coupling as function of the scalar field mass. As mentioned in Sec.~\ref{sec:action} a first observational constraint on the coupling is $\sqrt{\alpha}\lesssim 2.47 \textrm{km}$ for $10^{-15}\textrm{eV}\lesssim m\lesssim 10^{-13}\textrm{eV}$ based on the observed GWs from the first two observing runs of LVK \cite{Yamada:2019zrb}.\\
To compare this constraint on the coupling to our theoretical bound on the coupling constant found in Fig.~\ref{fig:phim}, we find the stated scalar mass range from \cite{Yamada:2019zrb} in terms of the dimensionless mass~\eqref{eq:mhatdef};
\begin{itemize}
    \item To convert $10^{-15}\textrm{eV}\lesssim m\lesssim 10^{-13}\textrm{eV}$ to the dimensionless $\hat{m}$, we combine \eqref{eq:mhatdef} and \eqref{eq:scalarmassparam} resulting in 
    \begin{equation}\label{eq:mevtomhat}
      \hat{m}=m_{eV} \frac{r_h e}{\hbar c}  ,
    \end{equation}
    with $m_{eV}$ the scalar field mass in $eV$.
    \item{Next we approximate the black hole horizons by the Schwarzschild radii which we obtain noting that the observed black holes from GWTC-1 with LVK lie in the range of stellar mass black holes $5 M_{\odot} \lesssim M \lesssim 150 M_{\odot}$, hence $r_h\sim r_s=2 G M/c^2$ lies in the range of $14.8\textrm{km}\lesssim r_h\lesssim 443\textrm{km}$ }
    \item{Substituting above horizon range in \eqref{eq:mevtomhat} we obtain the constraint on the coupling constant in  \cite{Yamada:2019zrb} corresponds to the dimensionless mass range $10^{-5}\lesssim \hat{m}\lesssim 10^{-1}$}
\end{itemize}
Comparing this mass range to Fig.~\ref{fig:phim} we find our theoretical bound on the coupling extends to masses larger than $\hat{m}\sim 10^{-1}$, hence the theory bound is the strongest constraint on the coupling in the large dimensionless mass range. \\

We can also use the results of the left panel of Fig.~\ref{fig:phim} to make a rough estimate of the possible scalar field mass range that would be interesting in relation to observation. From Fig.~\ref{fig:phim} we find that beyond $\hat{m}\sim 1$ the scalar field becomes highly suppressed, which decreases the likelihood for detection by probing the black hole environment. Hence $\hat{m}\sim 1$ seems the largest scalar field mass for which there is still significant scalar hair around the black hole. Then we consider a back of the envelope calculation similar to what was done in~\cite{Creci:2020mfg}. We assume no restriction on the type of black holes and take all astrophysical black holes which lie in the mass range $5 M_{\odot} \lesssim M \lesssim 10^{10} M_{\odot}$ again using $r_h \sim r_S$. Substituting this horizon range and $\hat{m}\sim 1$ in \eqref{eq:mevtomhat} inverting the equation to compute $m_{eV}$ we obtain
\begin{equation}
1.3\times 10^{-11} \mathrm{eV} \gtrsim m_{\varphi} \gtrsim 6.7\times 10^{-21} \mathrm{eV}\ .
\end{equation}
Massive sGB black holes thus enable exploring a large swath of parameter space of ultralight dark matter models as this mass range lies within the current bound for these models, see e.g.~\cite{Ferreira:2020fam} for a review.

\subsection{Dependencies of black hole properties}

\subsubsection{Innermost stable circular orbit and light ring}\label{sec:ISCOprsc}
Next, we use the full numerical solutions to analyze the dependence of gauge-invariant quantities such as the orbital frequency of a test particle at the innermost stable circular orbit (ISCO) and a photon at the unstable circular orbit (light ring) on the parameters of the theory. 

\begin{figure}[h!]
\centering
\begin{subfigure}{0.49\textwidth}
   \includegraphics[width=1\linewidth]{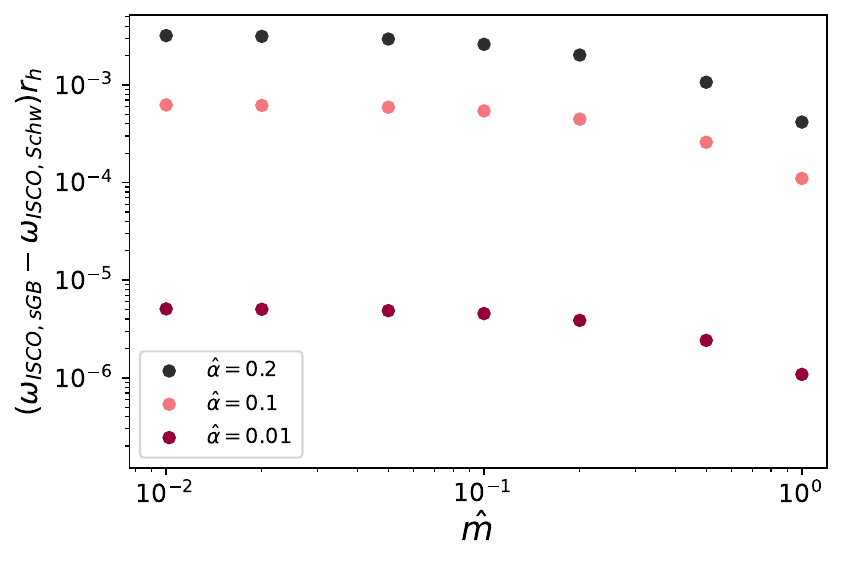}
       \centering
\end{subfigure}
\begin{subfigure}{0.49\textwidth}
    \includegraphics[width=1\linewidth]{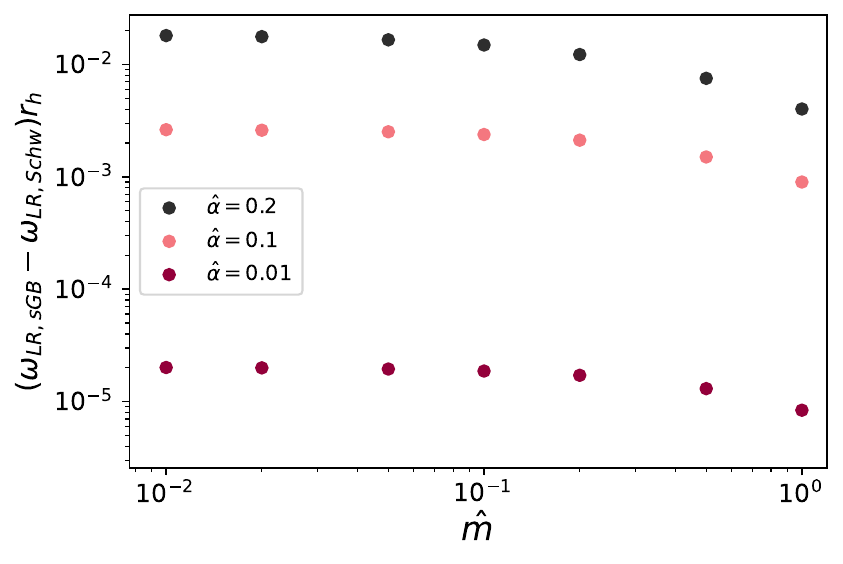}
        \centering
 \end{subfigure}
\caption[]{\emph{The difference in orbital angular frequency at the ISCO (left panel) and LR (right panel)} rescaled by the horizon radius from the Schwarzschild results as a function of the scalar mass for three different values of the coupling.}
       \label{fig:ISCOLR}
\end{figure}

In Appendix~\ref{AppD} we compute the ISCO and light ring (LR) radii from considering geodesic motions of test particles and photons, and formulating the dynamics in terms of an effective potential whose maximum determines the ISCO and LR. Specifically, we calculated the roots of the second derivative of~\eqref{eq:Veff} and~\eqref{eq:photonsphereeq} numerically after substituting the solutions for $A(r)$ and $B(r)$. We convert all expressions to functions of the orbital frequency as it is a coordinate-independent quantity by contrast to the radius, by using the relationship between the radial coordinate and frequency from~\eqref{eq:angularfreq}. In Fig.~\ref{fig:ISCOLR} we show the difference between the orbital frequency $\omega$ at the ISCO/LR in massive sGB and Schwarzschild spacetimes for different scalar field masses. Note that we give the results in terms of the dimensionless quantity $\omega r_h$, therefore the Schwarzschild frequencies $\omega_{ISCO}r_S=1/3\sqrt{6}$, $\omega_{LR}r_S=2/3\sqrt{3}$, need to be rescaled to $r_h$ in the same way as described in Sec.~\ref{sec:showsolpert}.

From both panels of Fig.~\ref{fig:ISCOLR} we conclude that, as the differences are positive, the orbital frequencies in massive sGB are larger (corresponding to the ISCO/LR radii being smaller) than for a Schwarzschild black hole with the same ADM mass. When comparing this to our result for the behavior of the curvature in the right panel of Fig.~\ref{fig:curv},  we see that for radii around $r_{ISCO}/r_h\sim 3$, $r_{LR}/r_h\sim 3/2$ the curvature in massive sGB is less strong than for a Schwarzschild black hole, and a stable orbit for a test particle/photon can therefore lie closer to the horizon. This also corresponds to the findings for the ISCO/LR frequencies in the literature on massless sGB, e.g. in \cite{Yunes:2011we, Heydari-Fard:2020iiu}. Furthermore, Fig.~\ref{fig:ISCOLR} shows that the difference in the orbital frequencies becomes smaller for smaller coupling and larger masses, as expected in these limits. As for other quantities, massless sGB gives the strongest deviations from a Schwarzschild blackhole. 
\subsubsection{ADM mass and scalar charge}\label{sec:MADMQ}

\begin{figure}[h]
\centering
\begin{subfigure}{0.49\textwidth}
   \includegraphics[width=1\linewidth]{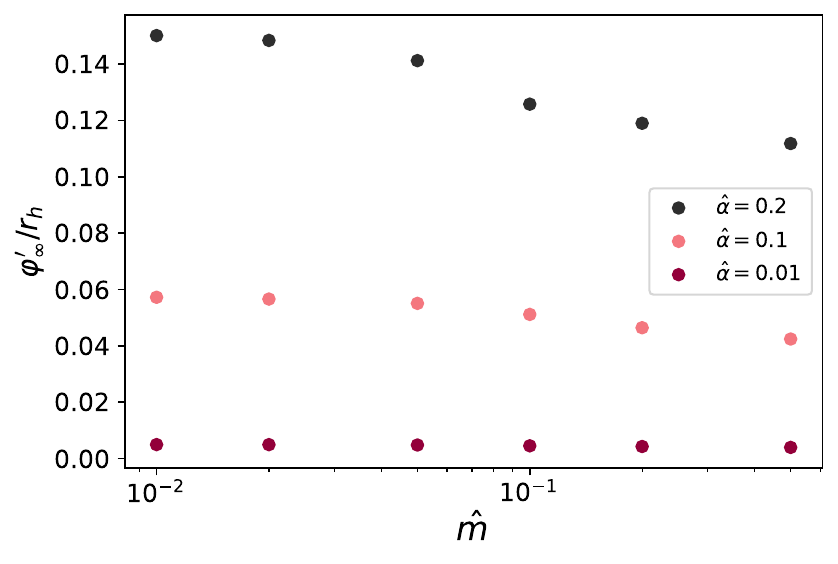}
       \centering 
\end{subfigure}
\begin{subfigure}{0.49\textwidth}
    \includegraphics[width=1\linewidth]{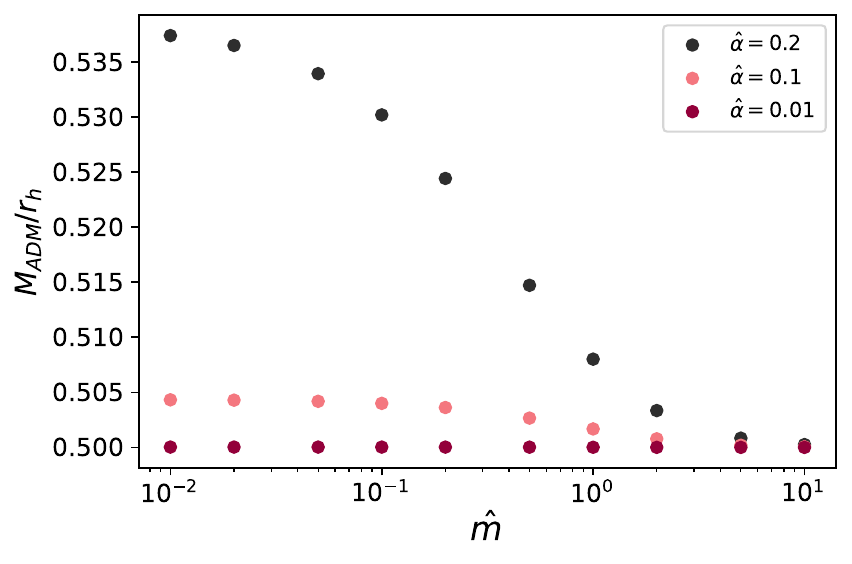}
        \centering
 \end{subfigure}
\caption[]{\emph{The amount of scalar charge (left panel) and ADM mass (right panel)} as function of the dimensionless scalar field mass $\hat{m}$ for three different values of the coupling.}
\label{fig:scalarcharge}
\end{figure}

Lastly we consider the analysis of the obtained ADM mass $M_{\rm ADM}=1/2 A_{\infty}^{\prime}$ and scalar monopole charge $\varphi_{\infty}^{\prime}$ defined in~\eqref{eq:asymplim}. We obtain these quantities from the numerical solutions as described in Sec.~\ref{sec:numfullsol} for different masses and coupling.
The ADM mass and scalar charge are relevant e.g. in effective action descriptions for black hole binary systems~\cite{Julie:2022huo}, where the two bodies are reduced to center-of-mass worldlines augmented with additional parameters 
that are matched to physical properties of the full configuration and capture its coarse-grained effects. 
In the massless case, the scalar charge is defined to be the coefficient of the $1/r$ term in the asymptotic falloff of the scalar profile. However, for massive scalar fields the asymptotic limit has an exponential decay~\eqref{eq:asymplim} and the 
definition of the charge must be adapted. We consider here the convention of~\cite{Barsanti:2022vvl}, which is still based on the decaying tail of the scalar field solution and defines the charge to be the prefactor of the exponential $\varphi_{\infty}^{\prime}$ as given in~\eqref{eq:asymplim}.
As described in Sec.~\ref{sec:numfullsol}, matching the solution to the asymptotic limit for the scalar field is more susceptible to the choice of integral region used for the matching than the metric functions. In practice we therefore limited the construction of $\varphi_{\infty}^{\prime}$ to $\hat{m}\leq 1$, as for larger values the solution outside the black hole horizon has already fallen off to nearly zero and it is not possible to unambiguously match to~\eqref{eq:asymplim} to determine $\varphi_{\infty}^{\prime}$. In principle one could obtain the charge in this regime by working with the solution in the interior of the horizon, however as we mentioned in Sec.~\ref{sec:phihanalysis} the $\hat{m}\lesssim1$ regime is the most interesting, therefore we limited our analysis to this regime. We show the results of these calculations for the ADM mass and scalar charge as function of the scalar field mass in Fig.~\ref{fig:scalarcharge}.
From the Fig.~\ref{fig:scalarcharge} we see that both the scalar charge and ADM mass become less sensitive to the scalar field mass for smaller values of the coupling. Both are also proportional to the coupling, where for vanishing coupling constant, the scalar charge vanishes and the ADM mass goes to $1/2 r_h$ as expected.

\section{Conclusion}
In this paper we performed a systematic study of various features of static, spherically symmetric black holes in sGB with a massive scalar field. This is a more natural scenario than assuming a massless field, as has been the focus of the majority of previous work, except for a few numerical examples. The scalar field mass introduces an additional scale in the problem and gives rise to richer features of the spacetime and scalar condensate. 
For the first time, we calculated perturbative solutions in a small coupling expansion up to third order in $\hat{\alpha}$ and compared this to full numerical solutions for the spacetime and scalar field. The small-coupling approximation yields more direct analytical insights into intriguing features that arise, while the numerical solutions capture fully nonlinear regimes. To compute numerical solutions, we used a bisection method to approach the scalar field solution with the desired asymptotic fall-off behavior at spatial infinity and a shooting method to obtain the metric potentials with the correct near-horizon and asymptotic behaviors. By extending the full numerical solutions inside the horizon, we found that the metric potentials and scalar field diverge at a finite radius. From analyzing the Kretschmann and contracted Weyl tensor curvature invariants we concluded that these divergences coincide with a genuine
curvature singularity. The location of this singularity depends on the coupling constant and the scalar field mass, where for smaller couplings and higher masses the singularity moves closer to the center of the black hole. The location of the singularity also impacts the performance of the small-coupling perturbative solution, which we found to be viable for small couplings, large scalar masses, and large distances from the black hole. These trends can be attributed to the fact that at large distances and for small coupling, the scalar and nonlinear curvature effects decrease, and for large scalar masses the scalar field decouples from the metric and the GB contribution to the action becomes a total derivative with no dynamical impact.  \\
For finding the black hole solution, in addition to the condition of asymptotic flatness one requires the scalar field to be regular at the horizon. This leads to conditions relating the amount of scalar field at the horizon to the coupling constant, black hole radius and scalar field mass. We discussed these conditions in the massless, linearized, and exact cases and interpreted them for the maximum value of the dimensionless coupling constant allowed for each choice of parameters. Inverting this argument led to a lower bound on the mass of the black hole. We found that introducing the scalar field mass leads to the existence of black hole solutions for larger couplings or conversely, for a fixed coupling, the domain of black holes extends to lower masses, see also~\cite{Doneva:2019vuh}. By comparing these near-horizon conditions for the scalar field to the parameters corresponding to the finite radius singularity being located at the black hole horizon, we found that both lead to bounds on the maximum dimensionless coupling constant.
We found that the near horizon condition and the singularity bound agree, concluding that requiring regularity of the scalar field on the horizon is coupled to censoring the finite radius singularity behind the horizon. Comparing these theoretical bounds on the coupling with the first observational constraint on massive sGB \cite{Yamada:2019zrb} we concluded that for stellar mass black holes, for masses $\hat{m}>10^{-1}$, our theoretical results provide the most stringent bounds to date. Additionally, in the case of the scalar Compton wavelength being larger than the black hole radius, the numerical matching between the near horizon and asymptotic limits shows that the actual values lie well below these bounds. On the other hand, for Compton wavelengths smaller than the black hole radius, the amount of scalar hair at the horizon in the exact calculations turns out to be identical to the maximum value determined by the regularity constraint. \\
Using the results for the amount of scalar hair near the horizon we could make a rough estimation of the scalar field mass range that could be promising in the light of observation. We found that this mass range includes the current scalar particle models.
Lastly we analysed how the ISCO radius, light ring radius, ADM mass and scalar monopole charge depend on the scalar field mass and coupling constant.\\
For future work, the study of black holes in sGB could be extended to include next to the scalar field mass, also the self interaction term in the scalar potential or one could add different interactions e.g. study the optical channel, including the interaction with light. Considering massive scalar fields in other promising modified gravity contexts would be interesting as well. Furthermore, obtaining rotating black hole solutions would be a next exploration. This would lead to the opportunity to study the possibility of superradiance instability in the context where the massive scalar field is also coupled to the spacetime curvature.
Our work contributed to the first exploratory studies of sGB black holes with massive scalar fields. The full and perturbative numerical solutions can be used in further studies of black holes in massive sGB and in modelling compact binary systems in these theories, extending the work of \cite{Shiralilou:2020gah,Shiralilou:2021mfl,vanGemeren:2023rhh} to include massive scalar fields. Furthermore our analysis and numerical method related to the massive scalar cloud configuration can be applied to massive scalar fields in a broader context and our results on the ADM mass and scalar monopole charge can be useful in the effective field theory description of compact binaries. In direct continuation of this work the analysis of the gravitational radiation from compact objects in massive sGB can be explored, contributing to the efforts of probing the strong field environments of black holes in the search for beyond GR signatures. \\

\textbf{\emph{Acknowledgments.}} We thank Gastón Creci, Peter Jonker and Ramon Wakelkamp for useful discussions. Furthermore, we would like to thank Thomas Sotiriou, Lik Hang Harry Shum, Miguel Bezares Figueroa  and  Llibert Aresté Saló for their questions on the first version of the manuscript, leading to an additional consistency check of our results with the existing literature. This publication is part of the Dutch Black Hole Consortium with project number NWA.1292.19.202 of the research programme NWA which is (partly) financed by the Dutch Research Council (NWO).
\clearpage

\appendix
\renewcommand{\theequation}{A.\arabic{equation}}

\setcounter{equation}{0}

\section{Explicit expressions for equations}\label{AppA}

In this appendix we show the explicit equations that were not given in the main text for the sake of readability.

\subsubsection{Field equations}
The explicit components of the modified Einstein equations~\eqref{eq:FE1} with~\eqref{eq:EMtensor} in terms of the metric~\eqref{eq:SSSmetric} are given by

\begin{equation}\label{eq:GmunuTmunu}
    \begin{aligned}
        G_t^t =& -\frac{1}{r^2} + \frac{e^{-B(r)}}{r^2 } - \frac{e^{-B(r)} B'(r)}{r },\\
        G_r^r =& -\frac{1}{r^2} + \frac{e^{-B(r)}}{r^2 } + \frac{e^{-B(r)} B'(r)}{r },\\
        G_{\theta}^{\theta}=& G_{\varphi}^{\varphi} = \frac{e^{-B(r)}\bigg(r A'(r)^2 - 2 B'(r) + A'(r)(2-r B'(r) + 2 r A''(r))\bigg)}{4 r },\\
        T_t^t =&-\frac{e^{-2B(r)}}{r^2 }\left[\varphi^{\prime 2}\left(r^2 e^{B(r)}+4 \alpha f''(\varphi)\left(e^{B(r)}-1\right)\right)-2\alpha f'(\varphi)\left(B^{\prime}(r) \varphi^{\prime}(r)\left(e^{B(r)}-3\right)\right.\right.\\
        &\left.\left.-2 \varphi^{\prime \prime}(r)\left(e^{B(r)}-1\right)\right)+ \left(e^{B(r)}m r  \varphi(r) \right)^2\right],\\
        T_r^r =& \frac{e^{-B(r)} \varphi^{\prime}(r)}{}\left[\varphi^{\prime}(r)-\frac{2 e^{-B(r)}\left(e^B(r)-3\right) \alpha f'(\varphi) A^{\prime}(r)}{ r^2}\right] - \left(m\varphi(r)\right)^2,\\
        T_\theta^\theta=&T_{\varphi}^{\varphi}=-\frac{e^{-2 B(r)}}{ r^2}\left[\varphi^{\prime 2}(r)\left(r  e^{B(r)}-2 \alpha f''(\varphi) A^{\prime}(r)\right)-4 \alpha f'(\varphi)\left(A^{\prime 2} (r)\varphi^{\prime}(r)\right.\right.\\
        &\left.\left.+2 \varphi^{\prime}(r) A^{\prime \prime}(r)+A^{\prime}(r)\left(2 \varphi^{\prime \prime}(r)-3 B(r)^{\prime} \varphi^{\prime}(r)\right)\right)+e^{2B(r)}m^2r^2\varphi(r)^2\right].
    \end{aligned}
\end{equation}

The scalar field equation~\eqref{eq:FE2} becomes
\begin{equation}
\begin{aligned}\label{eq:scalareomexpl}
&2 r \varphi^{\prime \prime}(r)+\left(4+r A^{\prime}(r)-r B^{\prime}(r)\right) \varphi^{\prime}(r)+\frac{ \alpha f'(\varphi) e^{-B(r)}}{ r}\left[\left(e^{B(r)}-3\right) A^{\prime}(r) B^{\prime}(r)\right.\\
&\left.-\left( e^{B(r)}-1\right)\left(2 A^{\prime \prime}(r)+A^{\prime 2}(r)\right)\right]-2e^{B(r)}m^2r\varphi(r)=0.
\end{aligned}
\end{equation}

\subsubsection{Master equations in $A$ and $\varphi$}
In Sec.~\ref{sec:fullsol}, we rewrote the modified Einstein equations as a system of second order differential equations~\eqref{eq:ddAddphi}, where the right hand sides are given by the functions

\begin{equation}\label{eq:f}
\begin{aligned}
&f(r, \varphi(r),\varphi'(r), A'(r)) =\\
&\Bigg(4 e^{4 B(r)} m^4 \varphi (r)^3 \left(e^{B(r)} r-4 \alpha  f'(\varphi ) \varphi '(r)\right) r^4-8 e^{4 B(r)} \left(-1+e^{B(r)}\right)
   m^4 \alpha  \varphi (r)^4 f'(\varphi ) r^3\\
   &-e^{2 B(r)} m^2 \varphi (r) \left(e^{B(r)} r-4 \alpha  f'(\varphi ) \varphi '(r)\right) \left(4
   e^{B(r)} \left(r^2 \varphi '(r)^2+e^{B(r)}-1\right)-3 A'(r)\right.\\
   &\left.\left(e^{B(r)} r+2 \left(-3+e^{B(r)}\right) \alpha  f'(\varphi ) \varphi '(r)\right)\right)
   r^2+8 e^{2 B(r)} \varphi '(r) \left(e^{B(r)} r+\left(-5+e^{B(r)}\right) \alpha  f'(\varphi ) \varphi '(r)\right) \\
   &\left(r^2 \varphi '(r)^2+e^{B(r)}-1\right)
   r+2 e^{2 B(r)} m^2 \varphi (r)^2 \left(-e^{2 B(r)} \left(r A'(r)+4\right) \varphi '(r) r^3-6\right.\\
   &\left.\left(3-4 e^{B(r)}+e^{2 B(r)}\right) \alpha ^2
   A'(r) f'(\varphi )^2 \varphi '(r)+\alpha  f'(\varphi ) \left(4 e^{B(r)} \left(\left(-1+e^{B(r)}\right)^2+4 r^2 \varphi '(r)^2\right)\right.\right.\\
   &\left.\left.+r A'(r) \left(4 \varphi
   '(r)^2 \left(e^{B(r)} r^2+2 \left(-1+e^{B(r)}\right) \alpha  f''(\varphi )\right)-5 e^{B(r)} \left(-1+e^{B(r)}\right)\right)\right)\right)
   r\\
   &+e^{B(r)} \alpha  A'(r)^3 f'(\varphi ) \left(r-4 \alpha  f'(\varphi ) \varphi '(r)\right) \left(e^{B(r)} r+2 \left(-3+e^{B(r)}\right) \alpha  f'(\varphi )
   \varphi '(r)\right)\\
   &+e^{B(r)} A'(r) \left(e^{B(r)} \varphi '(r) \left(\left(e^{B(r)} r^2-4 \alpha  f''(\varphi )\right) \varphi '(r)^2+2 e^{B(r)}
   \left(-4+e^{B(r)}\right)\right) r^3\right.\\
   &\left.-12 \left(15-8 e^{B(r)}+e^{2 B(r)}\right) \alpha ^2 f'(\varphi )^2 \varphi '(r)^3 r+4 \alpha  f'(\varphi ) \left(-r^2
   \left(e^{B(r)} r^2+4 \left(-2+e^{B(r)}\right) \alpha  f''(\varphi )\right) \varphi '(r)^4\right.\right.\\
   &\left.\left.-2 \left(3 e^{B(r)} \left(-3+e^{B(r)}\right) r^2+\left(1-3
   e^{B(r)}+2 e^{2 B(r)}\right) \alpha  f''(\varphi )\right) \varphi '(r)^2+e^{B(r)} \left(-1+e^{B(r)}\right)^2\right)\right)\\
   &-A'(r)^2 \left(e^{3 B(r)}
   \varphi '(r) r^4+2 e^{B(r)} \alpha  f'(\varphi ) \left(\left(e^{B(r)} \left(-4+e^{B(r)}\right) r^2-2 \left(-5+3 e^{B(r)}\right) \alpha  f''(\varphi
   )\right) \varphi '(r)^2\right.\right.\\
   &\left.\left.+e^{B(r)} \left(-1+e^{B(r)}\right)\right) r+4 \alpha ^2 f'(\varphi )^2 \varphi '(r) \left(e^{B(r)} \left(3-4 e^{B(r)}+e^{2
   B(r)}\right)-2 \varphi '(r)^2\right. \right. \\
   & \left.\left.\left(e^{B(r)} \left(-3+2 e^{B(r)}\right) r^2+\left(9-8 e^{B(r)}+3 e^{2 B(r)}\right) \alpha  f''(\varphi
   )\right)\right)\right)\Bigg)\bigg/ \Bigg(-4 e^{2 B(r)} \\
   &\left(e^{B(r)} r-4 \alpha  f'(\varphi ) \varphi '(r)\right) \left(r^2 \varphi '(r)^2+e^{B(r)}-1\right) r^2+4 e^{2
   B(r)} m^2 \varphi (r)^2 \left(e^{B(r)} r^2\right.\\
   &\left.\left(e^{B(r)} r-4 \alpha  f'(\varphi ) \varphi '(r)\right)-4 \left(-1+e^{B(r)}\right) \alpha ^2
   A'(r) f'(\varphi )^2\right) r^2-8 \left(-1+e^{B(r)}\right) \\
   &\alpha ^2 A'(r)^2 f'(\varphi )^2 \left(e^{B(r)} r+\left(-9+5 e^{B(r)}\right) \alpha 
   f'(\varphi ) \varphi '(r)\right)+e^{B(r)} A'(r) \left(3 e^{2 B(r)} r^4+8 e^{B(r)}\right.\\
   &\left.\left(-4+e^{B(r)}\right) \alpha  f'(\varphi ) \varphi '(r) r^3+4 \alpha ^2
   f'(\varphi )^2 \left(5 \left(-1+e^{B(r)}\right)^2-4 \left(-4+e^{B(r)}\right) r^2 \varphi '(r)^2\right)\right)\Bigg)
\end{aligned}
\end{equation}
and
\begin{equation}\label{eq:h}
\begin{aligned}
&h(r, \varphi(r),\varphi'(r), A'(r)) =\\
&\Bigg(4 e^{4 B(r)} m^4 \varphi (r)^3 \left(e^{B(r)} r-4 \alpha  f'(\varphi ) \varphi '(r)\right) r^4-8 e^{4 B(r)} \left(-1+e^{B(r)}\right)
   m^4 \alpha  \varphi (r)^4 f'(\varphi ) r^3\\
   &-e^{2 B(r)} m^2 \varphi (r) \left(e^{B(r)} r-4 \alpha  f'(\varphi ) \varphi '(r)\right) \left(4
   e^{B(r)} \left(r^2 \varphi '(r)^2+e^{B(r)}-1\right)-3 A'(r)\right.\\
   &\left.\left(e^{B(r)} r+2 \left(-3+e^{B(r)}\right) \alpha  f'(\varphi ) \varphi '(r)\right)\right)
   r^2+8 e^{2 B(r)} \varphi '(r) \left(e^{B(r)} r+\left(-5+e^{B(r)}\right) \alpha  f'(\varphi ) \varphi '(r)\right)\\
   &\left(r^2 \varphi '(r)^2+e^{B(r)}-1\right)
   r+2 e^{2 B(r)} m^2 \varphi (r)^2 \left(-e^{2 B(r)} \left(r A'(r)+4\right) \varphi '(r) r^3-6 \left(3-4 e^{B(r)}\right.\right.\\
   &\left.\left.+e^{2 B(r)}\right) \alpha ^2
   A'(r) f'(\varphi )^2 \varphi '(r)+\alpha  f'(\varphi ) \left(4 e^{B(r)} \left(\left(-1+e^{B(r)}\right)^2+4 r^2 \varphi '(r)^2\right)+r A'(r)\right.\right. \\
   &\left.\left.\left(4 \varphi
   '(r)^2 \left(e^{B(r)} r^2+2 \left(-1+e^{B(r)}\right) \alpha  f''(\varphi )\right)-5 e^{B(r)} \left(-1+e^{B(r)}\right)\right)\right)\right)
   r+e^{B(r)}\\
   &\alpha  A'(r)^3 f'(\varphi ) \left(r-4 \alpha  f'(\varphi ) \varphi '(r)\right) \left(e^{B(r)} r+2 \left(-3+e^{B(r)}\right) \alpha  f'(\varphi )
   \varphi '(r)\right)+e^{B(r)} A'(r) \\
   &\left(e^{B(r)} \varphi '(r) \left(\left(e^{B(r)} r^2-4 \alpha  f''(\varphi )\right) \varphi '(r)^2+2 e^{B(r)}
   \left(-4+e^{B(r)}\right)\right) r^3-12 \left(15-8 e^{B(r)} \right.\right.\\
   &\left.\left.+e^{2 B(r)}\right) \alpha ^2 f'(\varphi )^2 \varphi '(r)^3 r+4 \alpha  f'(\varphi ) \left(-r^2
   \left(e^{B(r)} r^2+4 \left(-2+e^{B(r)}\right) \alpha  f''(\varphi )\right) \varphi '(r)^4 \right.\right.\\
   &\left.\left.-2 \left(3 e^{B(r)} \left(-3+e^{B(r)}\right) r^2+\left(1-3
   e^{B(r)}+2 e^{2 B(r)}\right) \alpha  f''(\varphi )\right) \varphi '(r)^2+e^{B(r)} \left(-1+e^{B(r)}\right)^2\right)\right)\\
   &-A'(r)^2 \left(e^{3 B(r)}
   \varphi '(r) r^4+2 e^{B(r)} \alpha  f'(\varphi ) \left(\left(e^{B(r)} \left(-4+e^{B(r)}\right) r^2-2 \left(-5+3 e^{B(r)}\right)\right.\right.\right.\\
   &\left.\left.\left.\alpha  f''(\varphi
   )\right) \varphi '(r)^2+e^{B(r)} \left(-1+e^{B(r)}\right)\right) r+4 \alpha ^2 f'(\varphi )^2 \varphi '(r) \left(e^{B(r)} \left(3-4 e^{B(r)}+e^{2 B(r)}\right)\right.\right.\\
   &\left.\left.-2 \varphi '(r)^2 \left(e^{B(r)} \left(-3+2 e^{B(r)}\right) r^2+\left(9-8 e^{B(r)}+3 e^{2 B(r)}\right) \alpha  f''(\varphi
   )\right)\right)\right)\Bigg)\bigg/\Bigg(-4 e^{2 B(r)}\\
   &\left(e^{B(r)} r-4 \alpha  f'(\varphi ) \varphi '(r)\right) \left(r^2 \varphi '(r)^2+e^{B(r)}-1\right) r^2+4 e^{2
   B(r)} m^2 \varphi (r)^2\\
   &\left(e^{B(r)} r^2 \left(e^{B(r)} r-4 \alpha  f'(\varphi ) \varphi '(r)\right)-4 \left(-1+e^{B(r)}\right) \alpha ^2
   A'(r) f'(\varphi )^2\right) r^2-8 \left(-1+e^{B(r)}\right)\\
   &\alpha ^2 A'(r)^2 f'(\varphi )^2 \left(e^{B(r)} r+\left(-9+5 e^{B(r)}\right) \alpha 
   f'(\varphi ) \varphi '(r)\right)+e^{B(r)} A'(r) \left(3 e^{2 B(r)} r^4\right.\\
   &\left.+8 e^{B(r)} \left(-4+e^{B(r)}\right) \alpha  f'(\varphi ) \varphi '(r) r^3+4 \alpha ^2
   f'(\varphi )^2 \left(5 \left(-1+e^{B(r)}\right)^2\right.\right.\\
   &\left.\left.-4 \left(-4+e^{B(r)}\right) r^2 \varphi '(r)^2\right)\right)\Bigg)
\end{aligned}
\end{equation}

with $e^{B(r)}$ given by~\eqref{eq:expBexpl}.

\subsubsection{Near-horizon expansion in $1/A^\prime$}

In the near horizon limit, the expansion of the second order differential equations~\eqref{eq:ddAddphi} in terms of $1/A'(r)$ resulted in \eqref{eq:ddAddphiexp} with the coefficients given by

\begin{equation}\label{eq:abc}
\begin{aligned}
a= &-6 \alpha ^2 m^4 r^4 \varphi (r)^4 f'(\varphi )^2-2 \alpha  m^2 r^2 \varphi (r) f'(\varphi ) \left(2 \alpha  f'(\varphi ) \varphi '(r)+r\right)^2-m^2 \varphi (r)^2\\
&\left(4
   \alpha  r^5 f'(\varphi ) \varphi '(r)+4 \alpha ^2 r^2 f'(\varphi )^2 \left(r^2 \varphi '(r)^2-4\right)-16 \alpha ^4 f'(\varphi )^4 \varphi '(r)^2\right.\\
   &\left.-16 \alpha ^3 r f'(\varphi )^3 \varphi '(r)+r^6\right)+4
   \alpha  r^3 f'(\varphi ) \varphi '(r)+2 \alpha ^2 f'(\varphi )^2 \left(2 r^2 \varphi '(r)^2-3\right)+r^4,\\
b= & \left(m^2 r^2 \varphi (r)^2-1\right) \left(4 \alpha ^2 f'(\varphi )^2 \left(2 m^2 r^2 \varphi (r)^2-3\right)-8 \alpha ^3 m^2 r \varphi (r)^2 f'(\varphi
   )^3 \varphi '(r)\right.\\
   &\left.+2 \alpha  r^3 f'(\varphi ) \varphi '(r)+r^4\right),\\
c=& \alpha  m^4 r^3 \varphi (r)^4 f'(\varphi ) \left(r-4 \alpha  f'(\varphi ) \varphi '(r)\right)-m^2 r^2 \varphi (r) \left(2 \alpha  f'(\varphi ) \varphi
   '(r)+r\right)^2\\
   &-m^2 \varphi (r)^2 \left(2 \alpha  r^2 f'(\varphi ) \left(r^2 \varphi '(r)^2+1\right)-8 \alpha ^3 f'(\varphi )^3 \varphi'(r)^2-12 \alpha ^2 r f'(\varphi
   )^2 \varphi '(r)\right.\\
   &\left.+r^5 \varphi '(r)\right)+\alpha  f'(\varphi ) \left(2 r^2 \varphi '(r)^2+3\right)+r^3 \varphi '(r) .
\end{aligned}
\end{equation}

\subsubsection{Regularity condition}
Requiring regularity of the scalar field at the horizon lead to \eqref{eq:phihprimem} for the derivative of the scalar field at the horizon with coefficients

\begin{equation}\label{eq:AB}
\begin{aligned}
&A=-4 \alpha ^2 m^4 r_h^3 \varphi_h^4 f'(\varphi_h )^2-m^2 r_h \varphi_h^2 \left(r_h^4-12 \alpha ^2 f'(\varphi_h )^2\right)-4 \alpha  m^2 r_h^3 \varphi_h f'(\varphi_h )+r_h^3\\
&B=4 \alpha  f'(\varphi_h ) \left(-m^2 \varphi_h^2 \left(r_h^4-4 \alpha ^2 f'(\varphi_h )^2\right)-2 \alpha  m^2 r_h^2 \varphi_h f'(\varphi_h )+r_h^2\right),
\end{aligned}
\end{equation}
\begin{equation}\label{eq:fullconstr}
\begin{aligned}
    C=&16 \alpha ^4 m^2 \varphi_h^2 f'(\varphi_h )^4 \left(m^2 r_h^2 \varphi_h^2-6\right)+48 \alpha ^3 m^2 r_h^2 \varphi_h f'(\varphi_h )^3\\
    &+8 \alpha ^2 r_h^2 f'(\varphi_h )^2 \left(2
   m^2 r_h^2 \varphi_h^2-3\right)+r_h^6.
   \end{aligned}
\end{equation}

\section{Theoretical arguments for a monotonically decreasing linearized scalar profile}\label{AppB}
\renewcommand{\theequation}{B.\arabic{equation}}

\setcounter{equation}{0}
With similar arguments as in the discussion in~\cite{Horne:1992bi}, one can deduce that the solution to~\eqref{eq:eomphilin} has to be a monotonically increasing function in terms of $u$ or decreasing in terms of $r$.
We start from the linear-in-coupling equation of motion~\eqref{eq:eomphilin} hence work in the dimensionless parameter $u$ defined in~\eqref{eq:udef}. We are searching for solutions with a finite behavior at the horizon $u=1$ and an asymptotically flat solution at infinity $u=0$ as found in~\eqref{eq:asymplimsflin}. 
This involves the following considerations:\\

\noindent \textit{1) Once the solution becomes negative it can only become more negative, and cannot increase to zero again }

Suppose that the solution for the scalar profile becomes negative. To change sign again to positive values requires the existence of a minimum at negative field values. Multiplying \eqref{eq:eomphilin} by $(-1)$ leads to

\begin{equation}
    (1-u)\varphi^{1}{}^{\prime\prime} - \varphi^{1}{}^{\prime} =\frac{\hat{m}^2 \varphi^{1}}{u^4} -3 f'(\varphi^{0})u^2\ . 
\end{equation}

If there is an extremum for negative $\varphi^{1}$, we have $\varphi^{1}<0$ and $ \varphi^{1}{}^{\prime}=0$ there. Hence at this location in between the boundaries
\begin{equation}\label{eq:lineomatextremum}
    (1-u)\varphi^{1}{}^{\prime\prime} =\frac{\hat{m}^2 \varphi^{1}}{u^4} -3 f'(\varphi^{0})u^2\ ,
\end{equation}

Now the right hand side is $<0$ and thus $\varphi^{1}{}^{\prime\prime}<0$, since $(1-u)\geq 0$.  Therefore, if there is an extremum for negative $\varphi^{1}$ it has to be a (local) maximum. This implies that field can only become more negative, which is incompatible with the required asymptotic behavior. Thus, for a positive coupling to have a solution that falls off to zero, the scalar field has to stay positive. \\

\noindent \textit{2) The positive scalar field cannot have an extremum }

Next, we consider the case where the scalar field starts out positive. At a local maximum for a positive scalar field we have $\varphi^{1}>0$ and $\varphi^{1}{}^{\prime}=0$. Evaluating~\eqref{eq:eomphilin} at this location results again in \eqref{eq:lineomatextremum}.
For a local maximum the second derivative should be negative, hence the right hand side should be negative as well. This implies the following inequality at the maximum
\begin{equation}
\label{eq:inequ2}
\frac{\hat{m}^2\varphi^{1}}{u^4} < 3 f'(\varphi^{0})u^2\ .
\end{equation}
Moving towards the horizon at $u = 1$ after a local maximum means $\varphi^{1}$ decreases and $u$ increases. Therefore the left hand side of the inequality~\eqref{eq:inequ2} decreases and the right hand side increases so the inequality holds. There cannot be an minimum because in that case~\eqref{eq:inequ2} would need to flip. Thus, the inequality holds up to the horizon. This further implies that the slope of the profile at the horizon is negative or zero. However the differential equation at the horizon is
\begin{equation}
\label{eq:horizonODE}
 -\varphi^{1}{}^{\prime}  =\hat{m}^2 \varphi^{1} -3 f'(\varphi^{0})\ .
 \end{equation}
 Because the inequality~\eqref{eq:inequ2} still holds at the horizon, the right hand side of~\eqref{eq:horizonODE} is negative. This implies from~\eqref{eq:horizonODE} a nonzero positive derivative at the horizon, which is in contradiction with the consequences of the inequality discussed above. Therefore, there cannot be a local maximum for positive field values. \\
 If the solution had a local minimum for the positive scalar field, it would require a local maximum as well to have an asymptotic fall off to 0, which we just argued cannot be the case. This means that having a local minimum would lead to a diverging solution at infinity.\\

\noindent \textit{3) The derivative of the scalar field at the horizon needs to be positive (or negative when working in $r$)}

From the arguments above, the scalar field at linear order in the coupling needs to be positive and cannot have local maxima or minima. Therefore the derivative of the scalar field at the horizon at $u=1$ needs to be positive to be able to connect to zero at infinity, because a negative derivative at $u=1$ leads to a ever increasing (or partly constant) function going inwards to infinity, never reaching zero.

\section{Numerical methods}\label{AppC}
\renewcommand{\theequation}{C.\arabic{equation}}

\setcounter{equation}{0}
In this appendix we describe in detail the two numerical methods used to obtain the perturbative and full solutions discussed in Sec.~\ref{sec:pertsol} and~\ref{sec:fullsol}. Additionally a discussion on numerical precision tests is given.

\subsection{Bisection method}
Firstly in section \ref{sec:numsollin} we describe solving the scalar field equation at linear order in the coupling. As described in this section the asymptotic limit of the solution for the linearized scalar field has an exponentially growing and decreasing mode \eqref{eq:asymplimsflin}. If not obtaining the initial condition for which this growing mode is exactly zero, there will always be a large radial distance at which the growing mode takes over and the solution diverges. Therefore a slight numerical inaccuracy already leads to a divergence. Obtaining the exact solution is hard, however approaching the right initial condition is relatively easy. In this section we describe how one can approach the right initial condition which corresponds to the exponentially decaying solution.

The differential equation \eqref{eq:eomphilin} is approached as an initial value problem, starting the integration at an infinitesimal distance from the horizon $u=1-10^{-5}$. The initial conditions are given by \eqref{eq:phi1prime}. For a fixed mass and coupling, we vary the constant $\varphi_h^{1}$ to find the solution that has an asymptotically flat limit. We obtain this by 
first determining an interval of $\varphi_h^{1}$ for which the asymptotic behavior switches from positive infinity to negative infinity. By decreasing this interval, the estimate of $\varphi_h^{1}$ corresponding to an asymptotically flat solution improves. We implement this through the following algorithm, for each choice of $\hat m$:
\begin{itemize}
    \item Make an initial guess  $\varphi_h^{1}$ obtained by extrapolating \eqref{eq:asymplimsflin} with $\bar{\varphi}_{\infty}^1=0$ and computing its value at the horizon. 
    \item Check if the solution corresponding to this guess diverges to positive or negative infinity.
    \item Incrementally increase (decrease) $\varphi_h^{1}$ if the solution with the initial guess diverges negatively (positively) and check the divergence behavior at each step.
    \item When reaching a step for which the divergence flips sign, defining this value as $\varphi_{h,flip}^{1}$, calculate $\varphi_{h,new}^{1} = (\varphi_{h,initial}^{1}+\varphi_{h,flip}^{1})/2$.
    \item Use this mean value $\varphi_{h,new}^{1}$ as the new initial guess, decrease the step size every iteration by one order of magnitude.
    \item Continue these iterations until the guess saturates, where more iterations result in more accurate solutions. 
\end{itemize}

In Fig.~\ref{fig:estimmethod} we show the solution of \eqref{eq:eomphilin} running the bisection method described above different number of times. One can see that for more cycles, the diverging behavior happens for smaller $u$/larger $r$. In principle, extending to infinite cycles, one would obtain the actual decaying solution. However the estimation for $\varphi_h^{1}$ would only differ infinitesimally, hence it is accurate enough to cut of the number of cycles at a finite value. In our analysis in section \ref{sec:numsollin} and \ref{sec:fullsol} we execute 15 cycles. This means that the estimation for $\varphi_h^{1}$ differs with an order of magnitude of $10^{-14}$ from the estimate at 14 cycles. This estimate therefore has very high accuracy, however the main reason applying this many cycles is to push the diverging behavior relatively close to $u=0$ without making the computational time too long. The linearized scalar field solution is substituted in the higher order field equations and therefore the diverging behavior works through in the solutions for the metric functions and higher order scalar field as well. Therefore to get an accurate perturbative solution for as largest range of $u$ as possible, around 15 cycles or more is advised.

\subsection{Shooting method}
For numerically calculating the perturbative solution to the metric functions in section \ref{sec:higherordercorr} and the full solution in section \ref{sec:fullsol}, we use the so called shooting method. In this section we describe in more detail what this method entails. 

The shooting method can be used as a numerical method to solve differential equations with a boundary value problem. This is the case for the modified Einstein equations for which we constructed the behavior of the metric functions at the boundaries; the near horizon and asymptotic limits. An additional requirement is that the solution does not have the instable behavior with respect to the initial conditions as is the case for the scalar field equation described in the previous section.

The shooting method is based on reframing the problem as an initial value problem with variable initial conditions. One integrates outwards to obtain the solution of this initial value problem for different guesses of the initial condition and evaluates the solution at infinity until these values at infinity agree with the boundary condition in the asymptotic limit. 
To describe this in more detail let us describe this for the specific case of solving the $tt$ component at second order in the coupling for the metric function $\bar{B}^2$ as is done in section \ref{sec:higherordercorr}. The boundary conditions are given by \eqref{eq:horlimB2} and \eqref{eq:asymplimB2}, where we can vary the near horizon constant $\bar{A}_h^{2}$.

\begin{itemize}
\item Construct a function $f[\bar{A}_h^{2},u]$ of the differential equation solver from the black hole horizon outwards to infinity, in this case for the $tt$ component of \eqref{eq:GmunuTmunu} with initial conditions at the horizon following \eqref{eq:horlimB2} with $\bar{A}_h^{2}$ as variable.
\item Define the function of the asymptotic limit $g[u]$ as in \eqref{eq:asymplimB2} .
\item Then define $h[\bar{A}_h^{2}] = f[\bar{A}_h^{2},0] - g[0]$ the difference between the solution to the initial value problem and the asymptotic limit evaluated at infinity $u=0$.
\item Find the root(s) of $h$, the value of $\bar{A}_h^{2}$ corresponding to the root is the correct initial condition to the boundary value problem and substituting this value for $\bar{A}_h^{2}$ in $f$ gives you the correct solution for $\bar{B}^2(u)$.
\end{itemize}

In our case the outer boundary lies in the asymptotic limit, however as one substitutes the linearized scalar field solution in the differential equations at higher orders in the coupling, the divergence behavior at finite $u$ of this scalar field also works through in the higher order equations. Therefore in practice instead of evaluating function $h$ at infinity, evaluate the functions at smallest possible $u$ before the divergence in $\varphi^{1}$ starts. This does slightly deteriorate the accuracy of the perturbative solution. 

Following this calculation results in the following solution for $\bar{B}^{2}$ for a mass of $\hat{m}=1$.
\begin{figure}[h]
\centering
\begin{subfigure}{0.55\textwidth}
   \includegraphics[width=1\linewidth]{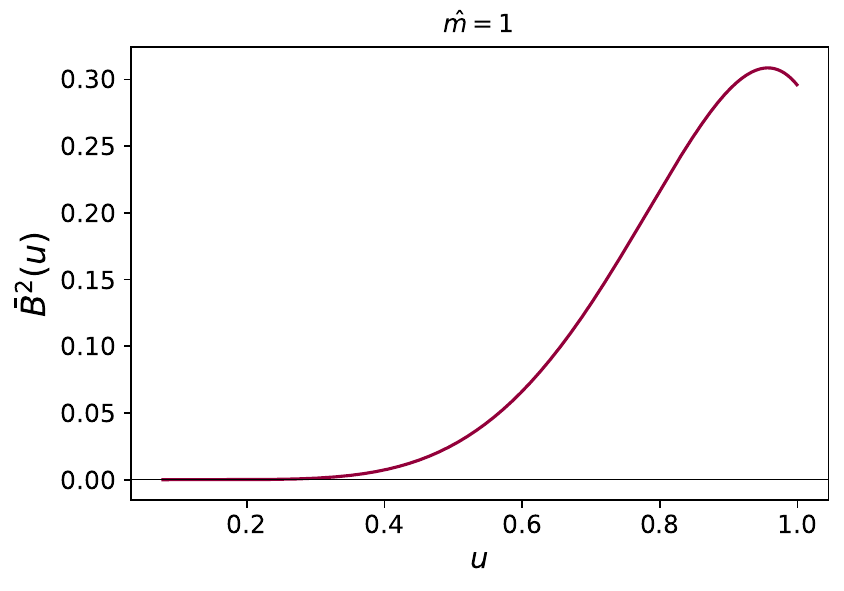}
       \centering 
\end{subfigure}
\caption[]{\emph{Solution for $\bar{B}^{2}$ }with $\hat{m}=1$. The shooting method resulted in $\bar{A}_h^{2} = 0.295$ as initial condition in \eqref{eq:horlimB2}.}
\label{fig:B2}
\end{figure}

\subsection{Numerical precision tests}
In this section we describe in more detail the used numerical method for the numerical integrator mentioned in Sec.~\ref{sec:pertsol}, \ref{sec:fullsol}. We used the {\sc Mathematica} numerical integrator \emph{NDSolve} for solving the boundary value problems in the perturbative and exact contexts. To obtain the numerical solutions for the metric functions and scalar field on itself the standard machine precision and \emph{"StiffnessSwitching"} method in the \emph{NDSolve} environment are sufficient, and no problems arise for the solutions and its derivatives. However, we encountered problems with numerical stability of the solutions in follow up calculations, more specifically when computing the percent difference of the Kretschmann scalar in Fig.~\ref{fig:curv}. This arose as oscillatory behaviour of the final numerical function describing this percent difference. We therefore set up a (non-exhaustive) sweep over the different methods and working precision for the \emph{NDSolve} environment to conclude which setting could mitigate this effect. Working from left to right we checked the following configurations, shown in Table~\ref{tab:tests}.
\begin{table}[]
\begin{tabular}{|l|lll}
\hline
\textbf{Methods} &
  \multicolumn{1}{l|}{\textbf{WorkingPrecision}} &
  \multicolumn{1}{l|}{\textbf{AccuracyGoal}} &
  \multicolumn{1}{l|}{\textbf{PrecisionGoal}} \\ \hline
"Adams"                                                                                    & \multicolumn{1}{l|}{5}  & \multicolumn{1}{l|}{20} & \multicolumn{1}{l|}{20} \\ \hline
"BDF"                                                                                      & \multicolumn{1}{l|}{15} & \multicolumn{1}{l|}{25} & \multicolumn{1}{l|}{25} \\ \hline
"ExplicitRungeKutta"                                                                       & \multicolumn{1}{l|}{25} & \multicolumn{1}{l|}{30} & \multicolumn{1}{l|}{30} \\ \hline
"ImplicitRungeKutta"                                                                       & \multicolumn{1}{l|}{30} &                         &                         \\ \cline{1-2}
"SymplecticPartitionedRungeKutta"                                                          & \multicolumn{1}{l|}{40} &                         &                         \\ \cline{1-2}
"MethodOfLines"                                                                            & \multicolumn{1}{l|}{50} &                         &                         \\ \cline{1-2}
"Extrapolation"                                                                            &                         &                         &                         \\ \cline{1-1}
"DoubleStep"                                                                               &                         &                         &                         \\ \cline{1-1}
"LocallyExact"                                                                             &                         &                         &                         \\ \cline{1-1}
"StiffnessSwitching"                                                                       &                         &                         &                         \\ \cline{1-1}
"Projection"                                                                               &                         &                         &                         \\ \cline{1-1}
"OrthogonalProjection"                                                                     &                         &                         &                         \\ \cline{1-1}
"IDA"                                                                                      &                         &                         &                         \\ \cline{1-1}
\begin{tabular}[c]{@{}l@{}}"StiffnessSwitching", Method $\rightarrow$\\  \{"ExplicitRungeKutta", Automatic\}\end{tabular} &
   &
   &
   \\ \cline{1-1}
\begin{tabular}[c]{@{}l@{}}"TimeIntegration" $\rightarrow$\\  \{"ExplicitRungeKutta", \\ "DifferenceOrder" $\rightarrow$ 8\}\end{tabular} &
   &
   &
   \\ \cline{1-1}
\begin{tabular}[c]{@{}l@{}}"TimeIntegration" $\rightarrow$ \\ "ExplicitEuler"\end{tabular} &                         &                         &                         \\ \cline{1-1}
\begin{tabular}[c]{@{}l@{}}"PDEDiscretization" $\rightarrow$ \\ \{"MethodOfLines", "SpatialDiscretization"\\  $\rightarrow$\{"TensorProductGrid", \\   "MinPoints" $\rightarrow$ 1000\}\}\end{tabular} &
   &
   &
   \\ \cline{1-1}
\begin{tabular}[c]{@{}l@{}}"PDEDiscretization" $\rightarrow$\\  \{"MethodOfLines", "SpatialDiscretization"\\  $\rightarrow$ \{"FiniteElement"\}\}\}\end{tabular} &
   &
   &
   \\ \cline{1-1}
\end{tabular}
\centering
\caption{Working from left to right, the different settings for the \emph{Method, WorkingPrecision, AccuracyGoal} and \emph{PrecisionGoal} within the \emph{NDSolve} function, for finding the configuration mitigating the effect of numerical inaccuracy.}
\label{tab:tests}
\end{table}

We executed the tests in the following manner. We set up a module function with the boundary value problem for the exact field equations as described in Sec.~\ref{sec:fullsol} that computes the solution for the metric function $A(r)$ and $\varphi(r)$, with the method, WorkingPrecision, AccuracyGoal and PrecisionGoal as variables. In the same module we compute the percentual difference of the Kretschmann scalar in massive sGB substituting the solutions, with the Schwarzschild curvature invariant. From random test we had already found the oscillations due to limited numerical precision to worsen for smaller choices of the coupling constant, hence we chose to do the tests for $\hat{\alpha}=0.01$, and to keep the running time manageable, we choose a small scalar field mass $\hat{m}=0.01$. Additionally from the sample tests we found that some of the methods in Table~\ref{tab:tests} that did improve on the numerical inprecision issues, did not give output for the default WorkingPrecision, therefore in general we set the WorkingPrecision and MachinePrecision to $30$. First we computed the percent difference function up to $r/r_H =10$ for the different methods in the first column of Table~\ref{tab:tests} with the AccuracyGoal and PrecisionGoal on default. We selected the method for which the function did not diverge at the horizon and which mitigated the oscillation the most, which resulted in \emph{"TimeIntegration" $\rightarrow$ \{"ExplicitRungeKutta", "DifferenceOrder" $\rightarrow$ 8\}}. \\
Then we repeated the calculation specifying to this method, now varying the WorkingPrecision found in the second column of Table~\ref{tab:tests}. For precision below $\mathrm{WorkingPrecision}=25$ in combination with above chosen method, the correct solution for the boundary value problem is not found, minimal precision of $\mathrm{WorkingPrecision}=25$ is required. The oscillations got damped for higher values of the precision as expected, from $\mathrm{WorkingPrecision}=30$ and onwards the oscillations up to $r/r_H =10$ are smoothed out completely. \\
Lastly we repeated the computation for the above mentioned method and $\mathrm{WorkingPrecision}=25$ for different values of the AccuracyGoald and PrecisionGoals given in the last two columns of Table~\ref{tab:tests}, choosing values comparable to the set WorkingPrecision. Both tested separate from each other and in the different combinations, checking what configuration of these settings mitigated the oscillatory behaviour that is still present at this WorkingPrecision. We found no observable improvement on the oscillatory behavior from these two settings. Hence specifying $\mathrm{WorkingPrecision}=30$ on itself results in sufficient numerical precision. The default setting for the AccuracyGoal and PrecisionGoal are both set as half the WorkingPrecision. For larger distances than $r/r_H =10$ the precision still might be too limited but in principle one could solve this issue by improving on the precision settings. Note we also did not explore every permutation of settings, however for our purposes computing the solutions with method \emph{"TimeIntegration" $\rightarrow$ \{"ExplicitRungeKutta", "DifferenceOrder" $\rightarrow$ 8\}}, $\mathrm{WorkingPrecision}=30$ and AccuracyGoal, PrecisionGoal on default, suffices. 

\section{Additional analysis of  perturbative solutions}\label{Apppert}
\renewcommand{\theequation}{D.\arabic{equation}}

\setcounter{equation}{0}
In addition to the analysis in Sec.~\ref{sec:showsol} we discuss in this appendix the perturbative solution in more detail, comparing the solution up to different orders in the coupling with the exact numerical case. The difference between the solutions is most noticeable in the near horizon region, where the spacetime curvature, see Fig.~\ref{fig:curv}, is strongest and the scalar field energy density the highest, see Fig.~\ref{fig:energydensity}. 

\begin{figure}[h!]
\centering
\begin{subfigure}{0.49\textwidth}
   \includegraphics[width=\linewidth]{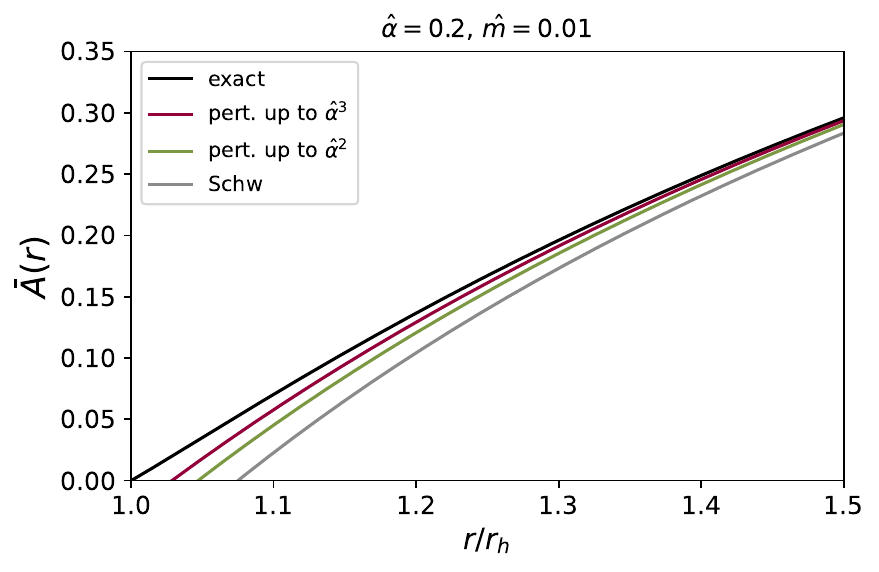}
       \centering
\end{subfigure}
\begin{subfigure}{0.49\textwidth}
    \includegraphics[width=\linewidth]{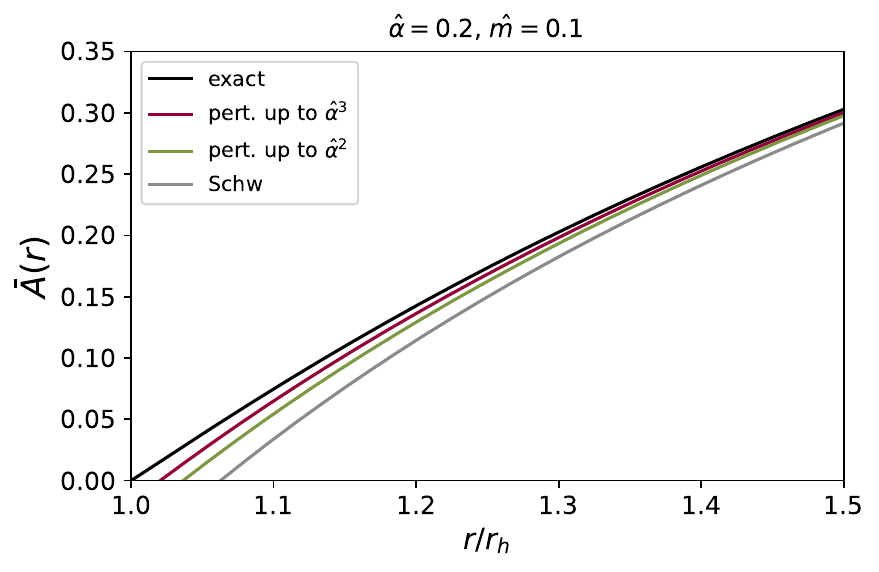}
        \centering
 \end{subfigure}
\caption[]{\emph{The solution of metric function $\bar{A}$} close to the horizon, comparing the exact, perturbative solution up to $\hat{\alpha}^3$, up to $\hat{\alpha}^2$ and the Schwarzschild solution respectively. }
\label{fig:pertbarA}
\end{figure}

Starting with the metric function $\bar{A}$ as defined in~\eqref{eq:rescaleAB} with the perturbative solution rescaled to variable $r$ with the method described in Sec.~\ref{sec:showsol}. The left and right panel of Fig.~\ref{fig:pertbarA} show the metric function near the horizon for $\hat{\alpha}=0.2$ and $\hat{m}=0.01$ and $\hat{m}=0.1$ respectively. We zoom in on the region near the horizon as there the differences between the curves is most noticeable, for larger radial distances the curves coincide in all cases below as expected. In both panels of Fig.~\ref{fig:pertbarA} one can see that the perturbative curves lie below the exact solution and above the Schwarzschild solution. Including corrections to higher order in the coupling for the perturbative solution results in the curve lying slightly closer to the exact solution as one would expect. The roots of the curves correspond to the respective horizon radii. Similar as we showed in~\ref{fig:Asol} the horizon radius for the exact solution is smaller than the perturbative and Schwarzschild horizons. Furthermore from both panels of Fig.~\ref{fig:pertbarA} we find the horizon radius shifts towards the horizon of the exact curve for higher corrections to the perturbative solution. \\

\begin{figure}[h!]
\centering
\begin{subfigure}{0.49\textwidth}
   \includegraphics[width=\linewidth]{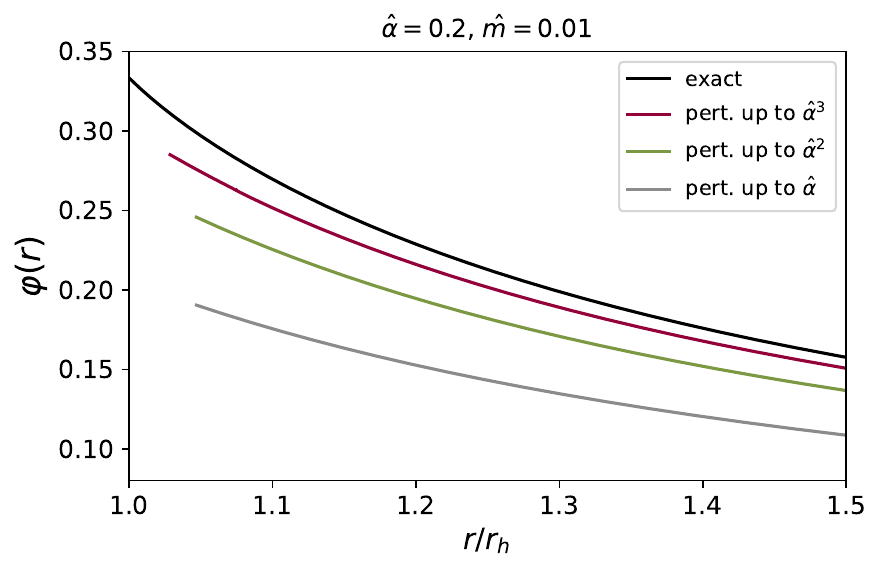}
       \centering
\end{subfigure}
\begin{subfigure}{0.49\textwidth}
    \includegraphics[width=\linewidth]{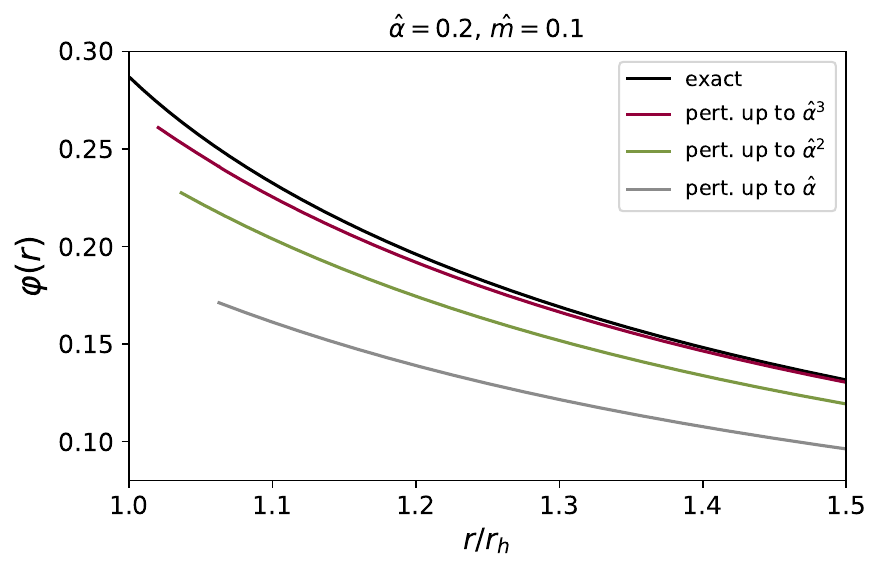}
        \centering
 \end{subfigure}
\caption[]{\emph{The solution of $\varphi$} close to the horizon, comparing the exact, perturbative solution up to $\hat{\alpha}^3$, up to $\hat{\alpha}^2$ and the linearized solution respectively.}
\label{fig:pertphi}
\end{figure}

In Fig.~\ref{fig:pertphi} we show the perturbative solution of the scalar field up to linear, quadratic and cubic order in the coupling compared to the exact solution for $\hat{\alpha}=0.2$ and $\hat{m}=0.01$ and $\hat{m}=0.1$ respectively.
In both panels we find that the perturbative solution approaches the exact solution from below and again becomes more accurate with increasing orders in $\hat{\alpha}$, as one would expect. Furthermore, comparing the left and right panel we find the a smaller difference with the exact solution for larger scalar field mass where the improvement is more noticeable than for the metric function in Fig.~\ref{fig:pertbarA}. For the case of the scalar field we do not find any particular change comparing the second order solution (green curves) to the cubic order solution (pink curves). This is interesting as for the latter, the corrections to the metric function first contribute to the scalar field solution, see Table~\ref{tab:dependencies}.

From the analysis in this appendix together with Sec.~\ref{sec:showsol} we can conclude that the difference of perturbative solution with the exact solution becomes smaller for small values of the coupling, large values of the scalar field mass and/or large distances from the horizon. The comparison does not show any qualitatively new non-perturbative behaviour that would not be captured by the perturbative solution when adding higher order corrections to increase accuracy. 

\section{Additional analysis near horizon constraint and finite radius singularity}\label{Appphimax}
\renewcommand{\theequation}{E.\arabic{equation}}

\setcounter{equation}{0}

In this appendix we discuss in more detail how we obtained the right panel of Fig.~\ref{fig:phim} and Fig.~\ref{fig:phimax} below. The aim of Sec.~\ref{sec:phihanalysis} and this appendix is to compare the near horizon constraint that ensures the regularity of the scalar field at the horizon and the requirement of censoring the finite radius singularity behind the black hole horizon.
\begin{figure}[h]
\centering
\begin{subfigure}{0.7\textwidth}
   \includegraphics[width=1\linewidth]{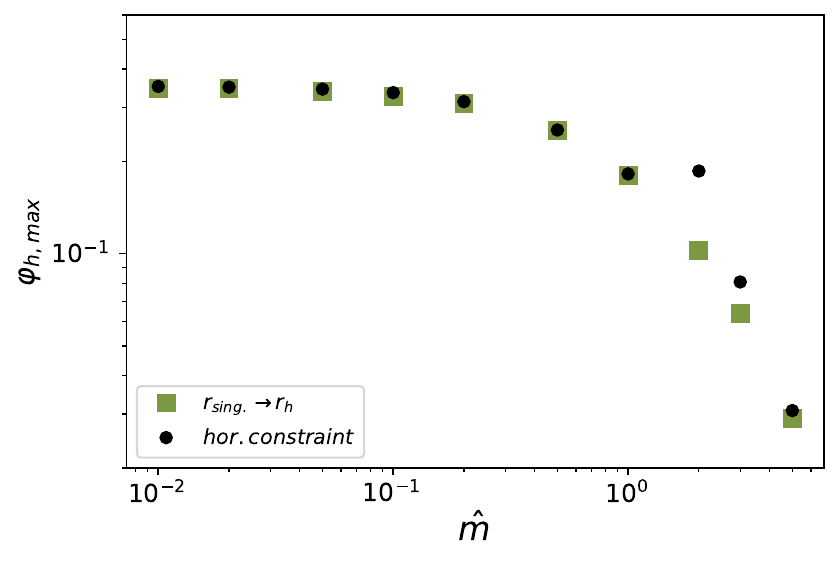}
       \centering 
\end{subfigure}
\caption[]{\emph{Maximum allowed value of the scalar field at the horizon } as a function of the scalar field mass from the requirement of preventing a naked singularity (squares) and the near horizon constraint~\eqref{eq:phihprimem} (dots).} 
\label{fig:phimax}
\end{figure}
In the following step-by-step procedure we explain how we obtained the maximum coupling and maximum $\varphi_h$ values for each value of the scalar field mass in Fig.~\ref{fig:phim} and Fig.~\ref{fig:phimax}. 
\begin{enumerate}
    \item For a fixed mass $\hat m$ and an initial guess $\hat \alpha=0.1$ , we computed the solution to \eqref{eq:ddAddphi} for the scalar field and metric functions. 
    \item Subsequently, we check the position of the finite radius singularity for the computed solutions by calculating the Kretchmann scalar and check where it diverges, see Sec.~\ref{sec:curvscal}. We find that for the studied scalar mass range and this initial guess for the coupling, the singularity lies inside the horizon radius. 
    \item We increase $\hat{\alpha}$ and repeat the steps above for each choice of the scalar field mass up to the solution for which the curvature singularity lies on the horizon. This identifies the largest possible  $\hat{\alpha}=\hat \alpha_{\rm max}^{\rm sing.}$ to prevent a naked singularity for a certain scalar field mass. \emph{These values give the green square points in the right panel of Fig.~\ref{fig:phim}. }
    \item Additionally, from the solution for the scalar field for this configuration of the parameters, we identify the maximum allowed amount of scalar hair at the horizon for which the singularity is censored $\varphi_h^{\hat \alpha_{\rm max}^{\rm sing.}}$. \emph{These values give the green square points in Fig.~\ref{fig:phimax}.}
    \item Now we substitute $\varphi_h^{\hat \alpha_{\rm max}^{\rm sing.}}$ at each choice for the scalar field mass in the near horizon constraint~\eqref{eq:phihprimem} and solve the inequality $\varphi_h'<0$ for the coupling obtaining $\hat\alpha_{\rm max}^{\rm hor.constraint}$. \emph{These values give the black dot points in the right panel of Fig.~\ref{fig:phim}. }
    \item Similarly we substitute $\hat\alpha_{\rm max}^{\rm hor.constraint}$ in the near horizon constraint~\eqref{eq:phihprimem} for each choice of the scalar field mass and solve the inequality $\varphi_h'<0$ for the scalar field value at the horizon obtaining  $\varphi_h^{\hat\alpha_{\rm max}^{\rm hor.constraint}}$. \emph{These values give the black dot points in the right panel of Fig.~\ref{fig:phimax}. }
\end{enumerate}


\noindent We see from Fig.~\ref{fig:phimax} that for masses $\hat{m}<1$, the constraints from the curvature singularity and regularity of the field at the horizon on the maximum $\varphi_h$ coincide. For slightly larger masses, the cusp feature arises in the horizon constraint. The origin of this cusp is the same as the cusps shown in the dashed curves in the left panel of Fig.~\ref{fig:phim}. As explained in Sec.~\ref{sec:phihanalysis} this cusp corresponds to a bifurcation point for value of $\varphi_h$ for which $\varphi_h'<0$, but we only consider the bottom branch to be physical.  In  the mass regime around the cusp, the requirement of not having a naked singularity is a stronger constraint than the near horizon requirement. For the maximum allowed coupling in the right panel of Fig.~\ref{fig:phim}, both cases agree for the range of masses we studied. Hence we conclude that the maximum coupling is not sensitive to the choice of $\varphi_h$ as this discrepancy between the squares and dots does not arise there.

\section{Calculation of the ISCO and light ring radii}\label{AppD}

\renewcommand{\theequation}{F.\arabic{equation}}

\setcounter{equation}{0}

In this appendix we show how one can determine the ISCO radius and light ring radius in Schwarzschild coordinates and the corresponding orbital frequencies. This is used in Sec.~\ref{sec:ISCOprsc}.\\
The ISCO radius can be determined from the effective potential. Starting from a static spherically symmetric metric~\eqref{eq:SSSmetric}, one can write down the normalization of the four velocity \\$g_{\mu\nu}\dot{x}^{\mu}\dot{x}^{\nu}=-1$. Before writing this down explicitly we can use the symmetries of the spacetime, e.g. as the metric components are independent of $\phi$ and $t$ there are two constants of motion $E=-e^{A(r)}\dot{t}$ and $L=r^2 \dot{\phi}$, the energy and angular momentum per unit mass. As the conservation of (the direction of) angular momentum requires the motion of a particle to be planar, together with rotational symmetry, one can fix the motion to be equatorial with $\theta=\frac{\pi}{2}$. Substituting these quantities in the normalization condition we obtain
\begin{equation}\label{eq:rdot}
\begin{aligned}
e^{B(r)}\dot{r}^2 &= -1 +e^{-A(r)}E^2-\frac{L^2}{r^2}\ ,\\
\dot{r}^2 &=V_{eff}(r)\ ,
\end{aligned}
\end{equation}
with
\begin{equation}\label{eq:Veff}
V_{eff}(r) = e^{-B(r)}( -1 +e^{-A(r)}E^2-\frac{L^2}{r^2})\ ,
\end{equation}
the effective potential as~\eqref{eq:rdot} now describes the equation for a classical particle moving in potential $V_{eff}(r)$\footnote{In the literature there are slightly different interpretations of $V_{eff}$ e.g. sometimes the $E^2$ term is treated separately or the sign might be opposite. The definition of~\eqref{eq:Veff} intuitively makes sense as for large $L$ there are two extrema which correspond to circular orbits. The extremum closest to the horizon correspond to a maximum and hence the unstable orbit and the outer extremum to a minimum, the stable orbit. The innermost stable orbit is found at the point where the two extrema coincide, hence for this purpose the different interpretations generally do not matter. However the interpretation of the extrema in this way of defining the effective potential makes most sense.}. Additionally to the effective potential we can write down the radial component of the geodesic equation
\begin{equation}\label{eq:geodesicr}
\dot{r}^2 + \frac{{e^{A(r)}}^{\prime}}{2 e^{B(r)}} \dot{t}^2 - \frac{{r^2}^{\prime}}{2 e^{B(r)}}\dot{\phi}^2 = 0\ .
\end{equation}
For finding the innermost stable circular orbit we are interested in circular orbits and therefore $\dot{r} = \ddot{r}=0$. Substituting these conditions in~\eqref{eq:geodesicr} and using this equation to construct the angular frequency $\omega = \frac{\dot{\phi}}{\dot{t}}$ results in 
\begin{equation}\label{eq:angularfreq}
\omega^2 = \frac{{e^{A(r)}}^{\prime}}{{r^2}^{\prime}}\ .
\end{equation}

Then combining~\eqref{eq:rdot}, the condition for circular orbits, the definitions of the constants of motion and~\eqref{eq:angularfreq} we find for the energy and angular momentum per unit mass for circular orbits
\begin{equation}\label{eq:EandL}
    \begin{aligned}
        &E = -\frac{-e^{A(r)}}{\sqrt{e^{A(r)} -r^2 \omega^2}}\ ,\\
        &L = \frac{r^2 \omega}{\sqrt{e^{A(r)} -r^2 \omega^2}}\ .
    \end{aligned}
\end{equation}
Now $V_{eff}(r)$ has two extrema, the inner extremum is a maximum corresponding to an unstable circular orbit and the outer with a minimum an thus a stable circular orbit. The minimum radius for this stable circular orbits happens when these two extrema coincide, this is when the second order derivative of $V_{eff}(r)$ has a root. Therefore taking the second order derivative to $r$ of~\eqref{eq:Veff} treating $E$ and $L$ as constants of motion, followed by substituting~\eqref{eq:EandL} and finding the radius that corresponds to the root results in $r_{ISCO}$. Substituting this radial coordinate in~\eqref{eq:angularfreq}, one obtains the orbital frequency at the ISCO radius, which is in contrary to the radius a coordinate independent quantity. \\

Finding the light ring is a bit more straight forward. Photons travel along null paths $ds=0$, additionally we are interested in circular orbits $dr=0$ and for similar arguments as before we can set $\theta=\pi/2$ hence $d\theta=0$. This simplifies the equation for null paths to
\begin{equation}\label{eq:ds0}
\dot{\phi}^2=\frac{e^{A(r)}}{r^2}\ .
\end{equation}
Additionally from the radial component of the geodesic equation in circular orbits we obtained \eqref{eq:angularfreq}, substituting this in \eqref{eq:ds0} gives
\begin{equation}\label{eq:photonsphereeq}
\frac{e^{A(r)}{}^{\prime}}{r^2{}^{\prime}}=\frac{e^{A(r)}}{r^2}\ ,
\end{equation}
solving for $r$ results in $r_{LR}$. Substituting this radial coordinate in \eqref{eq:angularfreq}, one obtains the orbital frequency at the light ring radius. \\


\bibliography{bib.bib}

\begin{thebibliography}{10}
\providecommand{\url}[1]{\texttt{#1}}
\providecommand{\urlprefix}{URL }
\expandafter\ifx\csname urlstyle\endcsname\relax
  \providecommand{\doi}[1]{doi:\discretionary{}{}{}#1}\else
  \providecommand{\doi}{doi:\discretionary{}{}{}\begingroup \urlstyle{rm}\Url}\fi
\providecommand{\eprint}[2][]{\url{#2}}

\bibitem{Will:2014kxa}
C.~M. Will,
\newblock \emph{{The Confrontation between General Relativity and Experiment}},
\newblock Living Rev. Rel. \textbf{17}, 4 (2014),
\newblock \doi{10.12942/lrr-2014-4},
\newblock \eprint{1403.7377}.

\bibitem{Turyshev:2008dr}
S.~G. Turyshev,
\newblock \emph{{Experimental Tests of General Relativity}},
\newblock Ann. Rev. Nucl. Part. Sci. \textbf{58}, 207 (2008),
\newblock \doi{10.1146/annurev.nucl.58.020807.111839},
\newblock \eprint{0806.1731}.

\bibitem{Berti:2015itd}
E.~Berti \emph{et~al.},
\newblock \emph{{Testing General Relativity with Present and Future Astrophysical Observations}},
\newblock Class. Quant. Grav. \textbf{32}, 243001 (2015),
\newblock \doi{10.1088/0264-9381/32/24/243001},
\newblock \eprint{1501.07274}.

\bibitem{Adelberger:2003zx}
E.~G. Adelberger, B.~R. Heckel and A.~E. Nelson,
\newblock \emph{{Tests of the gravitational inverse square law}},
\newblock Ann. Rev. Nucl. Part. Sci. \textbf{53}, 77 (2003),
\newblock \doi{10.1146/annurev.nucl.53.041002.110503},
\newblock \eprint{hep-ph/0307284}.

\bibitem{Wex:2014nva}
N.~{Wex},
\newblock \emph{{Testing Relativistic Gravity with Radio Pulsars}},
\newblock arXiv e-prints arXiv:1402.5594 (2014),
\newblock \doi{10.48550/arXiv.1402.5594},
\newblock \eprint{1402.5594}.

\bibitem{LIGOScientific:2016lio}
B.~P. Abbott \emph{et~al.},
\newblock \emph{{Tests of general relativity with GW150914}},
\newblock Phys. Rev. Lett. \textbf{116}(22), 221101 (2016),
\newblock \doi{10.1103/PhysRevLett.116.221101},
\newblock [Erratum: Phys.Rev.Lett. 121, 129902 (2018)],
\newblock \eprint{1602.03841}.

\bibitem{LIGOScientific:2019fpa}
B.~P. Abbott \emph{et~al.},
\newblock \emph{{Tests of General Relativity with the Binary Black Hole Signals from the LIGO-Virgo Catalog GWTC-1}},
\newblock Phys. Rev. D \textbf{100}(10), 104036 (2019),
\newblock \doi{10.1103/PhysRevD.100.104036},
\newblock \eprint{1903.04467}.

\bibitem{LIGOScientific:2020tif}
R.~Abbott \emph{et~al.},
\newblock \emph{{Tests of general relativity with binary black holes from the second LIGO-Virgo gravitational-wave transient catalog}},
\newblock Phys. Rev. D \textbf{103}(12), 122002 (2021),
\newblock \doi{10.1103/PhysRevD.103.122002},
\newblock \eprint{2010.14529}.

\bibitem{Nojiri:2018ouv}
S.~Nojiri, S.~D. Odintsov and V.~K. Oikonomou,
\newblock \emph{{Ghost-free Gauss-Bonnet Theories of Gravity}},
\newblock Phys. Rev. D \textbf{99}(4), 044050 (2019),
\newblock \doi{10.1103/PhysRevD.99.044050},
\newblock \eprint{1811.07790}.

\bibitem{Kovacs:2020pns}
A.~D. Kov\'acs and H.~S. Reall,
\newblock \emph{{Well-Posed Formulation of Scalar-Tensor Effective Field Theory}},
\newblock Phys. Rev. Lett. \textbf{124}(22), 221101 (2020),
\newblock \doi{10.1103/PhysRevLett.124.221101},
\newblock \eprint{2003.04327}.

\bibitem{Kovacs:2020ywu}
A.~D. Kov\'acs and H.~S. Reall,
\newblock \emph{{Well-posed formulation of Lovelock and Horndeski theories}},
\newblock Phys. Rev. D \textbf{101}(12), 124003 (2020),
\newblock \doi{10.1103/PhysRevD.101.124003},
\newblock \eprint{2003.08398}.

\bibitem{R:2022tqa}
A.~{Hegade K.~R.}, E.~R. {Most}, J.~{Noronha}, H.~{Witek} and N.~{Yunes},
\newblock \emph{{How do axisymmetric black holes grow monopole and dipole hair?}},
\newblock Phys. Rev. D \textbf{107}(10), 104047 (2023),
\newblock \doi{10.1103/PhysRevD.107.104047},
\newblock \eprint{2212.02039}.

\bibitem{Zwiebach:1985uq}
B.~Zwiebach,
\newblock \emph{{Curvature Squared Terms and String Theories}},
\newblock Phys. Lett. B \textbf{156}, 315 (1985),
\newblock \doi{10.1016/0370-2693(85)91616-8}.

\bibitem{Gross:1986mw}
D.~J. Gross and J.~H. Sloan,
\newblock \emph{{The Quartic Effective Action for the Heterotic String}},
\newblock Nucl. Phys. B \textbf{291}, 41 (1987),
\newblock \doi{10.1016/0550-3213(87)90465-2}.

\bibitem{Boulware:1985wk}
D.~G. Boulware and S.~Deser,
\newblock \emph{{String Generated Gravity Models}},
\newblock Phys. Rev. Lett. \textbf{55}, 2656 (1985),
\newblock \doi{10.1103/PhysRevLett.55.2656}.

\bibitem{Israel:1967wq}
W.~Israel,
\newblock \emph{{Event horizons in static vacuum space-times}},
\newblock Phys. Rev. \textbf{164}, 1776 (1967),
\newblock \doi{10.1103/PhysRev.164.1776}.

\bibitem{Israel:1967za}
W.~Israel,
\newblock \emph{{Event horizons in static electrovac space-times}},
\newblock Commun. Math. Phys. \textbf{8}, 245 (1968),
\newblock \doi{10.1007/BF01645859}.

\bibitem{Carter:1971zc}
B.~Carter,
\newblock \emph{{Axisymmetric Black Hole Has Only Two Degrees of Freedom}},
\newblock Phys. Rev. Lett. \textbf{26}, 331 (1971),
\newblock \doi{10.1103/PhysRevLett.26.331}.

\bibitem{Wald:1971iw}
R.~M. Wald,
\newblock \emph{{Final states of gravitational collapse}},
\newblock Phys. Rev. Lett. \textbf{26}, 1653 (1971),
\newblock \doi{10.1103/PhysRevLett.26.1653}.

\bibitem{Bekenstein:1971hc}
J.~D. Bekenstein,
\newblock \emph{{Nonexistence of baryon number for static black holes}},
\newblock Phys. Rev. D \textbf{5}, 1239 (1972),
\newblock \doi{10.1103/PhysRevD.5.1239}.

\bibitem{Chase}
J.~E. {Chase},
\newblock \emph{{Event horizons in static scalar-vacuum space-times}},
\newblock Communications in Mathematical Physics \textbf{19}(4), 276 (1970),
\newblock \doi{10.1007/BF01646635}.

\bibitem{Teitelboim:1972ps}
C.~Teitelboim,
\newblock \emph{{Nonmeasurability of the lepton number of a black hole}},
\newblock Lett. Nuovo Cim. \textbf{3S2}, 397 (1972),
\newblock \doi{10.1007/BF02826050}.

\bibitem{Bekenstein:1972ny}
J.~D. Bekenstein,
\newblock \emph{{Transcendence of the law of baryon-number conservation in black hole physics}},
\newblock Phys. Rev. Lett. \textbf{28}, 452 (1972),
\newblock \doi{10.1103/PhysRevLett.28.452}.

\bibitem{Hartle:1971qq}
J.~B. Hartle,
\newblock \emph{{Long-range neutrino forces exerted by kerr black holes}},
\newblock Phys. Rev. D \textbf{3}, 2938 (1971),
\newblock \doi{10.1103/PhysRevD.3.2938}.

\bibitem{Bekenstein:1995un}
J.~D. Bekenstein,
\newblock \emph{{Novel \textquoteleft{}\textquoteleft{}no-scalar-hair\textquoteright{}\textquoteright{} theorem for black holes}},
\newblock Phys. Rev. D \textbf{51}(12), R6608 (1995),
\newblock \doi{10.1103/PhysRevD.51.R6608}.

\bibitem{Hawking:1972qk}
S.~W. Hawking,
\newblock \emph{{Black holes in the Brans-Dicke theory of gravitation}},
\newblock Commun. Math. Phys. \textbf{25}, 167 (1972),
\newblock \doi{10.1007/BF01877518}.

\bibitem{Sotiriou:2011dz}
T.~P. Sotiriou and V.~Faraoni,
\newblock \emph{{Black holes in scalar-tensor gravity}},
\newblock Phys. Rev. Lett. \textbf{108}, 081103 (2012),
\newblock \doi{10.1103/PhysRevLett.108.081103},
\newblock \eprint{1109.6324}.

\bibitem{Kanti:1995vq}
P.~Kanti, N.~E. Mavromatos, J.~Rizos, K.~Tamvakis and E.~Winstanley,
\newblock \emph{{Dilatonic black holes in higher curvature string gravity}},
\newblock Phys. Rev. D \textbf{54}, 5049 (1996),
\newblock \doi{10.1103/PhysRevD.54.5049},
\newblock \eprint{hep-th/9511071}.

\bibitem{Pani:2009wy}
P.~Pani and V.~Cardoso,
\newblock \emph{{Are black holes in alternative theories serious astrophysical candidates? The Case for Einstein-Dilaton-Gauss-Bonnet black holes}},
\newblock Phys. Rev. D \textbf{79}, 084031 (2009),
\newblock \doi{10.1103/PhysRevD.79.084031},
\newblock \eprint{0902.1569}.

\bibitem{Sotiriou:2013qea}
T.~P. Sotiriou and S.-Y. Zhou,
\newblock \emph{{Black hole hair in generalized scalar-tensor gravity}},
\newblock Phys. Rev. Lett. \textbf{112}, 251102 (2014),
\newblock \doi{10.1103/PhysRevLett.112.251102},
\newblock \eprint{1312.3622}.

\bibitem{Benkel:2016rlz}
R.~Benkel, T.~P. Sotiriou and H.~Witek,
\newblock \emph{{Black hole hair formation in shift-symmetric generalised scalar-tensor gravity}},
\newblock Class. Quant. Grav. \textbf{34}(6), 064001 (2017),
\newblock \doi{10.1088/1361-6382/aa5ce7},
\newblock \eprint{1610.09168}.

\bibitem{Antoniou:2017hxj}
G.~Antoniou, A.~Bakopoulos and P.~Kanti,
\newblock \emph{{Black-Hole Solutions with Scalar Hair in Einstein-Scalar-Gauss-Bonnet Theories}},
\newblock Phys. Rev. D \textbf{97}(8), 084037 (2018),
\newblock \doi{10.1103/PhysRevD.97.084037},
\newblock \eprint{1711.07431}.

\bibitem{Antoniou_2018}
G.~Antoniou, A.~Bakopoulos and P.~Kanti,
\newblock \emph{Evasion of no-hair theorems and novel black-hole solutions in gauss-bonnet theories},
\newblock Physical Review Letters \textbf{120}(13) (2018),
\newblock \doi{10.1103/physrevlett.120.131102}.

\bibitem{Papageorgiou:2022umj}
A.~Papageorgiou, C.~Park and M.~Park,
\newblock \emph{{Rectifying no-hair theorems in Gauss-Bonnet theory}},
\newblock Phys. Rev. D \textbf{106}(8), 084024 (2022),
\newblock \doi{10.1103/PhysRevD.106.084024},
\newblock \eprint{2205.00907}.

\bibitem{Prabhu:2018aun}
K.~Prabhu and L.~C. Stein,
\newblock \emph{{Black hole scalar charge from a topological horizon integral in Einstein-dilaton-Gauss-Bonnet gravity}},
\newblock Phys. Rev. D \textbf{98}(2), 021503 (2018),
\newblock \doi{10.1103/PhysRevD.98.021503},
\newblock \eprint{1805.02668}.

\bibitem{Saravani:2019xwx}
M.~Saravani and T.~P. Sotiriou,
\newblock \emph{{Classification of shift-symmetric Horndeski theories and hairy black holes}},
\newblock Phys. Rev. D \textbf{99}(12), 124004 (2019),
\newblock \doi{10.1103/PhysRevD.99.124004},
\newblock \eprint{1903.02055}.

\bibitem{Silva:2017uqg}
H.~O. Silva, J.~Sakstein, L.~Gualtieri, T.~P. Sotiriou and E.~Berti,
\newblock \emph{{Spontaneous scalarization of black holes and compact stars from a Gauss-Bonnet coupling}},
\newblock Phys. Rev. Lett. \textbf{120}(13), 131104 (2018),
\newblock \doi{10.1103/PhysRevLett.120.131104},
\newblock \eprint{1711.02080}.

\bibitem{Dima:2020yac}
A.~Dima, E.~Barausse, N.~Franchini and T.~P. Sotiriou,
\newblock \emph{{Spin-induced black hole spontaneous scalarization}},
\newblock Phys. Rev. Lett. \textbf{125}(23), 231101 (2020),
\newblock \doi{10.1103/PhysRevLett.125.231101},
\newblock \eprint{2006.03095}.

\bibitem{Herdeiro:2020wei}
C.~A.~R. Herdeiro, E.~Radu, H.~O. Silva, T.~P. Sotiriou and N.~Yunes,
\newblock \emph{{Spin-induced scalarized black holes}},
\newblock Phys. Rev. Lett. \textbf{126}(1), 011103 (2021),
\newblock \doi{10.1103/PhysRevLett.126.011103},
\newblock \eprint{2009.03904}.

\bibitem{Berti:2020kgk}
E.~Berti, L.~G. Collodel, B.~Kleihaus and J.~Kunz,
\newblock \emph{{Spin-induced black-hole scalarization in Einstein-scalar-Gauss-Bonnet theory}},
\newblock Phys. Rev. Lett. \textbf{126}(1), 011104 (2021),
\newblock \doi{10.1103/PhysRevLett.126.011104},
\newblock \eprint{2009.03905}.

\bibitem{Collodel:2019kkx}
L.~G. Collodel, B.~Kleihaus, J.~Kunz and E.~Berti,
\newblock \emph{{Spinning and excited black holes in Einstein-scalar-Gauss\textendash{}Bonnet theory}},
\newblock Class. Quant. Grav. \textbf{37}(7), 075018 (2020),
\newblock \doi{10.1088/1361-6382/ab74f9},
\newblock \eprint{1912.05382}.

\bibitem{Doneva:2020nbb}
D.~D. Doneva, L.~G. Collodel, C.~J. Kr\"uger and S.~S. Yazadjiev,
\newblock \emph{{Black hole scalarization induced by the spin: 2+1 time evolution}},
\newblock Phys. Rev. D \textbf{102}(10), 104027 (2020),
\newblock \doi{10.1103/PhysRevD.102.104027},
\newblock \eprint{2008.07391}.

\bibitem{Doneva:2022ewd}
D.~D. Doneva, F.~M. Ramazano\u{g}lu, H.~O. Silva, T.~P. Sotiriou and S.~S. Yazadjiev,
\newblock \emph{{Spontaneous scalarization}},
\newblock Rev. Mod. Phys. \textbf{96}(1), 015004 (2024),
\newblock \doi{10.1103/RevModPhys.96.015004},
\newblock \eprint{2211.01766}.

\bibitem{Herdeiro:2015waa}
C.~A.~R. Herdeiro and E.~Radu,
\newblock \emph{{Asymptotically flat black holes with scalar hair: a review}},
\newblock Int. J. Mod. Phys. D \textbf{24}(09), 1542014 (2015),
\newblock \doi{10.1142/S0218271815420146},
\newblock \eprint{1504.08209}.

\bibitem{Ripley:2019irj}
J.~L. Ripley and F.~Pretorius,
\newblock \emph{{Gravitational collapse in Einstein dilaton-Gauss\textendash{}Bonnet gravity}},
\newblock Class. Quant. Grav. \textbf{36}(13), 134001 (2019),
\newblock \doi{10.1088/1361-6382/ab2416},
\newblock \eprint{1903.07543}.

\bibitem{Julie:2019sab}
F.-L. Juli\'e and E.~Berti,
\newblock \emph{{Post-Newtonian dynamics and black hole thermodynamics in Einstein-scalar-Gauss-Bonnet gravity}},
\newblock Phys. Rev. D \textbf{100}(10), 104061 (2019),
\newblock \doi{10.1103/PhysRevD.100.104061},
\newblock [Erratum: Phys.Rev.D 105, 109903 (2022)],
\newblock \eprint{1909.05258}.

\bibitem{Sotiriou:2014pfa}
T.~P. Sotiriou and S.-Y. Zhou,
\newblock \emph{{Black hole hair in generalized scalar-tensor gravity: An explicit example}},
\newblock Phys. Rev. D \textbf{90}, 124063 (2014),
\newblock \doi{10.1103/PhysRevD.90.124063},
\newblock \eprint{1408.1698}.

\bibitem{Sullivan:2019vyi}
A.~Sullivan, N.~Yunes and T.~P. Sotiriou,
\newblock \emph{{Numerical black hole solutions in modified gravity theories: Spherical symmetry case}},
\newblock Phys. Rev. D \textbf{101}(4), 044024 (2020),
\newblock \doi{10.1103/PhysRevD.101.044024},
\newblock \eprint{1903.02624}.

\bibitem{Ayzenberg:2014aka}
D.~Ayzenberg and N.~Yunes,
\newblock \emph{{Slowly-Rotating Black Holes in Einstein-Dilaton-Gauss-Bonnet Gravity: Quadratic Order in Spin Solutions}},
\newblock Phys. Rev. D \textbf{90}, 044066 (2014),
\newblock \doi{10.1103/PhysRevD.90.044066},
\newblock [Erratum: Phys.Rev.D 91, 069905 (2015)],
\newblock \eprint{1405.2133}.

\bibitem{Maselli:2015tta}
A.~Maselli, P.~Pani, L.~Gualtieri and V.~Ferrari,
\newblock \emph{{Rotating black holes in Einstein-Dilaton-Gauss-Bonnet gravity with finite coupling}},
\newblock Phys. Rev. D \textbf{92}(8), 083014 (2015),
\newblock \doi{10.1103/PhysRevD.92.083014},
\newblock \eprint{1507.00680}.

\bibitem{Kleihaus:2011tg}
B.~Kleihaus, J.~Kunz and E.~Radu,
\newblock \emph{{Rotating Black Holes in Dilatonic Einstein-Gauss-Bonnet Theory}},
\newblock Phys. Rev. Lett. \textbf{106}, 151104 (2011),
\newblock \doi{10.1103/PhysRevLett.106.151104},
\newblock \eprint{1101.2868}.

\bibitem{Kleihaus:2014lba}
B.~Kleihaus, J.~Kunz and S.~Mojica,
\newblock \emph{{Quadrupole Moments of Rapidly Rotating Compact Objects in Dilatonic Einstein-Gauss-Bonnet Theory}},
\newblock Phys. Rev. D \textbf{90}(6), 061501 (2014),
\newblock \doi{10.1103/PhysRevD.90.061501},
\newblock \eprint{1407.6884}.

\bibitem{Kleihaus:2015aje}
B.~Kleihaus, J.~Kunz, S.~Mojica and E.~Radu,
\newblock \emph{{Spinning black holes in Einstein\textendash{}Gauss-Bonnet\textendash{}dilaton theory: Nonperturbative solutions}},
\newblock Phys. Rev. D \textbf{93}(4), 044047 (2016),
\newblock \doi{10.1103/PhysRevD.93.044047},
\newblock \eprint{1511.05513}.

\bibitem{Kleihaus:2016dui}
B.~Kleihaus, J.~Kunz, S.~Mojica and M.~Zagermann,
\newblock \emph{{Rapidly Rotating Neutron Stars in Dilatonic Einstein-Gauss-Bonnet Theory}},
\newblock Phys. Rev. D \textbf{93}(6), 064077 (2016),
\newblock \doi{10.1103/PhysRevD.93.064077},
\newblock \eprint{1601.05583}.

\bibitem{Julie:2022huo}
F.-L. Juli\'e, H.~O. Silva, E.~Berti and N.~Yunes,
\newblock \emph{{Black hole sensitivities in Einstein-scalar-Gauss-Bonnet gravity}},
\newblock Phys. Rev. D \textbf{105}(12), 124031 (2022),
\newblock \doi{10.1103/PhysRevD.105.124031},
\newblock \eprint{2202.01329}.

\bibitem{Doneva:2017bvd}
D.~D. Doneva and S.~S. Yazadjiev,
\newblock \emph{{New Gauss-Bonnet Black Holes with Curvature-Induced Scalarization in Extended Scalar-Tensor Theories}},
\newblock Phys. Rev. Lett. \textbf{120}(13), 131103 (2018),
\newblock \doi{10.1103/PhysRevLett.120.131103},
\newblock \eprint{1711.01187}.

\bibitem{Cunha:2019dwb}
P.~V.~P. Cunha, C.~A.~R. Herdeiro and E.~Radu,
\newblock \emph{{Spontaneously Scalarized Kerr Black Holes in Extended Scalar-Tensor\textendash{}Gauss-Bonnet Gravity}},
\newblock Phys. Rev. Lett. \textbf{123}(1), 011101 (2019),
\newblock \doi{10.1103/PhysRevLett.123.011101},
\newblock \eprint{1904.09997}.

\bibitem{Peccei:1977hh}
R.~D. Peccei and H.~R. Quinn,
\newblock \emph{{CP Conservation in the Presence of Instantons}},
\newblock Phys. Rev. Lett. \textbf{38}, 1440 (1977),
\newblock \doi{10.1103/PhysRevLett.38.1440}.

\bibitem{Hui:2016ltb}
L.~Hui, J.~P. Ostriker, S.~Tremaine and E.~Witten,
\newblock \emph{{Ultralight scalars as cosmological dark matter}},
\newblock Phys. Rev. D \textbf{95}(4), 043541 (2017),
\newblock \doi{10.1103/PhysRevD.95.043541},
\newblock \eprint{1610.08297}.

\bibitem{Ferreira:2020fam}
E.~G.~M. Ferreira,
\newblock \emph{{Ultra-light dark matter}},
\newblock Astron. Astrophys. Rev. \textbf{29}(1), 7 (2021),
\newblock \doi{10.1007/s00159-021-00135-6},
\newblock \eprint{2005.03254}.

\bibitem{Arvanitaki:2010sy}
A.~Arvanitaki and S.~Dubovsky,
\newblock \emph{{Exploring the String Axiverse with Precision Black Hole Physics}},
\newblock Phys. Rev. D \textbf{83}, 044026 (2011),
\newblock \doi{10.1103/PhysRevD.83.044026},
\newblock \eprint{1004.3558}.

\bibitem{Kodama:2011zc}
H.~Kodama and H.~Yoshino,
\newblock \emph{{Axiverse and Black Hole}},
\newblock Int. J. Mod. Phys. Conf. Ser. \textbf{7}, 84 (2012),
\newblock \doi{10.1142/S2010194512004199},
\newblock \eprint{1108.1365}.

\bibitem{Boyadjiev:2002en}
T.~L. Boyadjiev and P.~P. Fiziev,
\newblock \emph{{Numerical modeling of charged black holes with massive dilaton}},
\newblock In \emph{{5th International Congress on Mathematical Modelling (V ICMM)}},
\newblock \doi{10.48550/arXiv.gr-qc/0311093} (2002), \eprint{gr-qc/0311093}.

\bibitem{Horne:1992bi}
J.~H. Horne and G.~T. Horowitz,
\newblock \emph{{Black holes coupled to a massive dilaton}},
\newblock Nucl. Phys. B \textbf{399}, 169 (1993),
\newblock \doi{10.1016/0550-3213(93)90621-U},
\newblock \eprint{hep-th/9210012}.

\bibitem{Barsanti:2022vvl}
S.~Barsanti, A.~Maselli, T.~P. Sotiriou and L.~Gualtieri,
\newblock \emph{{Detecting Massive Scalar Fields with Extreme Mass-Ratio Inspirals}},
\newblock Phys. Rev. Lett. \textbf{131}(5), 051401 (2023),
\newblock \doi{10.1103/PhysRevLett.131.051401},
\newblock \eprint{2212.03888}.

\bibitem{Brito:2015oca}
R.~Brito, V.~Cardoso and P.~Pani,
\newblock \emph{{Superradiance}: {New Frontiers in Black Hole Physics}},
\newblock Lect. Notes Phys. \textbf{906}, pp.1 (2015),
\newblock \doi{10.1007/978-3-319-19000-6},
\newblock \eprint{1501.06570}.

\bibitem{Herdeiro:2014goa}
C.~A.~R. Herdeiro and E.~Radu,
\newblock \emph{{Kerr black holes with scalar hair}},
\newblock Phys. Rev. Lett. \textbf{112}, 221101 (2014),
\newblock \doi{10.1103/PhysRevLett.112.221101},
\newblock \eprint{1403.2757}.

\bibitem{Staykov:2018hhc}
K.~V. Staykov, D.~Popchev, D.~D. Doneva and S.~S. Yazadjiev,
\newblock \emph{{Static and slowly rotating neutron stars in scalar\textendash{}tensor theory with self-interacting massive scalar field}},
\newblock Eur. Phys. J. C \textbf{78}(7), 586 (2018),
\newblock \doi{10.1140/epjc/s10052-018-6064-x},
\newblock \eprint{1805.07818}.

\bibitem{Ramazanoglu:2016kul}
F.~M. Ramazano\u{g}lu and F.~Pretorius,
\newblock \emph{{Spontaneous Scalarization with Massive Fields}},
\newblock Phys. Rev. D \textbf{93}(6), 064005 (2016),
\newblock \doi{10.1103/PhysRevD.93.064005},
\newblock \eprint{1601.07475}.

\bibitem{Macedo:2019sem}
C.~F.~B. Macedo, J.~Sakstein, E.~Berti, L.~Gualtieri, H.~O. Silva and T.~P. Sotiriou,
\newblock \emph{{Self-interactions and Spontaneous Black Hole Scalarization}},
\newblock Phys. Rev. D \textbf{99}(10), 104041 (2019),
\newblock \doi{10.1103/PhysRevD.99.104041},
\newblock \eprint{1903.06784}.

\bibitem{Doneva:2019vuh}
D.~D. Doneva, K.~V. Staykov and S.~S. Yazadjiev,
\newblock \emph{{Gauss-Bonnet black holes with a massive scalar field}},
\newblock Phys. Rev. D \textbf{99}(10), 104045 (2019),
\newblock \doi{10.1103/PhysRevD.99.104045},
\newblock \eprint{1903.08119}.

\bibitem{Bakopoulos_2020}
A.~Bakopoulos, P.~Kanti and N.~Pappas,
\newblock \emph{Large and ultracompact gauss-bonnet black holes with a self-interacting scalar field},
\newblock Physical Review D \textbf{101}(8) (2020),
\newblock \doi{10.1103/physrevd.101.084059}.

\bibitem{Zhang:2022kbf}
Y.-P. Zhang, Y.-Q. Wang, S.-W. Wei and Y.-X. Liu,
\newblock \emph{{Dynamics of scalar hair with self-interactions around a Schwarzchild black hole}},
\newblock Phys. Rev. D \textbf{106}(2), 024027 (2022),
\newblock \doi{10.1103/PhysRevD.106.024027},
\newblock \eprint{2203.10341}.

\bibitem{Brito:2017zvb}
R.~Brito, S.~Ghosh, E.~Barausse, E.~Berti, V.~Cardoso, I.~Dvorkin, A.~Klein and P.~Pani,
\newblock \emph{{Gravitational wave searches for ultralight bosons with LIGO and LISA}},
\newblock Phys. Rev. D \textbf{96}(6), 064050 (2017),
\newblock \doi{10.1103/PhysRevD.96.064050},
\newblock \eprint{1706.06311}.

\bibitem{Alsing:2011er}
J.~Alsing, E.~Berti, C.~M. Will and H.~Zaglauer,
\newblock \emph{{Gravitational radiation from compact binary systems in the massive Brans-Dicke theory of gravity}},
\newblock Phys. Rev. D \textbf{85}, 064041 (2012),
\newblock \doi{10.1103/PhysRevD.85.064041},
\newblock \eprint{1112.4903}.

\bibitem{Maselli:2020zgv}
A.~Maselli, N.~Franchini, L.~Gualtieri and T.~P. Sotiriou,
\newblock \emph{{Detecting scalar fields with Extreme Mass Ratio Inspirals}},
\newblock Phys. Rev. Lett. \textbf{125}(14), 141101 (2020),
\newblock \doi{10.1103/PhysRevLett.125.141101},
\newblock \eprint{2004.11895}.

\bibitem{Chen:2024ery}
M.-C. Chen, H.-T. Liu, Q.-Y. Zhang and J.~Zhang,
\newblock \emph{{Probing Massive Fields with Multi-Band Gravitational-Wave Observations}}  (2024),
\newblock \doi{10.48550/arXiv.2405.11583},
\newblock \eprint{2405.11583}.

\bibitem{Yamada:2019zrb}
K.~Yamada, T.~Narikawa and T.~Tanaka,
\newblock \emph{{Testing massive-field modifications of gravity via gravitational waves}},
\newblock PTEP \textbf{2019}(10), 103E01 (2019),
\newblock \doi{10.1093/ptep/ptz103},
\newblock \eprint{1905.11859}.

\bibitem{Shiralilou:2020gah}
B.~Shiralilou, T.~Hinderer, S.~Nissanke, N.~Ortiz and H.~Witek,
\newblock \emph{{Nonlinear curvature effects in gravitational waves from inspiralling black hole binaries}},
\newblock Phys. Rev. D \textbf{103}(12), L121503 (2021),
\newblock \doi{10.1103/PhysRevD.103.L121503},
\newblock \eprint{2012.09162}.

\bibitem{Shiralilou:2021mfl}
B.~Shiralilou, T.~Hinderer, S.~M. Nissanke, N.~Ortiz and H.~Witek,
\newblock \emph{{Post-Newtonian gravitational and scalar waves in scalar-Gauss\textendash{}Bonnet gravity}},
\newblock Class. Quant. Grav. \textbf{39}(3), 035002 (2022),
\newblock \doi{10.1088/1361-6382/ac4196},
\newblock \eprint{2105.13972}.

\bibitem{vanGemeren:2023rhh}
I.~van Gemeren, B.~Shiralilou and T.~Hinderer,
\newblock \emph{{Dipolar tidal effects in gravitational waves from scalarized black hole binary inspirals in quadratic gravity}},
\newblock Phys. Rev. D \textbf{108}(2), 024026 (2023),
\newblock \doi{10.1103/PhysRevD.108.024026},
\newblock [Erratum: Phys.Rev.D 109, 089901 (2024)],
\newblock \eprint{2302.08480}.

\bibitem{Gross1987}
D.~J. Gross and J.~H. Sloan,
\newblock \emph{The quartic effective action for the heterotic string},
\newblock Nuclear Physics B \textbf{291}, 41 (1987),
\newblock \doi{https://doi.org/10.1016/0550-3213(87)90465-2}.

\bibitem{Zwiebach1985}
B.~Zwiebach,
\newblock \emph{{Curvature Squared Terms and String Theories}},
\newblock Phys. Lett. B \textbf{156}, 315 (1985),
\newblock \doi{10.1016/0370-2693(85)91616-8}.

\bibitem{Moura:2006pz}
F.~Moura and R.~Schiappa,
\newblock \emph{{Higher-derivative corrected black holes: Perturbative stability and absorption cross-section in heterotic string theory}},
\newblock Class. Quant. Grav. \textbf{24}, 361 (2007),
\newblock \doi{10.1088/0264-9381/24/2/006},
\newblock \eprint{hep-th/0605001}.

\bibitem{Saffer:2021gak}
A.~Saffer and K.~Yagi,
\newblock \emph{{Tidal deformabilities of neutron stars in scalar-Gauss-Bonnet gravity and their applications to multimessenger tests of gravity}},
\newblock Phys. Rev. D \textbf{104}(12), 124052 (2021),
\newblock \doi{10.1103/PhysRevD.104.124052},
\newblock \eprint{2110.02997}.

\bibitem{Bertotti:2003rm}
B.~Bertotti, L.~Iess and P.~Tortora,
\newblock \emph{{A test of general relativity using radio links with the Cassini spacecraft}},
\newblock Nature \textbf{425}, 374 (2003),
\newblock \doi{10.1038/nature01997}.

\bibitem{Yagi:2012gp}
K.~Yagi,
\newblock \emph{{A New constraint on scalar Gauss-Bonnet gravity and a possible explanation for the excess of the orbital decay rate in a low-mass X-ray binary}},
\newblock Phys. Rev. D \textbf{86}, 081504 (2012),
\newblock \doi{10.1103/PhysRevD.86.081504},
\newblock \eprint{1204.4524}.

\bibitem{Yagi:2015oca}
K.~Yagi, L.~C. Stein and N.~Yunes,
\newblock \emph{{Challenging the Presence of Scalar Charge and Dipolar Radiation in Binary Pulsars}},
\newblock Phys. Rev. D \textbf{93}(2), 024010 (2016),
\newblock \doi{10.1103/PhysRevD.93.024010},
\newblock \eprint{1510.02152}.

\bibitem{Pani:2011xm}
P.~Pani, E.~Berti, V.~Cardoso and J.~Read,
\newblock \emph{{Compact stars in alternative theories of gravity. Einstein-Dilaton-Gauss-Bonnet gravity}},
\newblock Phys. Rev. D \textbf{84}, 104035 (2011),
\newblock \doi{10.1103/PhysRevD.84.104035},
\newblock \eprint{1109.0928}.

\bibitem{East:2022rqi}
W.~E. East and F.~Pretorius,
\newblock \emph{{Binary neutron star mergers in Einstein-scalar-Gauss-Bonnet gravity}},
\newblock Phys. Rev. D \textbf{106}(10), 104055 (2022),
\newblock \doi{10.1103/PhysRevD.106.104055},
\newblock \eprint{2208.09488}.

\bibitem{Yordanov:2024lfk}
P.~Y. Yordanov, K.~V. Staykov, S.~S. Yazadjiev and D.~D. Doneva,
\newblock \emph{{The power of binary pulsars in testing Gauss-Bonnet gravity}},
\newblock Astron. Astrophys. \textbf{687}, A17 (2024),
\newblock \doi{10.1051/0004-6361/202449679},
\newblock \eprint{2402.06305}.

\bibitem{Xu:2021kfh}
R.~Xu, Y.~Gao and L.~Shao,
\newblock \emph{{Neutron stars in massive scalar-Gauss-Bonnet gravity: Spherical structure and time-independent perturbations}},
\newblock Phys. Rev. D \textbf{105}(2), 024003 (2022),
\newblock \doi{10.1103/PhysRevD.105.024003},
\newblock \eprint{2111.06561}.

\bibitem{Perkins:2021mhb}
S.~E. Perkins, R.~Nair, H.~O. Silva and N.~Yunes,
\newblock \emph{{Improved gravitational-wave constraints on higher-order curvature theories of gravity}},
\newblock Phys. Rev. D \textbf{104}(2), 024060 (2021),
\newblock \doi{10.1103/PhysRevD.104.024060},
\newblock \eprint{2104.11189}.

\bibitem{Lyu:2022gdr}
Z.~Lyu, N.~Jiang and K.~Yagi,
\newblock \emph{{Constraints on Einstein-dilation-Gauss-Bonnet gravity from black hole-neutron star gravitational wave events}},
\newblock Phys. Rev. D \textbf{105}(6), 064001 (2022),
\newblock \doi{10.1103/PhysRevD.105.064001},
\newblock [Erratum: Phys.Rev.D 106, 069901 (2022), Erratum: Phys.Rev.D 106, 069901 (2022)],
\newblock \eprint{2201.02543}.

\bibitem{Wang:2021jfc}
H.-T. Wang, S.-P. Tang, P.-C. Li, M.-Z. Han and Y.-Z. Fan,
\newblock \emph{{Tight constraints on Einstein-dilation-Gauss-Bonnet gravity from GW190412 and GW190814}},
\newblock Phys. Rev. D \textbf{104}(2), 024015 (2021),
\newblock \doi{10.1103/PhysRevD.104.024015},
\newblock \eprint{2104.07590}.

\bibitem{Creci:2020mfg}
G.~Creci, S.~Vandoren and H.~Witek,
\newblock \emph{{Evolution of black hole shadows from superradiance}},
\newblock Phys. Rev. D \textbf{101}(12), 124051 (2020),
\newblock \doi{10.1103/PhysRevD.101.124051},
\newblock \eprint{2004.05178}.

\bibitem{Yunes:2011we}
N.~Yunes and L.~C. Stein,
\newblock \emph{{Non-Spinning Black Holes in Alternative Theories of Gravity}},
\newblock Phys. Rev. D \textbf{83}, 104002 (2011),
\newblock \doi{10.1103/PhysRevD.83.104002},
\newblock \eprint{1101.2921}.

\bibitem{Heydari-Fard:2020iiu}
M.~Heydari-Fard and H.~R. Sepangi,
\newblock \emph{{Thin accretion disk signatures of scalarized black holes in Einstein-scalar-Gauss-Bonnet gravity}},
\newblock Phys. Lett. B \textbf{816}, 136276 (2021),
\newblock \doi{10.1016/j.physletb.2021.136276},
\newblock \eprint{2009.13748}.

\end{thebibliography}
\end{document}